\documentclass[twoside,11pt]{article}
\usepackage{a4wide}
\usepackage{cite,latexsym,amsfonts,amssymb,exscale,epsfig,bbm}
\usepackage[centertags,sumlimits,intlimits,namelimits,reqno]{amsmath}
\usepackage{amsthm}
\theoremstyle{definition}
\newtheorem{theorem}{Theorem}[section]
\newtheorem{lemma}[theorem]{Lemma}
\newtheorem{corollary}[theorem]{Corollary}
\newtheorem{proposition}[theorem]{Proposition}
\newtheorem{definition}[theorem]{Definition}
\newtheorem{remark}[theorem]{Remark}
\newtheorem{example}[theorem]{Example}
\newenvironment{proofof}[1]{\paragraph{Proof of #1:}}{\qed}
\newcommand{\ontop}[2]{\genfrac{}{}{0pt}{2}{\scriptstyle #1}{\scriptstyle #2}}
\def\nn{\notag}

\def\address#1{\date{{\sl #1}\\\ \\\theversion}\gdef\date##1{}}%
\def\version#1{\gdef\theversion{#1}}%

\def\dopreprint{\hfill{\small\thepreprint}\\}%
\def\preprint#1{\def\thepreprint{#1}}%
\def\thepreprint#1{}%

\def\sym#1{{\mathcal #1}}
\def\emph#1{{\sl #1\/}}

\let\phi=\varphi
\let\theta=\vartheta
\let\epsilon=\varepsilon

\def\SO{{SO}}

\def\SU{{SU}}

\def\O{{O}}
\def\U{{U}}

\def\dim{\mathop{\rm dim}\nolimits}

\def\Hom{\mathop{\rm Hom}\nolimits}
\def\Aut{\mathop{\rm Aut}\nolimits}

\def\Rep{\tilde{\sym{R}}}%
\def\Irrep{\sym{R}}%
\def\Calg{C_{\rm alg}}%
\def\del{\partial}
\def\Inv{\mathop{\rm Inv}\nolimits}
\def\RP{{\mathbbm{RP}}}
\def\CP{{\mathbbm{CP}}}
\def\C{{\mathbbm C}}
\def\N{{\mathbbm N}}
\def\R{{\mathbbm R}}
\def\Z{{\mathbbm Z}}

\def\openone{\mathbbm{1}}%
\def\pacs#1{\noindent PACS: #1\par}%
\def\keywords#1{\noindent key words: #1\par}%
\def\acknowledgements{\section*{Acknowledgements}}%

\def\mycaption#1#2{%
  \begin{quote}
  \caption{\label{#1}#2}
  \end{quote}}

\let\hat=\widehat
\let\tilde=\widetilde

\def\ie{{\sl i.e.\/}}
\def\eg{{\sl e.g.\/}}

\def\etc{{\sl etc.\/}}
\def\cf{{\sl cf.\/}}

\newcounter{mathletter}%
\newcommand{\bmathletter}{%
  \refstepcounter{equation}%
  \setcounter{mathletter}{\value{equation}}%
  \setcounter{equation}{0}%
    \renewcommand{\theequation}{%
      \mbox{\thesection.\arabic{mathletter}\alph{equation}}}}%
\newcommand{\emathletter}{\setcounter{equation}{\value{mathletter}}}%
\newenvironment{mathletters}{\bmathletter}{\emathletter}%

\makeatletter

\newfont{\@aidxte}{cmsy10}
\newfont{\@aidxel}{cmsy10 scaled 1095}
\newfont{\@aidxtw}{cmsy10 scaled 1200}
\newlength\@aidxtexvi
\newlength\@aidxtexvii
\newlength\@aidxelxvi
\newlength\@aidxelxvii
\newlength\@aidxtwxvi
\newlength\@aidxtwxvii
\newcommand{\alignidx}[1]{%
  \@aidxtexvi=\fontdimen16\@aidxte
  \@aidxtexvii=\fontdimen17\@aidxte
  \@aidxelxvi=\fontdimen16\@aidxel
  \@aidxelxvii=\fontdimen17\@aidxel
  \@aidxtwxvi=\fontdimen16\@aidxtw
  \@aidxtwxvii=\fontdimen17\@aidxtw
    {\mbox{$%
    \fontdimen16\@aidxte=2.9pt
    \fontdimen17\@aidxte=2.9pt
    \fontdimen16\@aidxel=3.1pt
    \fontdimen17\@aidxel=3.1pt
    \fontdimen16\@aidxtw=3.3pt
    \fontdimen17\@aidxtw=3.3pt
    #1$}}%
    \fontdimen16\@aidxte=\@aidxtexvi
    \fontdimen17\@aidxte=\@aidxtexvii
    \fontdimen16\@aidxel=\@aidxelxvi
    \fontdimen17\@aidxel=\@aidxelxvii
    \fontdimen16\@aidxtw=\@aidxtwxvi
    \fontdimen17\@aidxtw=\@aidxtwxvii}

\renewcommand{\theequation}{\thesection.\arabic{equation}}
\@addtoreset{equation}{section}

\makeatother

\newenvironment{myenumerate}{%
  \begin{enumerate}
  \setlength{\partopsep}{0pt}
  \setlength{\parskip}{0pt}}{\end{enumerate}}

\version{1 April 2003}%
\preprint{DAMTP-2002-49}

\begin{document}
%

\title{\dopreprint Exact duality transformations for sigma models\\ and gauge theories}
\author{Hendryk Pfeiffer\thanks{e-mail: H.Pfeiffer@damtp.cam.ac.uk}}
\address{Emmanuel College, St.~Andrew's Street, Cambridge CB2 3AP\\
         and\\
         Department of Applied Mathematics and Theoretical Physics,\\
         Wilberforce Road, Cambridge CB3 0WA\\
         England, UK}
\date{\version}
\maketitle

%
\begin{abstract}
%

  We present an exact duality transformation in the framework of
  Statistical Mechanics for various lattice models with non-Abelian
  global or local symmetries. The transformation applies to sigma
  models with variables in a compact Lie group $G$ with global
  $G\times G$-symmetry (the chiral model) and with variables in coset
  spaces $G/H$ and a global $G$-symmetry (for example, the non-linear
  $O(N)$ or $\RP^N$ models) in any dimension $d\geq 1$. It is also
  available for lattice gauge theories with local gauge symmetry in
  dimensions $d\geq 2$ and for the models obtained from minimally
  coupling a sigma model of the type mentioned above to a gauge
  theory. The duality transformation maps the strong coupling regime
  of the original model to the weak coupling regime of the dual
  model. Transformations are available for the partition function, for
  expectation values of fundamental variables (correlators and
  generalized Wilson loops) and for expectation values in the dual
  model which correspond in the original formulation to certain ratios
  of partition functions (free energies of dislocations, vortices or
  monopoles). Whereas the original models are formulated in terms of
  compact Lie groups $G$ and $H$, coset spaces $G/H$ and integrals
  over them, the configurations of the dual model are given in terms
  of representations and intertwiners of $G$ and $H$. They are spin
  networks and spin foams. The partition function of the dual model
  describes the group theoretic aspects of the strong coupling
  expansion in a closed form.
\end{abstract}


%
\section{Introduction}
%

The most prominent example of an exact duality transformation in
Statistical Mechanics is the transformation for the two-dimensional
Ising model~\cite{KrWa41}. It is an exact transformation which changes
the variables of the full partition function of the model and maps the
high temperature regime of the original model to the low temperature
regime of the dual model and conversely (for the Ising model, the
original and the dual model coincide).

In the following, we make use of the correspondence of Quantum Field
Theory in the Euclidean (imaginary time) formulation in $d$ space
dimensions plus time with Equilibrium Statistical Mechanics in $d+1$
dimensions and often use the words \emph{path integral}, \emph{action}
and \emph{coupling} for \emph{partition function}, \emph{energy} and
\emph{temperature}, respectively.

The duality transformation of the Ising model was subsequently
generalized to more general lattice systems with $\Z_2$
symmetries~\cite{We71}, namely systems in $d$-dimensions whose
variables are $\Z_2$-valued $k$-forms, $0\leq k\leq d$, \ie\ spin
models with global $\Z_2$ symmetry, pure lattice gauge theories with
local $\Z_2$ gauge symmetry, theories for $\Z_2$-valued antisymmetric
tensor fields and so on, and to their counterparts with
$\U(1)$-symmetries~\cite{Sa77}, in particular to the $XY$-model and
pure $\U(1)$ gauge theory on the lattice~\cite{Sa77,Pe78}. For lattice
models with Abelian symmetries, there exists an essentially complete
picture~\cite{Sa80}, and the systems to which the duality
transformation applies, include even some Higgs models~\cite{EiSa78},
namely $\U(1)$-lattice gauge theory minimally coupled to a
$\U(1)$-valued scalar field, \ie\ a Higgs field with frozen radial
component.

All these examples of the Abelian duality transformation have some
features in common. They map the strong coupling regime of the
original model to the weak coupling regime of the dual model. This is
a consequence of the change of variables employed in the
transformation which essentially involves the Fourier decomposition of
the interaction terms $e^{-S}$ for some action $S$. For example, this
replaces $\U(1)$-variables by integers $\Z$ and maps Boltzmann weights
with narrow peaks to weights with wide peaks and conversely. The
structure of the dual model can be sketched as follows. If, say in a
sigma model, the variables are originally associated with the points of
the lattice and the interaction terms with the bonds, then the
variables of the dual model which are introduced by the Fourier
expansion, are located at the bonds. In a second step one removes the
old variables by performing the relevant sums or integrals which
yields additional Boltzmann weights, often constraints, for each
point.

As a consequence of the strong-weak nature of the duality
transformation, the dual partition function contains essential
information on the strong coupling expansion of the original model. In
fact, if one understands how the Fourier coefficients of the Boltzmann
weight depend on the coupling, the summands of the dual partition
function are precisely the terms of the strong coupling expansion and
can be sorted by the magnitude of their contribution at strong
coupling.

A systematic generalization of these transformations to systems with
non-Abelian symmetries proved to be difficult. The calculation of
strong coupling expansions of pure non-Abelian lattice gauge theory
(see, for example~\cite{DrZu83}) already exhibits some features of the
dual model which one wishes to construct. Fourier expansion is
generalized to character decomposition, and the dual variables are
irreducible representations of the symmetry group, generalizing the
wave numbers of the Fourier series. The main technical difficulties
are firstly to solve the integrals over the variables of the original
model in a systematic way, and secondly to disentangle the lattice
combinatorics in order to make the structure of the dual model
transparent. Both problems can be overcome if one deals with the
non-Abelian group variables at a sufficiently abstract level and if
one uses an efficient diagrammatic notation.

The first examples of non-Abelian generalizations were an explicit
calculation for pure $\SU(2)$ lattice gauge theory in $d=3$
dimensions~\cite{AnCh93} and, much less obviously, the equivalence of
lattice $BF$-theory (similar to pure lattice Yang--Mills theory, but
with $\delta$-functions as the Boltzmann weights) to certain
topological state sum models~\cite{Oo92}. This correspondence was
developed in a non-perturbative approach to quantum gravity. For
review articles, see, for example~\cite{Ba99,Or01a}. The approach to
quantum gravity by quantizing a discrete version of the gauge theory
formulation of general relativity, has lead to the definition of spin
foams~\cite{Ba98a,Re94,ReRo97}. A spin foam is an abstract
two-complex, consisting of vertices, edges and faces whose faces are
labelled with irreducible representations of some symmetry group while
the edges are labelled with compatible intertwiners. Spin foams can be
understood as a higher dimensional analogue of spin networks. A spin
network is a graph whose edges are labelled with representations and
whose vertices are labelled with compatible intertwiners (precise
definitions of spin networks and spin foams are given below in
Section~\ref{sect_spinnet}).

Spin foams provide the appropriate language for a generalization of
the exact duality transformation to pure non-Abelian lattice gauge
theory in arbitrary dimension $d\geq 2$ whose gauge group is a generic
compact Lie group $G$. See~\cite{OePf01,PfOe02} for lattice gauge
theory on hypercubic lattices and~\cite{Pf01,Oe02} for the
generalization to more general lattices and quantum groups rather than
Lie groups.

In this article, we extend the non-Abelian duality transformation to a
large class of sigma models\footnote{The author is grateful to Alan
Sokal who suggested to study this generalization.} whose variables
take values in $G$ or $G/H$, $G$ a compact Lie group and $H$ a Lie
subgroup, and which have certain global or local symmetries. This
includes, for example, the chiral, the $\O(N)$ and $\RP^N$ models,
and the models that are obtained from minimally coupling such a sigma
model to a non-Abelian lattice gauge theory, for example, some
generalized Higgs models with frozen radial degree of freedom.

The duality transformation retains its key properties, namely that it
provides a strong-weak relation, that it yields a closed form of the
strong coupling expansion of the original model and that it maps
expectation values of the dual model to ratios of partition functions
(free energies of dislocations, vortices or monopoles) in the original
formulation and conversely. It therefore relates the fundamental
variables of one formulation with some topological defects (collective
properties) in the other.

The transformation maps the original model which is formulated in
terms of compact Lie groups $G$ and $H$, functions on $G$ and
integrals over $G$ or $G/H$, to the dual model which is given in terms
of the irreducible representations and intertwiners of $G$ and
$H$. The transformation can be understood as a particular application
of a Tannaka--Kre\v\i n like duality relating groups to their
representation categories. That these categories will appear in the
dual formulation, had already been proposed in~\cite{GrSc98}. The dual
model can be formulated using merely the language of category
theory. In the simplest case, it uses the category of
finite-dimensional representations of the symmetry group $G$. This can
be extended to more general categories that do not arise as the
representation categories of compact Lie groups. The generalization of
lattice gauge theory to quantum groups~\cite{Pf01,Oe02} is one
example. For more details on the relation of groups and quantum groups
with certain tensor categories, see, for example~\cite{Ma95a}. In this
article, we do not explicitly use the language of category theory, but
rather present diagrams in addition to the explicit formulas so that
one can easily infer the categorial formulation from these diagrams.

While the configurations of the model dual to lattice gauge theory are
spin foams~\cite{OePf01}, one obtains spin networks as the
configurations of the model dual to a sigma model. We thus call the
dual models \emph{spin foam models} and \emph{spin network models},
respectively. As the notions of spin networks and spin foams have been
developed in an approach to quantum gravity, but might not be familiar
to the reader working on Statistical Mechanics, we try to make this
article self-contained and therefore review all relevant definitions
and also some background material on the representation theory of
compact Lie groups.

The present article is organized as follows. In
Section~\ref{sect_groups}, we summarize some background material on
the representation theory of compact Lie groups and introduce a
convenient diagrammatical notation. In Section~\ref{sect_definitions},
we present our notation for the lattices we use, namely graphs and
abstract two-complexes, and we recall the definitions of spin networks
and spin foams. In Section~\ref{sect_chiralmodel}, we present the
duality transformation for the lattice chiral model with symmetry
group $G$. This is generalized in Section~\ref{sect_nonlinearmodel} to
the non-linear sigma model with variables in coset spaces $G/H$ and in
Section~\ref{sect_higgs} to the non-linear sigma model for $G/H$
coupled to a lattice gauge theory with gauge group $G$. We conclude in
Section~\ref{sect_discussion} where we discuss applications,
directions for future research and open questions.

%
\section{Mathematical background}
%
\label{sect_groups}

In this section, we review some basic concepts and results from the
representation theory of compact Lie groups. The material presented
here is largely textbook knowledge, see, for
example~\cite{CaSe95,ViKl93} where most of the proofs can be
found. The purely algebraic evaluation of the group integrals was
first given in~\cite{OePf01}, our diagrammatic language
follows~\cite{Pf01,Oe02}.

\subsection{Representation functions}

Let $G$ be a compact Lie group. This notion includes in particular any
finite group (with the discrete topology). We denote
finite-dimensional complex vector spaces on which $G$ is represented
by $V_\rho$ and by $\rho\colon G\to\Aut V_\rho$ the corresponding
group homomorphisms. Since each finite-dimensional complex
representation of $G$ is equivalent to a unitary representation, we
select a set $\Rep_G$ containing one unitary representation of $G$ for
each equivalence class of finite-dimensional representations. The
tensor product, the direct sum and taking the dual are supposed to be
closed operations on this set. This amounts to a particular choice of
representation isomorphisms $\rho_1\otimes\rho_2\leftrightarrow\rho_3$
\etc, $\rho_j\in\Rep_G$, which is implicit in our formulas. We
furthermore denote by $\Irrep_G\subseteq\Rep_G$ the subset of irreducible
representations.

For a representation $\rho\in\Rep_G$, the dual representation is denoted
by $\rho^\ast$, and the dual vector space of $V_\rho$ by
$V_\rho^\ast$. The dual representation is given by $\rho^\ast\colon
G\mapsto \Aut V_\rho^\ast$, where
\begin{equation}
\label{eq_dualrep}
  \rho^\ast(g)\colon V_\rho^\ast\to V_\rho^\ast,\quad
    \eta\mapsto \eta\circ\rho(g^{-1}),
\end{equation}
\ie\ $(\rho^\ast(g)\eta)(v)=\eta(\rho(g^{-1})v)$ for all $v\in
V_\rho$. There exists a one-dimensional `trivial' representation of
$G$ which is isomorphic to $\C$.

For the unitary representations $V_\rho$, $\rho\in\Rep_G$, we have
standard (sesquilinear) scalar products $\left<\cdot;\cdot\right>$ and
orthonormal bases ${\{e_j\}}_j$. Therefore, we can define a bijective
antilinear map $\ast\colon V_\rho\to V_\rho^\ast$ induced by the
scalar product,
\begin{equation}
  \ast(v):=(w\mapsto\left<v;w\right>),\qquad v\in V_\rho,
\end{equation}
and construct the dual bases ${\{\eta^j\}}_j$ by
$\eta^j:=\ast(e_j)$. Identifying $(V_\rho^\ast)^\ast=V_\rho$, this
yields $\left<e_j;e_k\right>=\eta^j(e_k)=\delta_{jk}$ and furthermore
induces a scalar product on $V_\rho^\ast$, namely
$\bigl<\eta^j;\eta^k\bigr>=\eta^k(e_j)$, $1\leq j,k\leq\dim V_\rho$.

The matrix elements of the representation matrices $\rho(g)$ define
complex valued functions,
\begin{equation}
  t_{jk}^{(\rho)}\colon G\to\C,\qquad g\mapsto t_{jk}^{(\rho)}(g)
  :=\eta^j(\rho(g)e_k)={(\rho(g))}_{jk},
\end{equation}
where $\rho\in\Rep_G$, $1\leq j,k\leq\dim V_\rho$. They are called
\emph{representation functions} of $G$ and form a commutative and
associative unital algebra over $\C$,
\begin{equation}
  \Calg(G) := \{\,t_{jk}^{(\rho)}\colon\quad
    \rho\in\Rep_G, 1\leq j,k\leq\dim V_\rho\,\},
\end{equation}
whose product is given by the matrix elements of the tensor product of
representations,
\begin{eqnarray}
\label{eq_operationprod}
  (t_{jk}^{(\rho)}\cdot t_{\ell m}^{(\sigma)})(g)
  := t_{j\ell,km}^{(\rho\otimes\sigma)}(g),
\end{eqnarray}%
where $\rho,\sigma\in\Rep_G$, $1\leq j,k\leq\dim V_\rho$ and
$1\leq\ell,m\leq\dim V_\sigma$.

We find the following expressions involving the group unit $e\in G$,
\begin{equation}
  t_{jk}^{(\rho)}(e) = \delta_{jk},
\end{equation}
products of group elements,
\begin{equation}
\label{eq_copro}
  t_{jk}^{(\rho)}(g\cdot h)=\sum_{\ell=1}^{\dim V_\rho} 
    t_{j\ell}^{(\rho)}(g)\cdot t_{\ell k}^{(\rho)}(h),
\end{equation}
and inverse group elements,
\begin{equation}
  t_{jk}^{(\rho)}(g^{-1})={(\rho(g)^{-1})}_{jk}
    =\overline{{(\rho(g))}_{kj}}=\overline{t_{kj}^{(\rho)}(g)},
\end{equation}
as well as,
\begin{equation}
\label{eq_inverse_dual}
  t_{jk}^{(\rho)}(g^{-1})= \eta^j({\rho(g)}^{-1}e_k) 
    = (\rho^\ast(g)\eta^j)(e_k)
    = \bigl<\eta^k;\rho^\ast(g)\eta^j\bigr> 
    = t_{kj}^{(\rho^\ast)}(g),
\end{equation}
so that for unitary representations, the dual representation is just
the conjugate one. The bar denotes complex conjugation.

\subsection{Peter--Weyl decomposition and theorem}

The structure of the algebra $\Calg(G)$ can be understood if
$\Calg(G)$ is considered as a representation of $G\times G$ by
combined left and right translation of the function argument,
\begin{equation}
  (G\times G)\times\Calg(G)\to\Calg(G),\quad
    ((g_1,g_2),f)\mapsto (h\mapsto f(\alignidx{g_1^{-1}hg_2})).
\end{equation}
It can then be decomposed into its irreducible components as a
representation of $G\times G$.

\begin{theorem}[Peter--Weyl decomposition]
\label{thm_peterweyl}
Let $G$ be a compact Lie group.
\begin{myenumerate}
\item 
  There is an isomorphism
\begin{equation}
\label{eq_structure_calg}
  \Calg(G)\cong
    \bigoplus_{\rho\in\Irrep_G}(V_\rho\otimes V_\rho^\ast),
\end{equation}
  of representations of $G\times G$. Here the direct sum runs over the
  equivalence classes of finite-dimensional irreducible
  representations of $G$. The direct summands $V_\rho\otimes
  V_\rho^\ast$ are irreducible as representations of $G\times G$.
\item
  The direct sum in~\eqref{eq_structure_calg} is orthogonal with
  respect to the $L^2$-scalar product on $\Calg(G)$ which is formed
  using the Haar measure of $G$ on the left hand side, and using the
  standard scalar products on the right hand side,
\begin{equation}
\label{eq_l2measure}
  {\bigl<t_{jk}^{(\rho)};t_{\ell m}^{(\sigma)}\bigr>}_{L^2}
    := \int_G\overline{t_{jk}^{(\rho)}(g)}\cdot t_{\ell m}^{(\sigma)}(g)\,dg
    = \frac{1}{\dim V_{\rho}}\delta_{\rho\sigma}\delta_{j\ell}\delta_{km},
\end{equation}
  where $\rho,\sigma\in\Irrep_G$ are irreducible. The Haar measure is
  denoted by $\int_G$ and normalized so that $\int_G\,dg=1$.
\end{myenumerate}
\end{theorem}

If $G$ is finite, the Haar measure coincides with the normalized
summation over all group elements. The
decomposition~\eqref{eq_structure_calg} directly corresponds to our
notation of the representation functions $t^{(\rho)}_{jk}$ for
irreducible $\rho\in\Irrep_G$.

\begin{corollary}
Each representation function $f\in\Calg(G)$ can be decomposed
according to~\eqref{eq_structure_calg},
\begin{equation}
\label{eq_peterweyl_series}
  f(g) = \sum_{\rho\in\Irrep_G}\sum_{j,k=1}^{\dim V_\rho}
    \hat f^{(\rho)}_{jk}\,t^{(\rho)}_{jk}(g),\quad\mbox{where}\quad
  \hat f^{(\rho)}_{jk}=\dim V_\rho\,\int_G\overline{t^{(\rho)}_{jk}(g)}f(g)\,dg.  
\end{equation}
\end{corollary}

For any algebraic $f\in\Calg(G)$, all except finitely many
coefficients $f^{(\rho)}_{jk}$ are zero. The analytical aspects of
$\Calg(G)$ are given by the Peter--Weyl theorem.

\begin{theorem}[Peter--Weyl Theorem]
Let $G$ be a compact Lie group. Then $\Calg(G)$ is dense in $L^2(G)$
with respect to the $L^2$-norm.
\end{theorem}

We use the Peter--Weyl theorem in order to complete $\Calg(G)$ with
respect to the $L^2$-norm to $L^2(G)$. Functions $f\in L^2(G)$ then
correspond to square summable series
in~\eqref{eq_peterweyl_series}. These series are invariant under a
reordering of summands, and their limits commute with group
integrations. We make use of these invariances in the duality
transformation. If $G$ is a finite group, $\Calg(G)$ is a
finite-dimensional vector space so that the corresponding results hold
trivially.

We can summarize these ideas and state that the algebraic structure of
$\Calg(G)$ is sufficient to determine the structure of the larger
function space $L^2(G)$.

\subsection{Character decomposition}

The \emph{characters} of $G$ are the algebraic class functions, \ie\
those functions $f\in\Calg(G)$ that satisfy $f(hgh^{-1})=f(g)$ for all
$g,h\in G$.

\begin{proposition}
For class functions $f\in\Calg(G)$, the Peter--Weyl
decomposition~\eqref{eq_peterweyl_series} specializes to the
\emph{character decomposition}
\begin{equation}
\label{eq_charexp}
  f(g) = \sum_{\rho\in\Irrep_G}\hat f_\rho\,\chi^{(\rho)}(g),
    \qquad\mbox{where}\qquad 
  \hat f_\rho = \int_G\overline{\chi^{(\rho)}(g)}f(g)\,dg.
\end{equation}
Here
\begin{equation}
  \chi^{(\rho)}:=\sum_{j=1}^{\dim V_\rho}t^{(\rho)}_{jj}
\end{equation}
denotes the character of the representation $\rho\in\Rep_G$. For
irreducible $\rho,\sigma\in\Irrep_G$, the orthogonality
relation~\eqref{eq_l2measure} implies,
\begin{equation}
  \bigl<\chi^{(\rho)};\chi^{(\sigma)}\bigr>_{L^2}
  = \int_G\overline{\chi^{(\rho)}(g)}\chi^{(\sigma)}(g)\,dg
  = \delta_{\rho\sigma}.
\end{equation}
\end{proposition}

\subsection{Algebraic evaluation of group integrals}
\label{sect_algintegral}

For the duality transformation, it is important to understand the Haar
measure of $G$ in the picture of the Peter--Weyl
decomposition~\eqref{eq_structure_calg}. First we decompose a generic
representation function into representation functions of irreducible
representations.

\begin{proposition}
\label{prop_fulldecompose}
Let $G$ be a compact Lie group and $\rho\in\Rep_G$ be a
finite-dimensional unitary representation of $G$ with the complete 
decomposition
\begin{equation}
\label{eq_repdecompose}
  V_\rho\cong\bigoplus_{j=1}^k V_{\tau_j},\qquad \tau_j\in\Irrep_G, k\in\N,
\end{equation}
into irreducible components $\tau_j$. Let $P^{(j)}\colon V_\rho\to
V_{\tau_j}\subseteq V_\rho$ be the $G$-invariant orthogonal projectors
associated with the above decomposition. Then
\begin{equation}
\label{eq_fulldecompose}
  t^{(\rho)}_{mn}(g)=\sum_{j=1}^k\sum_{p,q=1}^{\dim V_{\tau_j}}
    \overline{P^{(j)}_{pm}}\,t^{(\tau_j)}_{pq}(g)\,P^{(j)}_{qn},
\end{equation}
where $P^{(j)}_{qn}=\bigl<w_q^{(j)};v_n\bigr>$. Here ${\{v_i\}}_i$
denotes an orthonormal basis of $V_\rho$ and ${\{w^{(j)}_i\}}_i$ an
orthonormal basis of $V_{\tau_j}\subseteq V_\rho$.
\end{proposition}

\begin{proof}
The representation function is Peter--Weyl decomposed by inserting
$\openone=\sum_{j=1}^k P^{(j)}$ twice into the right hand side of
$t_{mn}^{(\rho)}(g)=\left<v_m;\rho(g)v_n\right>$. We use
$G$-invariance $[P^{(j)},\rho(g)]=0$ and transversality
$P^{(i)}P^{(j)}=\delta_{ij}P^{(j)}$ in order to obtain
\begin{equation}
\label{eq_haardecomp}
  t^{(\rho)}_{mn}(g)=\sum_{j=1}^k\bigl<v_m;P^{(j)}\rho(g)P^{(j)}v_n\bigr>.
\end{equation}
Here $\rho(g)P^{(j)}=\tau_j(g)P^{(j)}$ and 
\begin{equation}
\label{eq_haarproj}
  P^{(j)}=\sum_{p=1}^{\dim V_{\tau_j}}w^{(j)}_p\cdot\theta^{(j)p},
\end{equation}
where ${\{\theta^{(j)i}\}}_i$ denotes a basis dual to
${\{w^{(j)}_i\}}_i$. Inserting~\eqref{eq_haarproj}
into~\eqref{eq_haardecomp}, we obtain~\eqref{eq_fulldecompose}.
\end{proof}

For representation functions of an irreducible representation
$\rho\in\Irrep_G$, the Haar measure is
\begin{equation}
  \int_G t^{(\rho)}_{jk}(g)\,dg = \left\{
    \begin{matrix}
      1,&\mbox{if $\rho$ is trivial},\\
      0,&\mbox{otherwise},
    \end{matrix}\right.
\end{equation}
as a consequence of its left-right translation invariance. This can be
applied to~\eqref{eq_fulldecompose} in order to derive an entirely
algebraic expression for the Haar measure.

\begin{corollary}
\label{cor_haaralg}
Let $G$ be a compact Lie group and $\rho\in\Rep_G$ be a
finite-dimensional unitary representation of $G$ with the
decomposition~\eqref{eq_repdecompose}.  Assume that precisely the
first $\ell$ components $\tau_1,\ldots,\tau_\ell$, $0\leq\ell\leq k$,
are equivalent to the trivial representation. Then the Haar measure of
a representation function $t_{mn}^{(\rho)}$, $1\leq m,n\leq\dim
V_\rho$, is given by
\begin{equation}
\label{eq_haaralg}
  \int_G t_{mn}^{(\rho)}(g)\,dg 
    = \sum_{j=1}^\ell\alignidx{\overline{P^{(j)}_m}P^{(j)}_n}.
\end{equation}
Here we have omitted the vector indices corresponding to the
one-dimensional representations.
\end{corollary}

In our calculations, we will refer to Corollary~\ref{cor_haaralg} in
a context in which the integrand is a product of representation
functions of irreducible representations. This motivates the following
definition. 

\begin{definition}
\label{def_haarinter}
Let $G$ be a compact Lie group and $\rho_1,\ldots,\rho_r\in\Irrep_G$,
$r\in\N$, be finite-dimensional irreducible representations of
$G$. The \emph{Haar intertwiner},
\begin{mathletters}
\label{eq_haarinter}
\begin{equation}
  T\colon\bigotimes_{\ell=1}^rV_{\rho_\ell}\to\bigotimes_{\ell=1}^rV_{\rho_\ell},
\end{equation}%
is the linear map defined by its matrix elements
\begin{equation}
\label{eq_haarintermatrix}
  T_{m_1m_2\ldots m_r;n_1n_2\ldots n_r} := \int_G
    t^{(\rho_1)}_{m_1n_1}(g)t^{(\rho_2)}_{m_2n_2}(g)\cdots t^{(\rho_r)}_{m_rn_r}(g)\,dg.
\end{equation}%
\end{mathletters}%
\end{definition}

\begin{proposition}
\label{prop_haarproperties}
The Haar intertwiner $T$ of~\eqref{eq_haarinter} satisfies
\begin{equation}
\label{eq_haardecompose}
  T_{m_1m_2\ldots m_r;n_1n_2\ldots n_r} 
  = \sum_{j}\alignidx{\overline{P^{(j)}_{m_1m_2\ldots m_r}}P^{(j)}_{n_1n_2\ldots n_r}},
\end{equation}
for the projectors
\begin{equation}
\label{eq_haarinterproj}
  P^{(j)}_{n_1n_2\ldots n_r}
  :=\bigl<w^{(j)};e_{n_1}^{(\rho_1)}\otimes e_{n_2}^{(\rho_2)}\otimes\cdots\otimes e_{n_r}^{(\rho_r)}\bigr>,
\end{equation}
with the definitions of Proposition~\ref{prop_fulldecompose}, as well
as for all $h\in G$,
\begin{gather}
\label{eq_interinv}
  T = (\rho_1(h)\otimes\cdots\otimes\rho_r(h))\circ T 
     = T\circ (\rho_1(h)\otimes\cdots\otimes\rho_r(h)),\\
\label{eq_interproj}
  T\circ T=T.
\end{gather}
The first equation~\eqref{eq_haardecompose} is a consequence of
Corollary~\ref{cor_haaralg} while~\eqref{eq_interinv}
and~\eqref{eq_interproj} follow from the translation invariance of the
Haar measure. In particular, $T$ forms a morphism of representations
(\emph{intertwiner}) of $G$. The map $T$ has been studied in a more
general context in~\cite{Oe02}.
\end{proposition}

\begin{figure}[t]
\begin{center}
\input{pstex/diagrams.pstex_t}
\end{center}
\mycaption{fig_diagrams}{%
  Diagrams to visualize the index structure in the calculation of
  group integrals. (a) The identity map of $V_\rho$; (b) a
  representation function $t^{(\rho)}_{mn}$; (c) a product of
  representation functions $t^{(\rho_1)}_{m_1n_1}\cdots
  t^{(\rho_r)}_{m_rn_r}$; (d) the Haar intertwiner; and (e) the
  calculation of the Haar intertwiner~\eqref{eq_haarintermatrix}.}
\end{figure}

In the subsequent sections, we will apply Corollary~\ref{cor_haaralg}
in rather complicated calculations. It is therefore convenient to
introduce diagrams which visualize the structure of the indices in
these formulas (Figure~\ref{fig_diagrams}).

The diagrams are read from top to bottom. We draw directed lines which
are labelled with representations $\rho\in\Rep_G$ of $G$. If the arrow
points down, the line denotes the identity map of $V_\rho$,
Figure~\ref{fig_diagrams}(a). If the arrow points up, it refers to the
identity map of the dual representation $V_\rho^\ast$. A
representation function $t^{(\rho)}_{mn}$ is denoted by a box with an
incoming and an outgoing line (b), and a product of representation
functions by boxes placed next to each other (c). The Haar intertwiner
is visualized by the box labelled $T$ in (d), and the calculation of
$T$ given by~\eqref{eq_haarintermatrix} is shown in diagram (e) where
the full dots represent the projectors, and the dotted line indicates
the simultaneous summation over them.

\subsection{Coset spaces and spherical functions}
\label{sect_coset}

In the study of coset spaces $G/H$, we allow $H$ to be any Lie
subgroup of $G$. First we recall some basic definitions.

\begin{definition}
Let $G$ be a compact Lie group and $H\leq G$ be a Lie subgroup. 
\begin{myenumerate}
\item
  A finite-dimensional irreducible representation $V_\rho$ of $G$ is
  said to be of \emph{class-1 with respect to} $H$ if $V_\rho$
  contains an $H$-invariant vector $0\neq v_0\in V_\rho$, \ie\
  $\rho(h)v_0=v_0$ for all $h\in H$. The subset
  $\Irrep^G_H\subseteq\Irrep_G$ denotes the set of class-1
  representations of $G$ with respect to $H$.
\item
  $H$ is called a \emph{massive} subgroup of $G$ if for each class-1
  representation $\rho\in\Irrep^G_H$, the subspace of $H$-invariant
  vectors,
\begin{equation}
  \Inv_H^{(\rho)}=\{\,v\in V_\rho\colon\quad \rho(h)v=v\quad
    \mbox{for all}\quad h\in H\,\},
\end{equation}
  is one-dimensional.
\end{myenumerate}
\end{definition}

\begin{proposition}
\label{prop_invvector}
Let $G$ be a compact Lie group, $H\leq G$ a Lie subgroup and
$\rho\in\Irrep_G$ a finite-dimensional irreducible representation of
$G$. The subspace of $H$-invariant vectors in $V_\rho$ is spanned by
the (not necessarily linearly independent) vectors
\begin{equation}
\label{eq_invvector}
  v^{(k)}:=\sum_{j=1}^{\dim V_\rho}\alignidx{v^{(k)}_je_j},\qquad\mbox{where}\qquad
  v^{(k)}_j:=\int_H t^{(\rho)}_{jk}(h)\,dh.
\end{equation}
Here ${\{e_j\}}_j$ denotes the standard orthonormal basis of
$V_\rho$, and $1\leq k\leq\dim V_\rho$.
\end{proposition}

The motivation for studying $H$-invariant vectors is given by the
following result which allows us to construct functions on the space
$G/H$ of left cosets, \ie\ functions on $G$ that are constant on the
cosets $gH$.

\begin{proposition}
\label{prop_spherical}
Let $0\neq v^{(k)}\in V_\rho$ be an $H$-invariant vector. Then the
\emph{generalized spherical functions}
\begin{equation}
  H^{(\rho)}_{jk}(g) := \sum_{\ell=1}^{\dim V_\rho}\alignidx{t^{(\rho)}_{j\ell}(g)\,v_\ell^{(k)}},
\end{equation}
$1\leq j\leq\dim V_\rho$, are constant on the cosets $gH\in G/H$ and
therefore induce functions $H^{(\rho)}_{jk}\colon G/H\to\C$, $x\mapsto
H^{(\rho)}_{jk}(x):=H^{(\rho)}_{jk}(g_x)$, where $g_x\in G$ is a
representative of the coset $x=g_xH\in G/H$.
\end{proposition}

Combining the Peter--Weyl decomposition~\eqref{eq_peterweyl_series} of
$\Calg(G)$ with the above ideas, we can construct the algebraic
functions $\Calg(G/H)$ on the coset space.

\begin{corollary}
\label{cor_spherical}
Let $G$ be a compact Lie group and $H\leq G$ be a Lie subgroup. Denote
the dimensions of the $H$-invariant subspaces by
$\kappa_\rho:=\dim\Inv_H^{(\rho)}$ and choose the orthonormal basis
${\{e_j\}}_j$ of each $V_\rho$ so that precisely
$e_1,\ldots,e_{\kappa_\rho}$ are $H$-invariant. Then the functions
\begin{equation}
  H^{(\rho)}_{jk}\colon G/H\to\C,\quad x\mapsto
  H^{(\rho)}_{jk}(x):=t^{(\rho)}_{jk}(g_x), 
\end{equation}
$\rho\in\Irrep^G_H$, $1\leq j\leq\dim V_\rho$, $1\leq
k\leq\kappa_\rho$, form a basis of $\Calg(G/H)$. These functions
satisfy the orthogonality relation,
\begin{equation}
  \bigl<H^{(\rho)}_{jk};H^{(\sigma)}_{\ell m}\bigr>_{L^2} =
  \int_{G/H}\overline{H^{(\rho)}_{jk}(x)}H^{(\sigma)}_{\ell m}(x)\,dx =
  \frac{1}{\dim V_\rho}\delta_{\rho\sigma}\delta_{j\ell}\delta_{km}.
\end{equation}
\end{corollary}

\begin{remark}
\begin{myenumerate}
\item
  Spherical functions exist only for class-1 representations as
  $\kappa_\rho\neq 0$ only there.
\item
  In the case of a massive subgroup $H$, there is $\kappa_\rho=1$ for
  all class-1 representations $\rho\in\Irrep^G_H$. The second index of
  the spherical functions can thus be omitted, \ie
\begin{equation}
  H^{(\rho)}_j\colon G/H\to\C,\qquad x\mapsto H^{(\rho)}_j(x):=t^{(\rho)}_{j1}(g_x),
\end{equation}
  where $1\leq j\leq\dim V_\rho$. 
\item
  If $H\unlhd G$ is a normal subgroup, there is $\kappa_\rho=\dim
  V_\rho$ for all class-1 representations. In other words, for a given
  irreducible representation $\rho\in\Irrep_G$ of $G$, either all
  representation functions $t^{(\rho)}_{jk}$, $1\leq j,k\leq\dim
  V_\rho$, are spherical functions, or none of them is.
\end{myenumerate}
\end{remark}

\begin{example}
\begin{myenumerate}
\item
  The spheres $S^N\cong\SO(N+1)/\SO(N)$ or $S^N\cong\O(N+1)/\O(N)$ are
  formed using massive subgroups.
\item
  Odd spheres can alternatively be obtained from
  $S^{2N+1}\cong\SU(N+1)/\SU(N)$ or $S^{2N+1}\cong\U(N+1)/\U(N)$, in
  particular $S^3\cong\SU(2)$. The spherical functions of $S^3$ can
  thus be constructed either as functions on $\SO(4)/\SO(3)$ using the
  construction sketched above or from the identification
  $S^3\cong\SU(2)$. For the latter approach, see the introductory part
  of~\cite{Pf02a}.
\item
  Other coset spaces which are of interest in the context of sigma
  models, are $\RP^{N-1}\cong\O(N)/(\O(N-1)\times\O(1))$ as a special
  case of the Grassmanian $G_{kN}^\R\cong\O(N)/(\O(N-k)\times\O(k))$
  and their complex counterparts
  $\CP^{N-1}\cong\U(N)/(\U(N-1)\times\U(1))$ and
  $G_{kN}^\C\cong\U(N)/(\U(N-k)\times\U(k))$.
\end{myenumerate}
\end{example}

\begin{remark}
\label{rem_integrals}
Any function $f\colon G/H\to\C$ naturally extends to a function
$\tilde f\colon G\to\C$ which is constant on the cosets, \ie\ $\tilde
f(gh)=\tilde f(g)$ for all $g\in G$, $h\in H$. Obviously $f(x)=\tilde
f(g_x)$ for all $x\in G/H$ and an arbitrary representative $g_x\in G$
of $x$. Integrals over $G/H$ can thus be evaluated using integrals
over $G$,
\begin{equation}
  \int_{G/H}f(x)\,dx = \int_G\tilde f(g)\,dg.
\end{equation}
As the context is usually clear, we omit the tilde ($\tilde{\ }$) from
now on.
\end{remark}

The analogue of the Haar intertwiner~\eqref{eq_haarinter} for coset
spaces can be defined as follows.

\begin{definition}
\label{def_cosethaar}
Let $G$ be a compact Lie group, $H\leq G$ be a Lie subgroup and
$\rho_1,\ldots,\rho_r\in\Irrep^G_H$, $r\in\N$, be of class-1 with
respect to $H$. The
\emph{coset space Haar map},
\begin{mathletters}
\label{eq_haarcoset}
\begin{equation}
  I\colon\bigotimes_{\ell=1}^rV_{\rho_\ell}\to
    \bigotimes_{\ell=1}^rV_{\rho_\ell},
\end{equation}%
is the linear map defined by its matrix elements
\begin{equation}
  I_{m_1m_2\ldots m_r;n_1n_2\ldots n_r}:=
  \int_{G/H}H^{(\rho_1)}_{m_1n_1}(x)H^{(\rho_2)}_{m_2n_2}(x)\cdots
    H^{(\rho_r)}_{m_rn_r}(x)\,dx,
\end{equation}%
\end{mathletters}
where $1\leq m_\ell\leq\dim V_{\rho_\ell}$ and $1\leq
n_\ell\leq\kappa_{\rho_\ell}$. 
\end{definition}

\begin{figure}[t]
\begin{center}
\input{pstex/diagcoset.pstex_t}
\end{center}
\mycaption{fig_diagcoset}{%
  (a) A spherical function $H^{(\rho)}_{mn}(x)$; (b) the coset space
  Haar map and (c) its calculation in terms of $G$-invariant
  projectors~\eqref{eq_haarcoset1}, see Proposition~\ref{prop_cosethaar}.}
\end{figure}

\begin{proposition}
\label{prop_cosethaar}
The coset space Haar map~\eqref{eq_haarcoset} satisfies
\begin{equation}
\label{eq_haarcoset1}
  I_{m_1m_2\ldots m_r;n_1n_2\ldots n_r}=\sum_j
    \overline{P^{(j)}_{m_1m_2\ldots m_r}}P^{(j)}_{n_1n_2\ldots n_r},
\end{equation}
with the notation of Proposition~\ref{prop_haarproperties}, as well as
for all $h\in G$,
\begin{equation}
\label{eq_haarcoset2}
  I=(\rho_1(h)\otimes\cdots\otimes\rho_r(h))\circ I,
\end{equation}
which makes use of the left action of $G$ on $G/H$. If in addition
$H\unlhd G$ is a normal subgroup, then $I$ satisfies for all $y\in G/H$,
\begin{equation}
\label{eq_haarcoset3}
  I=I\circ (\rho_1(y)\otimes\cdots\otimes\rho_r(y)),
\end{equation}
using the notation $\rho_\ell(y):=\rho_\ell(g_y)$ for any
representative $g_y$ of $y$.
\end{proposition}

\begin{proof}
Equations~\eqref{eq_haarcoset1} and~\eqref{eq_haarcoset2} follow from
Proposition~\ref{prop_haarproperties} and
Remark~\ref{rem_integrals}. The same is true for~\eqref{eq_haarcoset3}
if we write it for a representative of the coset $y\in G/H$.
\end{proof}

Observe that the coset space Haar map is in general not an intertwiner
of $G$. However, for any fixed choice of indices
$n_\ell\in\{1,\ldots,\kappa_{\rho_\ell}\}$, it defines a
$G$-invariant vector $\tilde I_{n_1n_2\ldots
n_r}\in\bigotimes_{\ell=1}^rV_{\rho_\ell}$. 

We visualize spherical functions $H^{(\rho)}_{mn}(x)$ and the coset
space Haar map as in Figure~\ref{fig_diagcoset}. The contraction of
indices whose range is restricted to $\{1,\ldots,\kappa_\rho\}$, is
represented by dashed lines. They do not correspond to representations
of $G$ and are therefore not labelled with any symbol such as
$\rho$. A thick line in the box for $H^{(\rho)}_{mn}$ and in the coset
space Haar map $I$ indicates that the indices on this side are
special. Figure~\ref{fig_diagcoset}(c) shows the
calculation~\eqref{eq_haarcoset1}. 

\begin{figure}[t]
\begin{center}
\input{pstex/diagmassive.pstex_t}
\end{center}
\mycaption{fig_diagmassive}{%
  The specialization of Figure~\ref{fig_diagcoset} to the case of a
  massive subgroup $H\leq G$, see Corollary~\ref{cor_massive}.}
\end{figure}

If $H\unlhd G$ is a normal subgroup, $\kappa_\rho=\dim V_\rho$ for all
class-1 representations so that the dashed lines become solid again as
they do correspond to representations of $G$. In particular
Definition~\ref{def_cosethaar} and Proposition~\ref{prop_cosethaar}
restrict to Definition~\ref{def_haarinter} and
Proposition~\ref{prop_haarproperties}, respectively, if $H=\{e\}$ is
the trivial group. The special case when $H$ is a massive subgroup, is
also of interest.

\begin{corollary}
\label{cor_massive}
If $H\leq G$ is a massive subgroup, then $\kappa_\rho=1$ for the
class-1 representations. Therefore all indices $n_\ell$ can be omitted
from the expressions so that the Haar map reduces to a map
\begin{equation}
  I\colon\C\to\bigotimes_{\ell=1}^rV_{\rho_\ell},
\end{equation}
defined by its matrix elements
\begin{equation}
  I_{m_1m_2\ldots m_r}:=\int_{G/H}
    H^{(\rho_1)}_{m_1}(x)H^{(\rho_2)}_{m_2}(x)\cdots H^{(\rho_r)}_{m_r}(x)\,dx.
\end{equation}
Equation~\eqref{eq_haarcoset1} specializes to
\begin{equation}
  I_{m_1m_2\ldots m_r}=\sum_j\overline{P^{(j)}_{m_1m_2\ldots m_r}},
\end{equation}
and~\eqref{eq_haarcoset2} indicates that $I$ defines a $G$-invariant
vector $I\in\bigotimes_{\ell=1}^rV_{\rho_\ell}$.
\end{corollary}

The situation for massive subgroups is illustrated in
Figure~\ref{fig_diagmassive}. Further properties of the diagrams used
in Figure~\ref{fig_diagmassive} can be deduced as in the introductory
section of~\cite{Pf02a}.

\subsection{The centre of the group}

If representation functions are restricted to the centre $Z(G)$, we
obtain representation functions of the Abelian group $Z(G)$. 

\begin{lemma}
\label{lemma_centre}
Let $G$ be a compact Lie group and $X\in Z(G)$. Then for any irreducible
unitary representation $\rho$ of $G$ and $1\leq i,j\leq\dim V_{\rho}$,
\begin{equation}
  t_{ij}^{(\rho)}(X) = \delta_{ij}\cdot{\tilde t}^{(\rho)}(X),
\end{equation}
where ${\tilde t}^{(\rho)}\colon Z(G)\to\C$ is a representation of the
centre $Z(G)$ which is induced by $\rho$.
\end{lemma}

\begin{proof}
By Schur's lemma, the centre is mapped to multiples of the unit
matrix.
\end{proof}

\subsection{Special properties of some groups}
\label{sect_special}

For $G=\U(1)$, all finite-dimensional irreducible representations are
one-dimensional. They are denoted by $V_k\cong\C$ and characterized by
integers $k\in\Z$ (wave numbers of the Fourier series). In the unitary
case, their representation functions are given by $t^{(k)}(g)=g^k$,
$g\in\U(1)$, and their duals by $t^{(k^\ast)}(g)=g^{-k}$. All
representation functions are characters,
$\chi^{(k)}(g)=t^{(k)}(g)=e^{ik\phi}$, where we write
$g=e^{i\phi}\in\U(1)$.

From the representation functions and the
definition~\eqref{eq_operationprod}, we can calculate the tensor
product which is again one-dimensional,
\begin{equation}
\label{eq_u1tensor2}
  V_{k_1}\otimes\cdots\otimes V_{k_n}\cong V_{\sum\limits_{\ell=1}^nk_\ell}.
\end{equation}
It is isomorphic to the trivial representation if and only if
\begin{equation}
\label{eq_u1tensor}
  \sum_{\ell=1}^n k_\ell=0.
\end{equation}
Since all irreducible representations are one-dimensional, the Haar
intertwiner~\eqref{eq_haarinter},
\begin{equation}
  T\colon\bigotimes_{\ell=1}^rV_{k_\ell}\to\bigotimes_{\ell=1}^rV_{k_\ell},
\end{equation}
is just multiplication by a number. We have $T=1$
if~\eqref{eq_u1tensor} holds and $T=0$ otherwise. The sum over
projectors~\eqref{eq_haardecompose} is either empty or contains a
single unique term.

We write the elements of the cyclic groups $G=\Z_N$ as roots of unity,
$e^{2\pi i\ell/N}$, $\ell\in\{0,\ldots,N-1\}$, and parameterize their
finite-dimensional irreducible unitary representations $V_k\cong\C$ by
$k\in\{0,\ldots,N-1\}$. Representation functions and characters are
$t^{(k)}(g)=g^k$, $t^{(k^\ast)}(g)=g^{-k}$, $\chi^{(k)}=t^{(k)}$,
and~\eqref{eq_u1tensor2} and~\eqref{eq_u1tensor} still hold if the
sums are taken modulo $N$.

For $G=\SU(2)$, we characterize the finite-dimensional irreducible
representations $V_j$, $\dim V_j=2j+1$, by non-negative half-integers
$j\in\frac{1}{2}\N_0$. Parameterizing elements of $\SU(2)$ by
\begin{equation}
  g(\theta,\underline{n})=\openone\,\cos\frac{\theta}{2} +
  i\underline{\sigma}\cdot\underline{n}\,\sin\frac{\theta}{2},
\end{equation}
where $\theta\in[0,4\pi)$, $\underline{n}\in S^2\subseteq\R^3$ and
$\underline{\sigma}=(\sigma_1,\sigma_2,\sigma_3)$ are the Pauli
matrices, the characters are given by
\begin{equation}
  \chi^{(j)}(g)=\frac{\sin\bigl((2j+1)\frac{\theta}{2}\bigr)}{\sin\frac{\theta}{2}},
\end{equation}
in particular for the fundamental representation
$\chi^{(\frac{1}{2})}(g)=\cos\frac{\theta}{2}$.

\begin{figure}[t]
\begin{center}
\input{pstex/diagsu2.pstex_t}
\end{center}
\mycaption{fig_diagsu2}{%
  (a) Simplification of~\eqref{eq_haardecompose} for $G=\SU(2)$ in the
  case of three tensor factors; (b)--(d) Even if we do not write the
  dotted line anymore, this does not mean that any conceivable
  symmetry holds.} 
\end{figure}

Since for $\SU(2)$, there are no higher multiplicities in the
decomposition of $V_{j_1}\otimes V_{j_2}$, the space of invariant
projectors $V_{j_1}\otimes V_{j_2}\otimes V_{j_3}\to\C$ has a
dimension of at most one. For three irreducible representations, we
can therefore omit the summation over projectors
from~\eqref{eq_haardecompose} as is illustrated in
Figure~\ref{fig_diagsu2}(a) and impose the conditions $|j_1-j_2|\leq
j_3\leq j_1+j_2$, \etc\ instead. This provides a substantial
simplification. However, the three-valent vertex that appears there,
has in general only a cyclic, but not a full symmetry
(Figure~\ref{fig_diagsu2}(b--d)) so that one still has to take the
ordering of the tensor factors into account. Neglecting this subtlety
is a common mistake.

\subsection{Some character decompositions}
\label{sect_charexample}

For the duality transformation, we will apply the character
decomposition to the Boltzmann weight $f(g)=\exp(-s(g))$ whose local
action $s\colon G\to\R$ is an $L^2$-integrable class function that is
bounded below. The most common example is the \emph{Wilson action},
\begin{equation}
  s(g)=-\frac{\beta}{2\dim V_\rho}\bigl(\chi^{(\rho)}(g)+\overline{\chi^{(\rho)}(g)}\bigr),
\end{equation}
where $\rho$ denotes the fundamental representation of $G$ and $\beta$
is the inverse temperature or inverse coupling constant.

For $G=\U(1)$, the Wilson action reads $s(g)=-\beta\cos\phi$,
$g=e^{i\phi}$. The character decomposition coincides with the Fourier
series,
\begin{equation}
\label{eq_charexp_u1}
  f(g)=\sum_{k=-\infty}^{\infty}\hat f_k\,e^{ik\phi},\qquad
  \hat f_k=\frac{1}{2\pi}\int_0^{2\pi}e^{-ik\phi}\exp(-s(e^{i\phi}))\,d\phi
  = I_k(\beta),
\end{equation}
where $I_k(\beta)$ denote modified Bessel functions.

For $G=\Z_N$, $g=e^{2\pi i\ell/N}$, we write this decomposition as
\begin{equation}
\label{eq_charexp_zn}
  f(g)=\sum_{k=1}^{N-1}\hat f_ke^{2\pi ik\ell/N},\qquad
  \hat f_k=\frac{1}{N}\sum_{\ell=0}^{N-1}e^{-2\pi ik\ell/N}
    \exp(-s(e^{2\pi i\ell/N})).
\end{equation}

For $G=\SU(2)$, we have the Wilson action
$s(g)=-\beta\cos\frac{\theta}{2}$, and the character expansion of
$f(g)=\exp(-s(g))$ is given by
\begin{equation}
  f(g)=\sum_{j\in\frac{1}{2}\N_0}\hat f_j\,
    \frac{\sin\bigl((2j+1)\frac{\theta}{2}\bigr)}{\sin\frac{\theta}{2}},
  \qquad \hat f_j=\frac{2j+1}{\beta}I_{2j+1}(\beta).
\end{equation}

Another common action is the \emph{heat kernel} or \emph{generalized
Villain action} which is given for any compact Lie group in terms of
the character decomposition of the corresponding Boltzmann weight,
\begin{equation}
\label{eq_heatkernel}
  f(g) = \sum_{\rho\in\Irrep_G}\hat f_\rho\chi^{(\rho)}(g),\qquad
  \hat f_\rho=\dim V_\rho\cdot\exp\bigl(-\frac{C_\rho}{2\beta}\bigr),
\end{equation}
where $C_\rho$ denotes the eigenvalue of a quadratic Casimir
operator in the representation $\rho$. For example, we have for
$G=\U(1)$,
\begin{equation}
  \hat f_k=\exp\bigl(-\frac{k^2}{2\beta}\bigr),
\end{equation}
and for $G=\SU(2)$,
\begin{equation}
  \hat f_j=(2j+1)\,\exp\bigl(-\frac{j(j+1)}{2\beta}\bigr).
\end{equation}

In all these cases, the Boltzmann weight $f(g)=\exp(-s(g))$ has a
sharp peak at the group unit if $\beta$ is large (weak coupling, low
temperature) which facilitates a perturbative treatment whereas the
peak is very wide for small $\beta$ (strong coupling, high
temperature). For small $\beta$, however, the dominant contribution to
the character expansions listed above originates from the `small'
representations of $G$. An expansion in terms of irreducible
representations of $G$ therefore provides us with a strong coupling
expansion. This is most obvious for the heat kernel action where at
small $\beta$ the representations with small Casimir eigenvalue
dominate.

More details on strong coupling expansion techniques can be found
in~\cite{DrZu83}. For spherical functions, see~\cite{ViKl93} and in
particular for $S^N$ also~\cite{CuPe97}.

%
\section{Notation and definitions}
%
\label{sect_definitions}

\subsection{Graphs and abstract $2$-complexes}
\label{sect_graphs}

In order to formulate sigma models and gauge theories on very general
lattices, it is sufficient to focus on the combinatorial structure of
the lattices rather than on the details of their embedding into some
space or space-time manifold. Therefore we employ the notions of
graphs and abstract two-complexes. Sigma models are defined on graphs
so that we obtain the same expressions for their partition function in
any dimension $d\geq 1$. Similarly, gauge theories are defined on
abstract two-complexes, and we obtain a uniform description of gauge
theories valid in any dimension $d\geq 2$.

\begin{figure}[t]
\begin{center}
\input{pstex/complex_boundary.pstex_t}
\end{center}
\mycaption{fig_boundary}{%
  The maps $\del_\pm$, $\del_j$ and $\epsilon_j$ and the
  conditions~\eqref{eq_boundary}. Here $N(f)=3$, $\epsilon_1f=+1$,
  $\epsilon_2f=+1$ and $\epsilon_3f=(-1)$.}
\end{figure}

The following definitions differ slightly from what is standard, but
will prove most convenient for the purpose of the duality
transformation.

\begin{definition}
An \emph{oriented} (or \emph{directed}) \emph{graph} $(V,E)$ consists
of finite sets $V$ (vertices) and $E$ (edges) together with maps
\begin{mathletters}
\begin{alignat}{2}
  \del_+&\colon E\to V,&\qquad&\mbox{(end point of an edge)}\\
  \del_-&\colon E\to V.&\qquad&\mbox{(starting point of an edge)}
\end{alignat}%
\end{mathletters}%
\end{definition}

The notion of an abstract two-complex extends this definition and also
includes faces whose boundary consists of a sequence of edges.

\begin{definition}
An \emph{oriented two-complex} $(V,E,F)$ is an oriented graph $(V,E)$
together with a finite set $F$ (\emph{faces}) and maps
\begin{mathletters}
\begin{alignat}{2}
  N     &\colon F\to\N,&\qquad&\mbox{(number of edges in the boundary
                        of a face)}\\
  \del_j&\colon F\to E,&\qquad&\mbox{(the $j$-th edge in the boundary
                        of a face)}\\
  \epsilon_j&\colon F\to\{-1,+1\},&\qquad&\mbox{(its orientation)}
\end{alignat}%
\end{mathletters}%
such that
\begin{mathletters}
\label{eq_boundary}
\begin{gather}
  \del_{-\epsilon_jf}\del_jf = \del_{\epsilon_{j+1}f}\del_{j+1}f,\qquad
    1\leq j\leq N(f)-1,\\
  \del_{-\epsilon_{N(f)}f}\del_{N(f)}f = \del_{\epsilon_1f}\del_1f,
\end{gather}%
\end{mathletters}%
for all $f\in F$.
\end{definition}

The conditions~\eqref{eq_boundary} state that the edges in the
boundary of a face $f\in F$ are in cyclic ordering from $\del_{N(f)}f$
to $\del_1f$ where one encounters the edges with the orientation given
by $\epsilon_jf$, see Figure~\ref{fig_boundary}. Observe
that~\eqref{eq_boundary} contains combinatorial information similar to
the condition $\del\circ\del=0$ on the boundary operator $\del$ in
Abelian simplicial homology.

In the subsequent calculations, it is convenient to use the following
abbreviations.

\begin{definition}
Let $(V,E,F)$ denote an oriented two-complex. For a given edge $e\in
E$, the sets
\begin{mathletters}
\begin{eqnarray}
e_+ &:=& \{f\in F\colon\quad e=\del_j f,\quad\epsilon_j f=(+1)\quad
           \mbox{for some $j$,}\quad 1\leq j\leq N(f)\},\\
e_- &:=& \{f\in F\colon\quad e=\del_j f,\quad\epsilon_j f=(-1)\quad
           \mbox{for some $j$,}\quad 1\leq j\leq N(f)\},
\end{eqnarray}%
\end{mathletters}%
contain all faces that have the edge $e$ in their boundary with
positive ($+$) or negative ($-$) orientation, and we write $\delta
e:=e_+\cup e_-$ for the coboundary of an edge $e\in E$. For a given
face $f\in F$, the set
\begin{equation}
  f_0 := \{v\in V\colon\quad v=\del_-\del_jf\quad
           \mbox{for some $j$,}\quad 1\leq j\leq N(f)\},
\end{equation}
denotes all vertices that belong to the boundary of the face
$f$. Finally, the sets 
\begin{mathletters}
\begin{eqnarray}
  f_+ &:=& \{e\in E\colon\quad e=\del_jf,\quad\epsilon_jf=(+1)\quad
             \mbox{for some $j$,}\quad 1\leq j\leq N(f)\},\\
  f_- &:=& \{e\in E\colon\quad e=\del_jf,\quad\epsilon_jf=(-1)\quad
             \mbox{for some $j$,}\quad 1\leq j\leq N(f)\},
\end{eqnarray}%
\end{mathletters}%
contain all edges in the boundary of the face $f$ that occur with
positive ($+$) or negative ($-$) orientation, and $\del f:=f_+\cup
f_-$ denotes the full boundary of $f\in F$.
\end{definition}

We have formulated our definitions of graphs and two-complexes so that
they have only a finite number of vertices, edges and faces. This
condition ensures that our partition functions are well defined. The
collections of points, links and plaquettes of standard hyper-cubic
lattices form a special case of oriented two-complexes in the obvious
manner.

\subsection{Spin networks and spin foams}
\label{sect_spinnet}

Spin networks were introduced by Penrose~\cite{Pe71} in the context of
a quantization of space-time geometry. A spin network with symmetry
group $G$ is a graph together with a colouring of its edges with
irreducible representations of $G$ and a colouring of its vertices
with compatible intertwiners (representation morphisms). For the
subsequent calculations it is most convenient to separate the notions
of graph and colouring and to speak of a \emph{spin network} that
\emph{lives on a graph}.

\begin{definition}
\label{def_spinnetwork}
Let $G$ be a compact Lie group and $(V,E)$ be an oriented graph. A
\emph{spin network} $(\tau,Q)$ with symmetry group $G$ on $(V,E)$ is a
colouring of the edges with irreducible representations of $G$,
\begin{mathletters}
\begin{equation}
  \tau\colon E\to \Irrep_G,\quad e\mapsto \tau_e,
\end{equation}%
together with a colouring of the vertices $v\in V$ with compatible
intertwiners,
\begin{equation}
\label{eq_spinnet_compat}
  Q^{(v)}\in \Hom_G\Bigl(\bigotimes_{\ontop{e\in E\colon}{v=\del_+e}}V_{\tau_e},
                   \bigotimes_{\ontop{e\in E\colon}{v=\del_-e}}V_{\tau_e}\Bigr).
\end{equation}%
\end{mathletters}%
The tensor product in the domain is over the `incoming' edges and that
in the image over the `outgoing' edges.
\end{definition}

Spin networks appeared first in the context of quantum gravity. There
they define, for example, the physical states in the loop formulation
of gauge theories and the kinematical states of loop quantum
gravity~\cite{RoSm95b}. The observables of non-Abelian lattice gauge
theory can also be constructed from spin
networks~\cite{OePf01,PfOe02}. They are given by the \emph{spin
network functions} (Definition~\ref{def_spinnetfunction} below).

The concept of a spin network can be generalized by introducing
additional representations at the vertices, called \emph{charges}, and
by modifying the compatibility condition~\eqref{eq_spinnet_compat}
accordingly.

\begin{definition}
\label{def_spinnet_charge}
Let $G$ be a compact Lie group, $(V,E)$ an oriented graph, and
$\rho\colon V\to\Irrep_G$, $v\mapsto\rho_v$ assign an irreducible
representation of $G$ to each vertex. A \emph{spin network
$(\tau,Q,\rho)$ with charges} $\rho$ is a colouring of the edges with
irreducible representations,
\begin{mathletters}
\begin{equation}
  \tau\colon E\to \Irrep_G,\quad e\mapsto \tau_e,
\end{equation}%
together with a colouring of the vertices $v\in V$ with compatible
intertwiners,
\begin{equation}
  Q^{(v)}\in\Hom_G\Bigl(
    \bigl(\bigotimes_{\ontop{e\in E\colon}{v=\del_+e}}V_{\tau_e}\bigr)
    \otimes V_{\rho_v},
       \bigotimes_{\ontop{e\in E\colon}{v=\del_-e}}V_{\tau_e}\Bigr).
\end{equation}%
\end{mathletters}%
\end{definition}

We show in this article that spin networks with charges appear as the
configurations in the dual expression for correlators in sigma models,
and that they characterize the observables of generalized Higgs models.

A higher dimensional analogue of spin networks is the concept of spin
foams. Spin foams also appeared first in the context of quantum
gravity~\cite{Ba98a,Re94,ReRo97}.

\begin{definition}
\label{def_spinfoam}
Let $G$ be a compact Lie group and $(V,E,F)$ be an oriented
two-complex. A \emph{spin foam} $(\rho,P)$ with symmetry group $G$ on
$(V,E,F)$ is a colouring of the faces with finite-dimensional
irreducible representations of $G$,
\begin{mathletters}
\begin{equation}
  \rho\colon F\to\Irrep_G,\quad f\mapsto \rho_f,
\end{equation}%
together with a colouring of the edges $e\in E$ with compatible
intertwiners,
\begin{equation}
  P^{(e)}\in\Hom_G\Bigl(\bigotimes_{f\in e_+}\rho_f,
                   \bigotimes_{f\in e_-}\rho_f\Bigr).
\end{equation}%
\end{mathletters}%
The tensor product in the domain is over the representations at the
`incoming' faces, that in the image over the `outgoing' ones. Incoming
and outgoing are here defined by the relative orientations of the
edges and faces.
\end{definition}

These spin foams are often called \emph{closed spin foams} as
opposed to \emph{open spin foams} which are bounded by a spin
network. Open spin foams can be understood as the higher dimensional
analogue of spin networks with charges and are defined as follows.

\begin{definition}
\label{def_spinfoam_bounded}
Let $G$ be a compact Lie group, $(V,E,F)$ define an oriented
two-complex and $(\tau,Q)$ be a spin network on $(V,E)$. A \emph{spin
foam $(\rho,P,\tau,Q)$ bounded by} the spin network $(\tau,Q)$ is a
colouring of the faces with finite-dimensional irreducible
representations,
\begin{mathletters}
\begin{equation}
  \rho\colon F\to\Irrep_G,\quad f\mapsto \rho_f,
\end{equation}%
together with a colouring of the edges $e\in E$ with compatible
intertwiners,
\begin{equation}
  P^{(e)}\in\Hom_G\Bigl(\bigl(\bigotimes_{f\in e_+}V_{\rho_f}\bigr)
    \otimes V_{\tau_e},
                   \bigotimes_{f\in e_-}V_{\rho_f}\Bigr).
\end{equation}%
\end{mathletters}%
\end{definition}

%
\section{The chiral model}
%
\label{sect_chiralmodel}

In this section, we develop the duality transformation for the chiral
model with a symmetry group $G$ that is a compact Lie group. This
model forms the basis for the generalizations to the non-linear sigma
model with variables in a coset space $G/H$ and to the generalized
Higgs models in which the chiral model or the non-linear sigma model
is coupled to a lattice gauge theory.

\subsection{Partition function}

\begin{definition}
Let $G$ be a compact Lie group and $(V,E)$ be an oriented graph. Let
$s\colon G\to\R$ be an $L^2$-integrable and bounded class function
that satisfies $s(g^{-1})=s(g)$. The \emph{lattice chiral model} with
action $s$ is defined by the partition function
\begin{equation}
\label{eq_chiral_partition}
  Z=\Bigl(\prod_{v\in V}\int_Gdg_v\Bigr)\prod_{e\in E}
    w(\alignidx{g_{\del_+e}\cdot g_{\del_-e}^{-1}}),
\end{equation}
whose \emph{Boltzmann weight} is given by $w(g)=\exp(-s(g))$.
\end{definition}

\begin{remark}
\begin{myenumerate}
\item
  The set of configurations is the product $G^V:=G\times\cdots\times
  G$ of one copy of $G$ per vertex $v\in V$. The partition sum is just
  the Haar measure of $G^V$. There is an interaction term for each
  edge $e\in E$ relating the variables at the two end points,
  $g_{\del_+e}$ and $g_{\del_-e}$.
\item
  It is possible to choose different actions $s_e\colon G\to\R$ for
  each edge $e\in E$ so that one obtains Boltzmann weights
  $w_e(g)=\exp(-s_e(g))$. This is useful, for example, if one wishes
  to study inhomogeneous or anisotropic systems or non-regular
  lattices for which one would introduce geometric factors in order to
  compensate for the different lengths of the various edges. All
  calculations presented below generalize to this case, too, but we
  try to keep the notation simple and do not write down the additional
  index $e$ in the following sections.
\end{myenumerate}
\end{remark}

\begin{lemma}
Orientation reversal of any edge $e\in E$ is a symmetry of the lattice
chiral model. The model therefore depends only on the unoriented
graph. Orientation reversal of an edge $e\in E$ maps
$g_{\del_-e}\mapsto g_{\del_+e}$ and conversely, which leaves the
action invariant since $s(g^{-1})=s(g)$.
\end{lemma}

Our subsequent calculations are most transparent for the generic case
in which the partition function can depend on the orientations even
though this generality is not required for the common examples.

\begin{proposition}
The lattice chiral model~\eqref{eq_chiral_partition} has got a global
left-right $G\times G$-symmetry. Let $h,\tilde h\in G$, then the
transformation
\begin{equation}
\label{eq_chiral_symmetry}
  g_v\mapsto h\cdot g_v\cdot{\tilde h}^{-1},
\end{equation}
for all $v\in V$, is a symmetry of the action
$s(\alignidx{g_{\del_+e}\cdot g_{\del_-e}^{-1}})$ for each edge $e\in
E$ because $s$ is a class function.
\end{proposition}

\subsection{Expectation values}
\label{sect_chiral_observable}

The observables of the lattice chiral model are all possible
expectation values of functions $G^V\to\C$ that are compatible with
the symmetries. With the help of the Peter--Weyl decomposition, it is
possible to calculate the generic form of these observables. For the
chiral model, one obtains the well-known $n$-point functions. We
present the full calculation here because it illustrates the method
and this method generalizes to the more complicated models for which
we derive new results in the subsequent sections.

\begin{theorem}
\label{thm_chiral_observable}
Each algebraic function $f\colon G^V\to\C$ that is compatible with the
global $G\times G$-symmetry~\eqref{eq_chiral_symmetry}, is a linear
combination of functions (\emph{observables}) of the following type,
\begin{equation}
\label{eq_chiral_observable}
  f_{\rho,P,Q}({\{g_v\}}_{v\in V})
    =\Bigl(\prod_{v\in V}\sum_{\ell_v,m_v=1}^{\dim V_{\rho_v}}\Bigr)
    P_{\underbrace{\scriptstyle\ell_v\cdots}_{v\in V}}
    Q_{\underbrace{\scriptstyle m_v\cdots}_{v\in V}}
    \prod_{v\in V}t^{(\rho_v)}_{\ell_v m_v}(g_v).
\end{equation}
Here
\begin{equation}
  \rho\colon V\to\Irrep_G,\qquad v\mapsto\rho_v
\end{equation}
associates an irreducible representation of $G$ with each vertex, and
\begin{equation}
\label{eq_chiral_obstensor}
  P\colon\bigotimes_{v\in V}V_{\rho_v}\to\C,\qquad
  Q\colon\bigotimes_{v\in V}V_{\rho_v}^\ast\to\C,
\end{equation}
are intertwiners of $G$.
\end{theorem}

\begin{figure}[t]
\begin{center}
\input{pstex/chiral_obs.pstex_t}
\end{center}
\mycaption{fig_chiral_obs}{%
  (a) An oriented graph with vertices $u,v,w$ and edges $e,f,h$. (b)
  The structure of the observable~\eqref{eq_chiral_observable} of the
  lattice chiral model on that graph. (c) The two-point
  function~\eqref{eq_chiral_twopoint}.} 
\end{figure}

\begin{remark}
\begin{myenumerate}
\item
  By the notation
\begin{equation}
  \prod_{v\in V}\sum_{\ell_v,m_v=1}^{\dim V_{\rho_v}} = 
    \underbrace{\sum_{\ell_v,m_v=1}^{\dim V_{\rho_v}}\cdots
      \sum_{\ell_v,m_v=1}^{\dim V_{\rho_v}}}_{v\in V}
\end{equation}
  we mean that there is one sum over $\ell_v$ and $m_v$ for each
  vertex $v\in V$. Similarly,
\begin{equation}
  P_{\underbrace{\scriptstyle \ell_v\ldots}_{v\in V}}
\end{equation}
  indicates that the symbol $P$ has got one index $\ell_v$ for each
  $v\in V$. It is also understood that the ordering of tensor factors
  in~\eqref{eq_chiral_obstensor} corresponds to the ordering of
  indices of $P$ and $Q$ in~\eqref{eq_chiral_observable}. We use this
  notation frequently in the subsequent calculations.
\item
  The structure of the observable~\eqref{eq_chiral_observable} is
  illustrated in Figure~\ref{fig_chiral_obs}(b).
\item
  The irreducible representations $\rho_v$ describe the charges that
  are located at the vertices $v\in V$. If there are precisely $k$
  vertices whose $\rho_v$ is non-trivial, the normalized expectation
  value of the observable is a $k$-point function. For each $v\in V$,
  there is a representation function $t^{(\rho_v)}_{\ell_vm_v}$ that
  describes the $G$-dependence of the observable, and the intertwiners
  $P$ and $Q$ involve its indices $\ell_v$ and $m_v$ and are used in
  order to obtain a globally $G\times G$-invariant expression. The
  well-known $2$-point function for two vertices $v,w\in V$ is the
  normalized expectation value of
\begin{equation}
\label{eq_chiral_twopoint}
  \chi^{(\rho)}(\alignidx{g_v\cdot g_w^{-1}})
    = \sum_{\ell_v,m_v=1}^{\dim V_\rho}
      t^{(\rho)}_{\ell_v m_v}(g_v)\cdot t^{(\rho^\ast)}_{\ell_w m_w}(g_w)
      \cdot\delta_{\ell_v\ell_w}\delta_{m_vm_w},
\end{equation}
  (Figure~\ref{fig_chiral_obs}(c)). It forms a special case
  of~\eqref{eq_chiral_observable} in which the only non-trivial
  representations are a charge $\rho$ at $v$ and an anti-charge
  $\rho^\ast$ at $w$, and the intertwiners are trivial,
  $P_{\ell_v\ell_w}=\delta_{\ell_v\ell_w}$, \etc.
\item
  As a consequence of the Peter--Weyl theorem, the $L^2$-integrable
  functions that are compatible with the global $G\times G$-symmetry,
  are in the closure of the set of all algebraic $f_{\rho,P,Q}$, \ie\
  they can be obtained as limits of square summable series of
  functions $f_{\rho,P,Q}$.
\end{myenumerate}
\end{remark}

\begin{proofof}{Theorem~\ref{thm_chiral_observable}}
Algebraic functions $f\colon G^V\to\C$ are elements
$f\in\bigotimes_{v\in V}\Calg(G)$ and therefore have the Peter--Weyl
decomposition
\begin{equation}
f({\{g_v\}}_{v\in V}) 
  = \Bigl(\prod_{v\in V}\sum_{\rho_v\in\Irrep_G}\Bigr)
    \Bigl(\prod_{v\in V}\sum_{j_v,k_v=1}^{\dim V_{\rho_v}}\Bigr)
    \hat f^{(\rho_v,\ldots)}_{j_v\ldots,k_v\ldots}\prod_{v\in V}t^{(\rho_v)}_{j_vk_v}(g_v).
\end{equation}
If $f$ satisfies the global $G\times G$-symmetry, we can
apply~\eqref{eq_chiral_symmetry} for each vertex, and as this holds
for arbitrary $h,\tilde h\in G$, we can integrate the result over $h$
and $\tilde h$,
\begin{eqnarray}
f({\{g_v\}}_{v\in V}) 
  &=& \int_{G\times G}\,dh\,d\tilde h\,\Bigl(\prod_{v\in V}\sum_{\rho_v\in\Irrep_G}\Bigr)
    \Bigl(\prod_{v\in V}\sum_{j_v,k_v=1}^{\dim V_{\rho_v}}\Bigr)
    \hat f^{(\rho_v,\ldots)}_{j_v\ldots,k_v\ldots}\nn\\
  &&\times \prod_{v\in V}\sum_{\ell_v,m_v=1}^{\dim V_{\rho_v}}
    t^{(\rho_v)}_{j_v\ell_v}(h)t^{(\rho_v)}_{\ell_vm_v}(g_v)t^{(\rho_v)}_{m_vk_v}({\tilde h}^{-1}).
\end{eqnarray}
We apply~\eqref{eq_inverse_dual}, writing
$t^{(\rho_v)}_{m_vk_v}({\tilde
h}^{-1})=t^{(\rho_v^\ast)}_{k_vm_v}(\tilde h)$, and move all
summations to the front of the expression. Then we sort the products
by the arguments $g_v,h$ and $\tilde h$,
\begin{eqnarray}
f({\{g_v\}}_{v\in V}) 
  &=& \Bigl(\prod_{v\in V}\sum_{\rho_v\in\Irrep_G}\Bigr)
      \Bigl(\prod_{v\in V}\sum_{j_v,k_v,\ell_v,m_v=1}^{\dim V_{\rho_v}}\Bigr)
      \hat f^{(\rho_v,\ldots)}_{j_v\ldots,k_v\ldots}
      \Bigl(\prod_{v\in V}t^{(\rho_v)}_{\ell_vm_v}(g_v)\Bigr)\nn\\
  &&\times\Bigl(\int_G\prod_{v\in V}t^{(\rho_v)}_{j_v\ell_v}(h)\,dh\Bigr)
      \Bigl(\int_G\prod_{v\in V}t^{(\rho_v^\ast)}_{k_vm_v}(\tilde h)\,d\tilde h\Bigr).
\end{eqnarray}
The integrals over $G$ can be evaluated using~\eqref{eq_haardecompose}
so that
\begin{eqnarray}
\label{eq_chiral_step1}
f({\{g_v\}}_{v\in V}) 
  &=& \Bigl(\prod_{v\in V}\sum_{\rho_v\in\Irrep_G}\Bigr)
      \sum_{P\in\sym{P}}\sum_{Q\in\sym{Q}}
      \Bigl(\prod_{v\in V}\sum_{\ell_v,m_v=1}^{\dim V_{\rho_v}}\Bigr)
      {\tilde f}^{(\rho_v,\ldots)}\,
      P_{\underbrace{\scriptstyle\ell_v\ldots}_{v\in V}}
      Q_{\underbrace{\scriptstyle m_v\ldots}_{v\in V}}\nn\\
  &&\times\Bigl(\prod_{v\in V}t^{(\rho_v)}_{\ell_vm_v}(g_v)\Bigr),
\end{eqnarray}
where
\begin{equation}
  {\tilde f}^{(\rho_v,\ldots)}
  =\Bigl(\prod_{v\in V}\sum_{j_v,k_v=1}^{\dim V_{\rho_v}}\Bigr)
   \hat f^{(\rho_v,\ldots)}_{j_v\ldots,k_v\ldots}\,
   \overline{P_{\underbrace{\scriptstyle j_v\ldots}_{v\in V}}}\,
   \overline{Q_{\underbrace{\scriptstyle k_v\ldots}_{v\in V}}}.
\end{equation}
Here $\sym{P}$ denotes a basis of the space of $G$-invariant projectors
\begin{equation}
  \bigotimes_{v\in V}V_{\rho_v}\to\C,
\end{equation}
whose elements $P\in\sym{P}$ are normalized so that $P^2=P$ where the
trivial representation is embedded as $\C\subseteq\bigotimes_{v\in
V}V_{\rho_v}$.  Similarly, $\sym{Q}$ denotes a basis of $G$-invariant
projectors
\begin{equation}
  \bigotimes_{v\in V}V_{\rho_v}^\ast\to\C,
\end{equation}
with the analogous normalization. The
expression~\eqref{eq_chiral_step1} is a linear combination of
observables of the form~\eqref{eq_chiral_observable}.
\end{proofof}

\begin{remark}
\label{rem_zeroexpectation}
The global $G\times G$-symmetry can be realized as the translation
symmetry of the multiple Haar measure because for each $v\in V$,
$h,\tilde h\in G$ and any function $u\in\Calg(G)$,
\begin{equation}
  \int_Gu(g_v)\,dg_v=\int_Gu(h\cdot g_v\cdot {\tilde h}^{-1})\,dg_v.
\end{equation}
As the Boltzmann weight is invariant, the expectation value of any
non-invariant function $f^\prime\colon G^V\to\C$ under the partition
function vanishes. Note that this holds for any finite graph $(V,E)$.

Similarly, the expectation value vanishes for any function that is
not invariant under simultaneous orientation reversal of all edges
which corresponds to taking the dual of all representations and which
is realized by the inversion symmetry of the Haar measure,
\begin{equation}
  \int_Gu(g_v)\,dg_v = \int_Gu(g_v^{-1})\,dg_v.
\end{equation}
Therefore all interesting observables are functions $G^V\to\R$.
\end{remark}

\subsection{Duality transformation}
\label{sect_chiral_dual}

The duality transformation consists of two steps. First, we character
expand the Boltzmann weight of the original partition
function~\eqref{eq_chiral_partition} of the lattice chiral model. This
introduces sums over all irreducible representations of $G$ for each
edge as the new dual variables. Furthermore, this step is responsible
for the strong-weak or high temperature-low temperature relation of
the duality transformation as we have explained in
Section~\ref{sect_charexample}.

The second step is to employ the methods outlined in
Section~\ref{sect_algintegral} in order to solve all integrals over
$G$ and therefore to remove the old variables from the partition
function.

We start with the partition function~\eqref{eq_chiral_partition} and
insert the character expansion~\eqref{eq_charexp} of the Boltzmann
weight for each edge $e\in E$,
\begin{equation}
  Z=\Bigl(\prod_{v\in V}\int_Gdg_v\Bigr)\prod_{e\in E}
    \sum_{\tau_e\in\Irrep_G}\hat w_{\tau_e}
    \chi^{(\tau_e)}(\alignidx{g_{\del_+e}\cdot g_{\del_-e}^{-1}}).
\end{equation}
The character can be simplified using~\eqref{eq_copro}
and~\eqref{eq_inverse_dual}, 
\begin{equation}
  \chi^{(\tau_e)}(\alignidx{g_{\del_+e}\cdot g_{\del_-e}^{-1}})
  = \sum_{p_e,q_e=1}^{\dim V_{\tau_e}}
    t^{(\tau_e)}_{p_eq_e}(g_{\del_+e})t^{(\tau_e^\ast)}_{p_eq_e}(g_{\del_-e}).
\end{equation}
The sums are moved to the front of the expression, and we sort the
product of representation functions by the vertex $v\in V$ of their
arguments $g_v$,
\begin{eqnarray}
\label{eq_chiral_step3}
  Z&=&\Bigl(\prod_{e\in E}\sum_{\tau_e\in\Irrep_G}\Bigr)
      \Bigl(\prod_{e\in E}\hat w_{\tau_e}\Bigr)
      \Bigl(\prod_{e\in E}\sum_{p_e,q_e=1}^{\dim V_{\tau_e}}\Bigr)\nn\\
   &&\times\prod_{v\in V}\int_G\,dg_v
      \Bigl(\prod_{\ontop{e\in E\colon}{v=\del_+e}}t^{(\tau_e)}_{p_eq_e}(g_v)\Bigr)
      \Bigl(\prod_{\ontop{e\in E\colon}{v=\del_-e}}t^{(\tau_e^\ast)}_{p_eq_e}(g_v)\Bigr).
\end{eqnarray}
Here the last two products are over all edges $e\in E$ that have
$v=\del_\pm e$. The integrals over $G$ can be evaluated
using~\eqref{eq_haardecompose},
\begin{eqnarray}
  Z&=&\Bigl(\prod_{e\in E}\sum_{\tau_e\in\Irrep_G}\Bigr)
      \Bigl(\prod_{e\in E}\hat w_{\tau_e}\Bigr)
      \Bigl(\prod_{e\in E}\sum_{p_e,q_e=1}^{\dim V_{\tau_e}}\Bigr)\nn\\
   &&\times\prod_{v\in V}
      \sum_{S^{(v)}\in\sym{S}^{(v)}}
      \overline{S^{(v)}_{\underbrace{\scriptstyle p_e\ldots}_{\ontop{e\in E\colon}{v=\del_+e}}
                         \underbrace{\scriptstyle p_e\ldots}_{\ontop{e\in E\colon}{v=\del_-e}}}}
      S^{(v)}_{\underbrace{\scriptstyle q_e\ldots}_{\ontop{e\in E\colon}{v=\del_+e}}
               \underbrace{\scriptstyle q_e\ldots}_{\ontop{e\in E\colon}{v=\del_-e}}},
\end{eqnarray}
where $\sym{S}^{(v)}$, $v\in V$, denotes a basis of $G$-invariant
projectors
\begin{equation}
\label{eq_chiral_step2}
  \Bigl(\bigotimes_{\ontop{e\in E\colon}{v=\del_+e}}V_{\tau_e}\Bigr)\otimes
  \Bigl(\bigotimes_{\ontop{e\in E\colon}{v=\del_-e}}V_{\tau_e}^\ast\Bigr)\to\C,
\end{equation}
which are normalized so that ${S^{(v)}}^2=S^{(v)}$ if the trivial
representation $\C$ is embedded in the big tensor product. We obtain
the following result.

\begin{theorem}[Dual partition function]
\label{thm_chiral_dualpart}
Let $G$ be a compact Lie group and $(V,E)$ denote an oriented
graph. The partition function of the lattice chiral
model~\eqref{eq_chiral_partition} is equal to 
\begin{eqnarray}
\label{eq_chiral_dualpartition}
  Z&=&\Bigl(\prod_{e\in E}\sum_{\tau_e\in\Irrep_G}\Bigr)
      \Bigl(\prod_{v\in V}\sum_{S^{(v)}\in\sym{S}^{(v)}}\Bigr)
      \Bigl(\prod_{e\in E}\hat w_{\tau_e}\Bigr)
      \Bigl(\prod_{e\in E}\sum_{p_e,q_e=1}^{\dim V_{\tau_e}}\Bigr)\nn\\
   &&\times\prod_{v\in V}
      \overline{S^{(v)}_{\underbrace{\scriptstyle p_e\ldots}_{\ontop{e\in E\colon}{v=\del_+e}}
                         \underbrace{\scriptstyle p_e\ldots}_{\ontop{e\in E\colon}{v=\del_-e}}}}
      S^{(v)}_{\underbrace{\scriptstyle q_e\ldots}_{\ontop{e\in E\colon}{v=\del_+e}}
               \underbrace{\scriptstyle q_e\ldots}_{\ontop{e\in E\colon}{v=\del_-e}}},
\end{eqnarray}
where $\sym{S}^{(v)}$ is a basis of $G$-invariant
projectors~\eqref{eq_chiral_step2}.
\end{theorem}

\begin{figure}[t]
\begin{center}
\input{pstex/chiral_dualpart.pstex_t}
\end{center}
\mycaption{fig_chiral_dualpart}{%
  (a) The spin network that appears in the dual partition
  function~\eqref{eq_chiral_dualpartition} of the lattice chiral model
  on the graph of Figure~\ref{fig_chiral_obs}(a). (b) The alternative 
  expression~\eqref{eq_chiral_dualpartition2} for the same graph.} 
\end{figure}

\begin{remark}
\begin{myenumerate}
\item
  This dual partition function can be described in words as
  follows. The partition sum consists of a sum over all colourings of
  the edges $e\in E$ with irreducible representations $\tau_e$ of $G$
  and over all colourings of the vertices $v\in V$ with compatible
  intertwiners $S^{(v)}$. Compatible here means that each $S^{(v)}$
  corresponds to a map from the tensor product of the representations
  at the incoming edges $e\in E\colon v=\del_+e$ to the tensor product
  of the outgoing edges $e\in E\colon v=\del_-e$,
\begin{equation}
  S^{(v)}\colon\Bigl(\bigotimes_{\ontop{e\in E\colon}{v=\del_+e}}V_{\tau_e}\Bigr)\to
  \Bigl(\bigotimes_{\ontop{e\in E\colon}{v=\del_-e}}V_{\tau_e}\Bigr).
\end{equation}
  Indeed, such an $S^{(v)}$ is related to the one appearing
  in~\eqref{eq_chiral_step2} by the canonical isomorphisms
  $\Hom_G(V\otimes W^\ast,\C)\cong\Hom_G(V,W)$. The Boltzmann weight
  of the dual partition function consists of the character expansion
  coefficients $\hat w_{\tau_e}$ for each edge $e\in E$ and of a spin
  network given by the $S^{(v)}$ whose indices are contracted by the
  summations over $p_e$ and $q_e$. This is illustrated in
  Figure~\ref{fig_chiral_dualpart}(a).
\item
  The dual partition function of the lattice chiral model is therefore
  given by a sum over spin networks. We call such a model a \emph{spin
  network model} in analogy to the spin foam models which arise as the
  dual formulation of lattice gauge theory. The two layers of
  Figure~\ref{fig_chiral_dualpart}(a) reflect the chiral structure
  given by the two-fold global $G$-symmetry. The fact that the spin
  networks extend over the entire graph is a consequence of the global
  nature of the symmetry.
\item
  We comment on the Abelian special case below in
  Section~\ref{sect_chiral_abelian}. 
\item
  There is an alternative form of the dual partition function which
  uses a diagrammatical formulation similar to that developed
  in~\cite{Oe02} for lattice gauge theory. This result is given in the
  following corollary and illustrated in
  Figure~\ref{fig_chiral_dualpart}(b). 
\end{myenumerate}
\end{remark}

\begin{corollary}
From the intermediate step~\eqref{eq_chiral_step3} of the proof, we
obtain the following slightly different expression which involves the
Haar intertwiner~\eqref{eq_haarinter} instead of the sum over
projectors $S^{(v)}$,
\begin{equation}
\label{eq_chiral_dualpartition2}
  Z=\Bigl(\prod_{e\in E}\sum_{\tau_e\in\Irrep_G}\Bigr)
      \Bigl(\prod_{e\in E}\hat w_{\tau_e}\Bigr)
      \Bigl(\prod_{e\in E}\sum_{p_e,q_e=1}^{\dim V_{\tau_e}}\Bigr)
    \prod_{v\in V}
      T^{(v)}_{\underbrace{\scriptstyle p_e\ldots}_{\ontop{e\in E\colon}{v=\del_+e}}
               \underbrace{\scriptstyle p_e\ldots}_{\ontop{e\in E\colon}{v=\del_-e}},
               \underbrace{\scriptstyle q_e\ldots}_{\ontop{e\in E\colon}{v=\del_+e}}
               \underbrace{\scriptstyle q_e\ldots}_{\ontop{e\in E\colon}{v=\del_-e}}},
\end{equation}
where for each $v\in V$, the Haar intertwiner $T^{(v)}$ is a map
\begin{equation}
  T^{(v)}\colon
    \Bigl(\bigotimes_{\ontop{e\in E\colon}{v=\del_+e}}V_{\tau_e}\Bigr)\otimes
    \Bigl(\bigotimes_{\ontop{e\in E\colon}{v=\del_-e}}V_{\tau_e}^\ast\Bigr)\to
    \Bigl(\bigotimes_{\ontop{e\in E\colon}{v=\del_+e}}V_{\tau_e}\Bigr)\otimes
    \Bigl(\bigotimes_{\ontop{e\in E\colon}{v=\del_-e}}V_{\tau_e}^\ast\Bigr).
\end{equation}
\end{corollary}

The next step is the generalization of
Theorem~\ref{thm_chiral_dualpart} to the expectation value of the
observable~\eqref{eq_chiral_observable},
\begin{eqnarray}
\label{eq_chiral_expobservable}
  \left<f_{\rho,P,Q}\right> &=& \frac{1}{Z}
    \Bigl(\prod_{v\in V}\int_Gdg_v\Bigr)
    \Bigl(\prod_{e\in E} w(\alignidx{g_{\del_+e}\cdot g_{\del_-e}^{-1}})\Bigr)\nn\\
  &&\times
    \Bigl(\prod_{v\in V}\sum_{\ell_v,m_v=1}^{\dim V_{\rho_v}}\Bigr)
    P_{\underbrace{\scriptstyle\ell_v\cdots}_{v\in V}}
    Q_{\underbrace{\scriptstyle m_v\cdots}_{v\in V}}
    \prod_{v\in V}t^{(\rho_v)}_{\ell_v m_v}(g_v).
\end{eqnarray}
Again, we character expand the Boltzmann weight, simplify the
characters that occur in the expression and reorganize everything. The
step that generalizes~\eqref{eq_chiral_step3} then reads
\begin{eqnarray}
\label{eq_chiral_step5}
  \left<f_{\rho,P,Q}\right> &=& \frac{1}{Z}
    \Bigl(\prod_{e\in E}\sum_{\tau_e\in\Irrep_G}\Bigr)
    \Bigl(\prod_{e\in E}\hat w_{\tau_e}\Bigr)
    \Bigl(\prod_{e\in E}\sum_{p_e,q_e=1}^{\dim V_{\tau_e}}\Bigr)
    \Bigl(\prod_{v\in V}\sum_{\ell_v,m_v=1}^{\dim V_{\rho_v}}\Bigr)
    P_{\underbrace{\scriptstyle\ell_v\cdots}_{v\in V}}
    Q_{\underbrace{\scriptstyle m_v\cdots}_{v\in V}}\nn\\
   &&\times\prod_{v\in V}\int_G\,dg_v
      \Bigl(\prod_{\ontop{e\in E\colon}{v=\del_+e}}t^{(\tau_e)}_{p_eq_e}(g_v)\Bigr)
      \Bigl(\prod_{\ontop{e\in E\colon}{v=\del_-e}}t^{(\tau_e^\ast)}_{p_eq_e}(g_v)\Bigr)
      t^{(\rho_v)}_{\ell_v m_v}(g_v).
\end{eqnarray}
Compared with~\eqref{eq_chiral_step3}, there is an additional factor
$t^{(\rho_v)}_{\ell_vm_v}$ for each $v\in V$ under the
integral. Solving the integrals, we obtain the following result.

\begin{figure}[t]
\begin{center}
\input{pstex/chiral_dualobs.pstex_t}
\end{center}
\mycaption{fig_chiral_dualobs}{%
  (a) The dual expression~\eqref{eq_chiral_dualobservable} for the
  expectation value of the observable shown in
  Figure~\ref{fig_chiral_obs}(b), \cf\ the dual partition function in
  Figure~\ref{fig_chiral_dualpart}(a). (b) The alternative
  formulation~\eqref{eq_chiral_dualobservable2}.}
\end{figure}

\begin{theorem}[Dual observable]
\label{thm_chiral_dualobservable}
Let $G$ be a compact Lie group, $(V,E)$ an oriented graph and
$f_{\rho,P,Q}$ denote an observable of the
form~\eqref{eq_chiral_observable}. Then the expectation
value~\eqref{eq_chiral_expobservable} of $f_{\rho,P,Q}$ in the lattice
chiral model is equal to
\begin{eqnarray}
\label{eq_chiral_dualobservable}
  \left<f_{\rho,P,Q}\right> &=& \frac{1}{Z}
    \Bigl(\prod_{v\in V}\sum_{\ell_v,m_v=1}^{\dim V_{\rho_v}}\Bigr)
    P_{\underbrace{\scriptstyle\ell_v\cdots}_{v\in V}}
    Q_{\underbrace{\scriptstyle m_v\cdots}_{v\in V}}\nn\\
  &&\times
    \Bigl(\prod_{e\in E}\sum_{\tau_e\in\Irrep_G}\Bigr)
    \Bigl(\prod_{v\in V}\sum_{S^{(v)}\in\tilde{\sym{S}}^{(v)}}\Bigr)
    \Bigl(\prod_{e\in E}\hat w_{\tau_e}\Bigr)
    \Bigl(\prod_{e\in E}\sum_{p_e,q_e=1}^{\dim V_{\tau_e}}\Bigr)\nn\\
   &&\times\prod_{v\in V}
      \overline{S^{(v)}_{\underbrace{\scriptstyle p_e\ldots}_{\ontop{e\in E\colon}{v=\del_+e}}
                         \underbrace{\scriptstyle p_e\ldots}_{\ontop{e\in E\colon}{v=\del_-e}}\ell_v}}
      S^{(v)}_{\underbrace{\scriptstyle q_e\ldots}_{\ontop{e\in E\colon}{v=\del_+e}}
               \underbrace{\scriptstyle q_e\ldots}_{\ontop{e\in E\colon}{v=\del_-e}}m_v}.
\end{eqnarray}
For each vertex $v\in V$, $\tilde{\sym{S}}^{(v)}$ denotes a basis of
$G$-invariant projectors
\begin{equation}
\label{eq_chiral_step4}
  \Bigl(\bigotimes_{\ontop{e\in E\colon}{v=\del_+e}}V_{\tau_e}\Bigr)\otimes
  \Bigl(\bigotimes_{\ontop{e\in E\colon}{v=\del_-e}}V_{\tau_e}^\ast\Bigr)
  \otimes V_{\rho_v}\to\C
\end{equation}
with the usual normalization.
\end{theorem}

\begin{remark}
\begin{myenumerate}
\item
  Compared with the dual partition
  function~\eqref{eq_chiral_dualpartition}, the new features are the
  sums over the $\ell_v$ and $m_v$ and the matrix elements of $P$ and
  $Q$ from the definition of the
  observable~\eqref{eq_chiral_observable}. The remainder of the
  expression has the same structure as the dual partition function
  except for the fact that the intertwiners $S^{(v)}$ have
  changed. They now include the charges $\rho_v$, $v\in V$, of the
  observable in the compatibility condition~\eqref{eq_chiral_step4},
  and the spin networks of the dual partition function are coupled to
  these charges --- the $\ell_v$ and $m_v$ appear as additional
  indices of the $S^{(v)}$. The numerator of the dual expression is
  therefore given by a sum over spin networks with charges
  (Definition~\ref{def_spinnet_charge}). This is illustrated in
  Figure~\ref{fig_chiral_dualobs}(a).
\item
  Equation~\eqref{eq_chiral_dualobservable} shows that an expectation
  value of an observable is mapped to a ratio of partition functions
  in the dual picture, say $\left<f_{\rho,P,Q}\right>=Z(\rho)/Z$. The
  numerator $Z(\rho)$ is similar to the partition function, but the
  spin networks appearing there are now coupled to the charges of the
  observable, \ie\ the numerator corresponds to the partition function
  in the presence of `background charges'.
\item
  Again there exists an alternative formulation based on the
  intermediate step~\eqref{eq_chiral_step5} and involving the Haar
  intertwiner. This is stated in the following corollary and shown in
  Figure~\ref{fig_chiral_dualobs}(b).
\end{myenumerate}
\end{remark}

\begin{corollary}
From the intermediate step~\eqref{eq_chiral_step5} of the proof, we
obtain,
\begin{eqnarray}
\label{eq_chiral_dualobservable2}
  \left<f_{\rho,P,Q}\right> &=& \frac{1}{Z}
    \Bigl(\prod_{v\in V}\sum_{\ell_v,m_v=1}^{\dim V_{\rho_v}}\Bigr)
    P_{\underbrace{\scriptstyle\ell_v\cdots}_{v\in V}}
    Q_{\underbrace{\scriptstyle m_v\cdots}_{v\in V}}\,
    \Bigl(\prod_{e\in E}\sum_{\tau_e\in\Irrep_G}\Bigr)
    \Bigl(\prod_{e\in E}\hat w_{\tau_e}\Bigr)
    \Bigl(\prod_{e\in E}\sum_{p_e,q_e=1}^{\dim V_{\tau_e}}\Bigr)\nn\\
   &&\times\prod_{v\in V}
    T^{(v)}_{\underbrace{\scriptstyle p_e\ldots}_{\ontop{e\in E\colon}{v=\del_+e}}
             \underbrace{\scriptstyle p_e\ldots}_{\ontop{e\in E\colon}{v=\del_-e}}\ell_v,
             \underbrace{\scriptstyle q_e\ldots}_{\ontop{e\in E\colon}{v=\del_+e}}
             \underbrace{\scriptstyle q_e\ldots}_{\ontop{e\in E\colon}{v=\del_-e}}m_v},
\end{eqnarray}
where the Haar intertwiner $T^{(v)}$ for any given $v\in V$ is a map
\begin{equation}
  T^{(v)}\colon
    \Bigl(\bigotimes_{\ontop{e\in E\colon}{v=\del_+e}}V_{\tau_e}\Bigr)\otimes
    \Bigl(\bigotimes_{\ontop{e\in E\colon}{v=\del_-e}}V_{\tau_e}^\ast\Bigr)
    \otimes V_{\rho_v}\to
    \Bigl(\bigotimes_{\ontop{e\in E\colon}{v=\del_+e}}V_{\tau_e}\Bigr)\otimes
    \Bigl(\bigotimes_{\ontop{e\in E\colon}{v=\del_-e}}V_{\tau_e}^\ast\Bigr)
    \otimes V_{\rho_v}.   
\end{equation}
\end{corollary}

\subsection{The Abelian special case}
\label{sect_chiral_abelian}

In this section, we illustrate the specialization to the Abelian case
in detail.

For $G=\U(1)$, the partition function reads
\begin{equation}
  Z = \Bigl(\prod_{v\in V}\frac{1}{2\pi}\int_0^{2\pi}\,d\phi_v\Bigr)
      \prod_{e\in E}\exp\biggl(
        -s(e^{i(\phi_{\del_+e}-\phi_{\del_-e})})\biggr)
\end{equation}
for some some action $s\colon\U(1)\to\R$. For $G=\Z_N$ we have
\begin{equation}
  Z = \Bigl(\prod_{v\in V}\frac{1}{N}\sum_{\ell_v=0}^{N-1}\Bigr)
      \prod_{e\in E}\exp\biggl(
        -s\bigl(e^{2\pi i(\ell_{\del_+e}-\ell_{\del_-e})/N}\bigr)\biggr),
\end{equation}
\ie\ the chiral model restricts to the $XY$-model if $G=\U(1)$, to the
$\Z_N$-vector Potts model if $G=\Z_N$ and in particular to the Ising
model if $G=\Z_2$. The dual partition
function~\eqref{eq_chiral_dualpartition} contains a sum over
irreducible representations for each edge which becomes in the Abelian
case a sum over $\Z$ or $\Z_N$ (Section~\ref{sect_special}).

As all irreducible representations are one-dimensional, the indices of
$S^{(v)}$ in~\eqref{eq_chiral_dualpartition} are absent, and the sum
over projectors restricts to the constraint~\eqref{eq_u1tensor},
therefore for $G=\U(1)$,
\begin{equation}
\label{eq_spin_dualpartition}
  Z=\Bigl(\prod_{e\in E}\sum_{k_e=-\infty}^\infty\Bigr)
    \Bigl(\prod_{v\in V}\delta\bigl(
      \sum_{\ontop{e\in E\colon}{v=\del_+e}}k_e-
      \sum_{\ontop{e\in E\colon}{v=\del_-e}}k_e\bigr)\Bigr)
    \Bigl(\prod_{e\in E}\hat w_{k_e}\Bigr),
\end{equation}
where we write $\delta(x)$ for the constraint that $x=0$. For
$G=\Z_N$, the sum over the $k_e$ is over $\{0,\ldots,N-1\}$ and all
arithmetic involving the $k_e$ is modulo $N$. The coefficients $\hat
w_k$ are the character expansion coefficients of the Boltzmann weight
(see~\eqref{eq_charexp_u1} and~\eqref{eq_charexp_zn}).

Equation~\eqref{eq_spin_dualpartition} is the well-known dual
expression of the partition function at a stage before the constraints
are solved, see, for example~\cite{Sa77,Sa80}. The solution of these
constraints then depends on the dimension and on the topology of the
lattice. For $G=\U(1)$ one obtains the solid-on-solid model in $d=2$
and $\Z$-lattice gauge theory in $d=3$~\cite{Sa77} whereas for
$G=\Z_N$, one finds again the $\Z_N$-vector Potts model in $d=2$ with
the self-duality of the Ising model~\cite{KrWa41} as a special case
for $N=2$, and a $\Z_N$-lattice gauge theory in $d=3$ whose $N=2$ case
was studied in~\cite{We71}.

In the Abelian situation, the dual partition
function~\eqref{eq_spin_dualpartition} contains only a colouring at
one level, namely the sum over all colourings of edges with
irreducible representations (wave numbers). The generalization to
non-Abelian symmetry groups introduces as a second level the sum over
all colourings of the vertices with compatible intertwiners. This
second colouring restricts to the familiar constraint of the
form~\eqref{eq_u1tensor} if $G$ is Abelian.

The symmetry compatible functions~\eqref{eq_chiral_observable} read in
the Abelian case $G=\U(1)$ [or $G=\Z_N$],
\begin{equation}
  f_{\underbrace{\scriptstyle \ell_v\ldots}_{v\in V}}({\{g_v\}}_{v\in V})
  = \prod_{v\in V}g_v^{\ell_v},
\end{equation}
where the $\ell_v\in\Z$ [or $\ell_v\in\{0,\ldots,N-1\}$] specify the
charges located at the vertices. The dual of the expectation
value~\eqref{eq_chiral_dualobservable} is then given by
\begin{equation}
  \bigl<f_{\underbrace{\scriptstyle \ell_v\ldots}_{v\in V}}\bigr>
  = \frac{1}{Z}\Bigl(\prod_{e\in E}\sum_{k_e=-\infty}^\infty\Bigr)
    \Bigl(\prod_{v\in V}\delta\bigl(
      \sum_{\ontop{e\in E\colon}{v=\del_+e}}k_e-
      \sum_{\ontop{e\in E\colon}{v=\del_-e}}k_e+\ell_v\bigr)\Bigr)
    \Bigl(\prod_{e\in E}\hat w_{k_e}\Bigr)
\end{equation}
for $G=\U(1)$ and the obvious analogue for $G=\Z_N$.

Already in the Abelian case, the duality transformation maps the
expectation value to a ratio of partition functions whose numerator is
a modification of the partition function in which the presence of
background charges has modified the constraints or compatibility
conditions.

\subsection{Expectation values of the dual model}
\label{sect_chiral_dualexp}

As Theorem~\ref{thm_chiral_dualobservable} shows, the dual expression
for the expectation value of an observable is given by a ratio of
partition functions. In particular, this dual expression does not
coincide with any expectation value under the dual partition function.

It is therefore an interesting problem to study the natural
observables of the dual partition function and to transform these
expressions back to the original formulation. From the Abelian special
case it is familiar that the transformation maps expectation values to
ratios of partition functions and therefore correlators constructed
from fundamental variables to free energies of collective excitations
and conversely, see, for example~\cite{Sa80,FrMa87}.

For lattice gauge theory with gauge group $G=\U(1)$ in $d=4$
dimensions, for example, there exist particular expectation values of
the dual partition function which describe the correlators of
world-lines of magnetic monopoles~\cite{FrMa87}. If one transforms
these expressions back to the original picture, one obtains ratios of
partition functions $Z(X)/Z$. The numerator can be understood as the
partition function of the model in the presence of a background
magnetic field probing monopoles, and the ratio $Z(X)/Z=e^{-F}$ is
related to the free energy $F$ of this monopole configuration. A first
natural generalization to the non-Abelian case was given by the
correlation functions of centre monopoles in~\cite{PfOe02},
expressions which have been studied in lattice gauge theory for some
time, but which have not been seen in the context of the duality
transformation.

In the Abelian sigma models, the analogue of the magnetic monopoles is
given by dislocations, vortices or world lines of vortices, depending
on the dimension and on the precise model. In the following, we present
the analogous definition for the lattice chiral model with non-Abelian
symmetry group $G$ which we call \emph{centre dislocations} as it uses
the centre $Z(G)$ of the symmetry group $G$ similarly to the centre
monopoles in order to parameterize the observables and because it
specializes to the \emph{dislocations} studied in~\cite{We71} in the
case $G=\Z_2$.

\begin{definition}
Let $G$ be a compact Lie group, $(V,E)$ be an oriented graph and
$X\colon E\to Z(G),e\mapsto X_e$ assign an element of the centre
$Z(G)$ to each edge $e\in E$. The \emph{centre dislocation} is the
following function $\sym{O}_X\colon{(\Irrep_G)}^V\to\C$ of the
configurations of the dual partition
function~\eqref{eq_chiral_dualpartition},
\begin{equation}
\label{eq_chiral_centre}
  \sym{O}_X({\{\tau_e\}}_{e\in E}) := \prod_{e\in E}\tilde t^{(\tau_e)}(X_e),
\end{equation}
where $\tilde t^{(\tau_e)}$ denotes the representation functions of
$Z(G)$ induced from the representation $\tau_e\in\Irrep_G$
(Lemma~\ref{lemma_centre}). 
\end{definition}

We can now employ the techniques of Section~\ref{sect_chiral_dual} in
order to transform the expectation value of the centre dislocation
back to the original picture.

\begin{theorem}
The normalized expectation value of the centre
dislocation~\eqref{eq_chiral_centre} under the dual partition
function~\eqref{eq_chiral_dualpartition} is equal to
\begin{equation}
\label{eq_chiral_origcentre}
  {\left<\sym{O}_X\right>}_{\mathrm{dual}} =
    \frac{1}{Z}\Bigl(\prod_{v\in V}\int_G\,dg_v\Bigr)
    \prod_{e\in E}w(\alignidx{g_{\del_+e}\cdot g_{\del_-e}^{-1}\cdot X_e}).
\end{equation}
\end{theorem}

\begin{proof}
Start from~\eqref{eq_chiral_origcentre}, insert the expansion of
$w(g)$ and apply Lemma~\ref{lemma_centre}. The proof is entirely
analogous to that of Theorem~\ref{thm_chiral_dualpart} with one
additional factor $\tilde t^{(\tau_e)}(X_e)$ for each edge $e\in E$ in
the integrand.
\end{proof}

\begin{remark}
The expectation value of the dual observable takes the form of a ratio
of partition functions in the original formulation. This is essentially
the converse of Theorem~\ref{thm_chiral_dualobservable}.  The
numerator can again be viewed as the partition function in the
presence of a background field $X$.

In the Abelian case, we have $Z(G)=G$. The possible choices for fields
$X$ depend on the particular group and on the dimension and topology
of the lattice. They have been carefully studied for several models.
\begin{myenumerate}
\item
  If $G=\U(1)$ and $(V,E)$ is a two-dimensional cubic lattice, then the
  disorder parameter of the $XY$-model which is related to the free
  energy of a vortex-antivortex pair, is of the
  form~\eqref{eq_chiral_origcentre}. In higher dimensions, this
  generalizes to vortex strings, vortex sheets, \etc.
\item
  For $G=\Z_2$ we obtain the dislocations of~\cite{We71} as the
  simplest dual observables. Their expectation value is again related
  to their free energies.
\end{myenumerate}
\end{remark}

There are more general functions ${(\Irrep_G)}^V\to\C$ whose
expectation value under the dual partition function can be
calculated. Let $e_0\in E$ be an edge and $\sigma\in\Irrep_G$ an
irreducible representation of $G$. Then we can study the indicator
function,
\begin{equation}
\label{eq_chiral_indicator}
  \sym{O}_{e_0,\sigma}({\{\tau_e\}}_{e\in E})=\delta_{\tau_{e_0}\sigma},
\end{equation}
which probes whether the representation $\sigma$ is assigned to the
edge $e_0$. The centre dislocations can be expressed as linear
combinations of these indicator functions,
\begin{equation}
  \sym{O}_X({\{\tau_e\}}_{e\in E}) 
  = \Bigl(\prod_{e\in E}\sum_{\sigma_e\in\Irrep_G}\Bigr)
    \prod_{e\in E}\sym{O}_{e,\sigma_e}\tilde t^{(\sigma_e)}(X_e).
\end{equation}
The expectation value of an indicator
function~\eqref{eq_chiral_indicator} under the dual partition
function~\eqref{eq_chiral_dualpartition} is then equal to
\begin{equation}
  {\left<\sym{O}_{e_0,\sigma}\right>}_{\mathrm{dual}}
  = \frac{1}{Z}\Bigl(\prod_{v\in V}\int_G\,dg_v\Bigr)
  \prod_{e\in E}\tilde w^{(e_0,\sigma)}_e(\alignidx{g_{\del_+e}\cdot g_{\del_-e}^{-1}}),
\end{equation}
where the Boltzmann weight $w(g)$ is modified at the edge $e_0$,
\begin{equation}
  \tilde w^{(e_0,\sigma)}_e(g)=\left\{
    \begin{matrix}
      w(g),&\mbox{if $e\neq e_0$},\\
      \sum\limits_{\rho\in\Irrep_G}\delta_{\rho,\sigma}\hat w_\rho\,\chi^{(\rho)}(g),
        &\mbox{if $e=e_0$},
    \end{matrix}\right.
\end{equation}
In general, a function involving the indicator functions in the dual
formulation leads to a convolution of the Boltzmann weight in the
original picture.

\begin{remark}
\begin{myenumerate}
\item 
  The definition of dual expectation values presented here is
  restricted to functions of the irreducible representations at the
  edges. It is also conceivable to make use of functions of the
  intertwiners at the vertices.
\item
  Indicator functions similar to~\eqref{eq_chiral_indicator} have been
  used to construct geometrical observables in the spin foam model of
  three-dimensional quantum gravity~\cite{Ba02}.
\end{myenumerate}
\end{remark}

\subsection{The strong-weak relation}

The dual partition function~\eqref{eq_chiral_dualpartition} of the
lattice chiral model is strong-weak dual to the original
formulation~\eqref{eq_chiral_partition}. This follows from the
properties of the character expansion of the Boltzmann weight and is
most transparent for the heat kernel action~\eqref{eq_heatkernel}. The
only $\beta$-dependent term of the dual partition function is the
product
\begin{equation}
\label{eq_dualweight}
  \prod_{e\in E}\hat w_{\tau_e} = \exp\bigl(-\frac{1}{2\beta}
    \sum_{e\in E}C^{(2)}_{\tau_e}\bigr),
\end{equation}
where the inverse temperature $\beta$ appears in the denominator! The
result for the Wilson action of $G=\U(1)$ or $G=\SU(2)$ looks more
complicated and involves modified Bessel functions, but it is
qualitatively quite similar. In all these cases, the term
corresponding to~\eqref{eq_dualweight} has a sharp peak as a function
of the $C^{(2)}_\tau$ if $\beta$ is small.

The $\beta$-dependence~\eqref{eq_dualweight} of the dual partition
function also encodes essential information on the strong coupling
expansion of the lattice chiral model. For small $\beta$, the dominant
contribution to~\eqref{eq_dualweight} comes from spin networks
(assignments of representations to the edges of the graph) whose sum
of the quadratic Casimir eigenvalues over all edges is very small. It
is now possible to sort them by the value of this sum so that the
configurations of the dual partition function are precisely the terms
of the strong coupling expansion!

%
\section{The non-linear sigma model}
%
\label{sect_nonlinearmodel}

We construct the lattice non-linear sigma model with variables in some
coset space $G/H$, where $H\leq G$ is a Lie subgroup of $G$, starting
from the chiral model. One half of the $G\times G$-symmetry of the
chiral model is used to couple elements $h\in H$ to the action
term. Integration over $h$ then makes sure that the action is constant
on the cosets $gH$ and therefore defines a model with variables in
$G/H$.

\subsection{Partition function}

\begin{lemma}
\label{lemma_cosetave}
Let $G$ be a compact Lie group and $H\leq G$ be a Lie subgroup. Let
$f\in L^2(G)$ be a class function of $G$ with character expansion
\begin{equation}
  f(g)=\sum_{\rho\in\Irrep_G}\hat f_\rho\chi^{(\rho)}(g).
\end{equation}
\begin{myenumerate}
\item
  For any $g_1,g_2\in G$, 
\begin{equation}
\label{eq_nlin_form}
  \int_Hf(\alignidx{g_1\cdot h\cdot g_2^{-1}})\,dh
  = \sum_{\rho\in\Irrep^G_H}\hat f_\rho\sum_{j=1}^{\dim V_\rho}\sum_{k=1}^{\kappa_\rho}
    H^{(\rho)}_{jk}(g_1)H^{(\rho^\ast)}_{jk}(g_2),
\end{equation}
using the conventions of Section~\ref{sect_coset}. 
\item
  The function $f$ defines a map $\tilde f\colon G/H\times G/H\to\C$,
\begin{equation}
  \tilde f(x,y):=\int_Hf(\alignidx{g_x\cdot h\cdot g_y^{-1}})\,dh,
\end{equation}
  where $g_x,g_y\in G$ denote representatives of the cosets $x,y\in
  G/H$. 
\item
  The function $\tilde f(x,y)$ has a global left $G$-symmetry, \ie\
  for any $g\in G$, $x,y\in G/H$,
\begin{equation}
\label{eq_nlin_leftsym}
  \tilde f(g\cdot x,g\cdot y) = \tilde f(x,y).
\end{equation}
\item
  If in addition $f(g^{-1})=f(g)$, then $\tilde f(x,y)=\tilde
  f(y,x)$.
\end{myenumerate}
\end{lemma}

\begin{remark}
If $H$ is a massive subgroup of $G$, then $\kappa_\rho=1$ for the
class-1 representations. In this case, any $L^2$-function $G/H\times
G/H\to\C$ with the symmetry~\eqref{eq_nlin_leftsym} is of the
form~\eqref{eq_nlin_form}. This statement does, however, not extend to
the case of generic Lie subgroups $H\leq G$. We define the lattice
non-linear sigma model for Boltzmann weights of the special
form~\eqref{eq_nlin_form}.
\end{remark}

\begin{definition}
\label{def_nlin_partition}
Let $G$ be a compact Lie group, $H\leq G$ be a Lie subgroup and
$(V,E)$ denote an oriented graph. Let $s\colon G\to\R$ be an
$L^2$-integrable and bounded class function that satisfies
$s(g^{-1})=s(g)$. Construct $\tilde w\colon G/H\times G/H\to\R$ from
$w(g)=\exp(-s(g))$ as in Lemma~\ref{lemma_cosetave}. The \emph{lattice
non-linear sigma model} is defined by the partition function
\begin{equation}
\label{eq_nlin_partition}
  Z=\Bigl(\prod_{v\in V}\int_{G/H}\,dx_v\Bigr)
    \prod_{e\in E}\tilde w(x_{\del_+e},x_{\del_-e}).
\end{equation}
\end{definition}

\begin{proposition}
The lattice non-linear sigma model has got a global left-$G$
symmetry. For any fixed $g\in G$, the transformation
\begin{equation}
\label{eq_nlin_symmetry}
  x_v\mapsto g\cdot x_v,
\end{equation}
for all $v\in V$, is a symmetry of the weight $\tilde
w(x_{\del_+e},x_{\del_-e})$. In the special case in which $H\unlhd G$
is a normal subgroup, there is also a global right-$G/H$ symmetry. Let
$y\in G/H$. Then the transformation
\begin{equation}
\label{eq_nlin_symmetry2}
  x_v\mapsto x_v\cdot y^{-1},
\end{equation}
for all $v\in V$, is also a symmetry of the weight.
\end{proposition}

\begin{example}
The Boltzmann weights $\tilde w(x,y):=\exp(-\tilde s(x,y))$ of the
lattice $N$-vector model (the $\O(N)$ non-linear sigma model) and of
the $\RP^{N-1}$-model are of the type of
Lemma~\ref{lemma_cosetave}. For the $N$-vector model, $G=\O(N)$,
$H=\O(N-1)$ and
\begin{equation}
  \alignidx{\tilde s(\underline{x},\underline{y})=-\beta\underline{x}\cdot\underline{y}},
\end{equation}
where $\alignidx{\underline{x},\underline{y}}\in S^{N-1}\subseteq\R^N$, and the
dot denotes the standard scalar product. For the $\RP^{N-1}$-model,
$G=\O(N)$, $H=\O(N-1)\times\Z_2$, and
\begin{equation}
  \alignidx{\tilde s(\underline{x},\underline{y})=-\frac{\beta}{2}
    {(\underline{x}\cdot\underline{y})}^2},
\end{equation}
for representatives $\alignidx{\underline{x},\underline{y}}$ of
classes in $\RP^{N-1}\cong S^{N-1}/\Z_2$. On cubic lattices, there
exists in both cases a suitable na\"\i ve continuum (or weak field)
limit in which the lattice constant tends to zero and the lattice
action towards the action of the corresponding continuum model.
\end{example}

\begin{remark}
\begin{myenumerate}
\item
  The partition function again depends only on the unoriented graph.
\item
  If $H=\{e\}$ is the trivial group, then any representation function
  is a generalized spherical function. The non-linear sigma model for
  $G/H$ coincides in this case with the chiral model for $G$, and the
  global $G\times G$-symmetry is restored.
\end{myenumerate}
\end{remark}

\subsection{Expectation values}

The observables of the lattice non-linear sigma model can be found by
the same methods as for the chiral model
(Section~\ref{sect_chiral_observable}). The calculation is very
similar so that we just state the results.

\begin{figure}[t]
\begin{center}
\input{pstex/nlin_obs.pstex_t}
\end{center}
\mycaption{fig_nlin_obs}{%
  (a) The observable~\eqref{eq_nlin_observable} of the lattice
  non-linear sigma model on the graph of
  Figure~\ref{fig_chiral_obs}. (b) The same function for the case of
  a massive subgroup $H\leq G$. (c) The
  observable~\eqref{eq_nlin_observable2} if $H\unlhd G$ is a normal subgroup.}
\end{figure}

\begin{theorem}
Each algebraic function ${(G/H)}^V\to\C$ that is compatible with the
global left-$G$ symmetry~\eqref{eq_nlin_symmetry}, is a linear
combination of observables of the following type,
\begin{equation}
\label{eq_nlin_observable}
  f_{\rho,P,k_v\ldots}({\{x_v\}}_{v\in V})
    =\Bigl(\prod_{v\in V}\sum_{\ell_v=1}^{\dim V_{\rho_v}}\Bigr)
    P_{\underbrace{\scriptstyle\ell_v\cdots}_{v\in V}}
    \prod_{v\in V}H^{(\rho_v)}_{\ell_v,k_v}(x_v),
\end{equation}
where
\begin{equation}
  \rho\colon V\to\Irrep^G_H,\qquad v\mapsto\rho_v,
\end{equation}
assigns a class-1 representation of $G$ with respect to $H$ to each
vertex; $k_v\in\{1,\ldots,\kappa_{\rho_v}\}$ for all $v\in V$, and
\begin{equation}
  P\colon\bigotimes_{v\in V}V_{\rho_v}\to\C,
\end{equation}
is an intertwiner of $G$.
\end{theorem}

\begin{remark}
\begin{myenumerate}
\item
  The structure of the function~\eqref{eq_nlin_observable} is
  illustrated in Figure~\ref{fig_nlin_obs}(a).
\item
  The well-known two-point function for a charge-anticharge pair $\rho$,
  $\rho^\ast$ at $v,w\in V$, is a special case,
\begin{equation}
  f_{k_v,k_v}(x_v,x_w)=\sum_{j_v=1}^{\dim V_\rho}
    H^{(\rho)}_{j_vk_v}(x_v)H^{(\rho^\ast)}_{j_wk_w}(x_w)\delta_{j_vj_w},
\end{equation}
  for fixed $k_v,k_w\in\{1,\ldots,\kappa_\rho\}$.
\item
  If $H$ is a massive subgroup of $G$, we have $\kappa_\rho=1$ for the
  class-1 representations so that the indices $k_v$ can be omitted
  from all expressions (Figure~\ref{fig_nlin_obs}(b)).
\end{myenumerate}
\end{remark}

\begin{theorem}
If in addition $H\unlhd G$ is a normal subgroup, then the algebraic
functions ${(G/H)}^V\to\C$ that are compatible with both the global
left-$G$ and the global right-$G/H$ symmetry, are linear combinations
of observables of the following form,
\begin{equation}
\label{eq_nlin_observable2}
  f_{\rho,P,Q}({\{x_v\}}_{v\in V})
  =\Bigl(\prod_{v\in V}\sum_{\ell_v=1}^{\dim V_{\rho_v}}\sum_{k_v=1}^{\kappa_{\rho_v}}\Bigr)
   P_{\underbrace{\scriptstyle\ell_v\ldots}_{v\in V}}
   Q_{\underbrace{\scriptstyle k_v\ldots}_{v\in V}}
   \prod_{v\in V}H^{(\rho_v)}_{\ell_vk_v}(x_v).
\end{equation}
Here 
\begin{equation}
  \rho\colon V\to\Irrep^G_H,\qquad v\mapsto\rho_v,
\end{equation}
assigns a class-1 representation of $G$ with respect to $H$ to each
vertex and
\begin{equation}
  P\colon\bigotimes_{v\in V}V_{\rho_v}\to\C,\qquad
  Q\colon\bigotimes_{v\in V}V_{\rho_v}^\ast\to\C,
\end{equation}
are intertwiners of $G$. 
\end{theorem}

\begin{remark}
\begin{myenumerate}
\item
  Figure~\ref{fig_nlin_obs}(c) illustrates the structure of the
  observables~\eqref{eq_nlin_observable2} if $H\unlhd G$ is a normal
  subgroup. Here the indices $k_v$ of~\eqref{eq_nlin_observable} are
  no longer independent, but rather exhibit a $G/H$-symmetry under
  which invariance is required. Therefore we need the second
  intertwiner $Q$. Furthermore, $\kappa_\rho=\dim V_\rho$ for all
  class-1 representations so that the dashed lines have become solid.
\item
  In particular for $H=\{e\}$, we recover the
  observable~\eqref{eq_chiral_observable} of the chiral model.
\item 
  In order to have non-vanishing expectation values, the observable
  not only has to be invariant under the
  symmetries~\eqref{eq_nlin_symmetry} and~\eqref{eq_nlin_symmetry2}
  (if applicable), but also under orientation reversal
  (Remark~\ref{rem_zeroexpectation}).
\end{myenumerate}
\end{remark}

\subsection{Duality transformation}

The duality transformation for the non-linear sigma model is very
similar to that of the chiral model. We summarize the main steps which
differ from the calculation for the chiral model and focus directly on
the most general case, the dual of an expectation value, from which
the transformation of the partition function can be easily inferred.

We start with an observable $f_{\rho,P,Q}$ of the
form~\eqref{eq_nlin_observable2}. If $H\unlhd G$ is a normal subgroup,
then $Q$ is an intertwiner of $G$. Otherwise, $Q$ is arbitrary so that
we obtain the function~\eqref{eq_nlin_observable} for generic
$k_v\in\{1,\ldots,\kappa_{\rho_v}\}$, $v\in V$.

We start with the expectation value of the
observable~\eqref{eq_nlin_observable2} under the partition
function~\eqref{eq_nlin_partition},
\begin{eqnarray}
\label{eq_nlin_expobservable}
  \left<f_{\rho,P,Q}\right>&=&\frac{1}{Z}
    \Bigl(\prod_{v\in V}\int_{G/H}\,dx_v\Bigr)
    \Bigl(\prod_{e\in E}\tilde w(x_{\del_+e},x_{\del_-e})\Bigr)\nn\\
  &&\times\Bigl(\prod_{v\in V}\sum_{\ell_v=1}^{\dim V_{\rho_v}}\sum_{k_v=1}^{\kappa_{\rho_v}}\Bigr)
    P_{\underbrace{\scriptstyle\ell_v\ldots}_{v\in V}}
    \overline{Q_{\underbrace{\scriptstyle k_v\ldots}_{v\in V}}}
    \prod_{v\in V}H^{(\rho_v)}_{\ell_vk_v}(x_v),
\end{eqnarray}
and insert for each $e\in E$ the expansion of Lemma~\ref{lemma_cosetave},
\begin{equation}
  \tilde w(x_{\del_+e},x_{\del_-e})
  =\sum_{\tau_e\in\Irrep^G_H}\hat w_{\tau_e}\sum_{j_e=1}^{\dim V_{\tau_e}}\sum_{m_e=1}^{\kappa_{\tau_e}}
    H^{(\tau_e)}_{j_em_e}(x_{\del_+e})H^{(\tau_e^\ast)}_{j_em_e}(x_{\del_-e}),
\end{equation}
where the $\hat w_\tau$ are the character
expansion coefficients of the function $w(g)=\exp(-s(g))$ of
Definition~\ref{def_nlin_partition}. The reorganized expression then
reads
\begin{eqnarray}
\label{eq_nlin_step3}
  \left<f_{\rho,P,Q}\right>&=&\frac{1}{Z}
    \Bigl(\prod_{v\in V}\sum_{\ell_v=1}^{\dim V_{\rho_v}}\sum_{k_v=1}^{\kappa_{\rho_v}}\Bigr)
    P_{\underbrace{\scriptstyle\ell_v\ldots}_{v\in V}}
    \overline{Q_{\underbrace{\scriptstyle k_v\ldots}_{v\in V}}}
    \Bigl(\prod_{e\in E}\sum_{\tau_e\in\Irrep^G_H}\Bigr)
    \Bigl(\prod_{e\in E}\hat w_{\tau_e}\Bigr)
    \Bigl(\prod_{e\in E}\sum_{j_e=1}^{\dim V_{\tau_e}}\sum_{m_e=1}^{\kappa_{\tau_e}}\Bigr)\nn\\
  &&\times
    \prod_{v\in V}\int_{G/H}\,dx_v
    \Bigl(\prod_{\ontop{e\in E\colon}{v=\del_+e}}H^{(\tau_e)}_{j_em_e}(x_v)\Bigr)
    \Bigl(\prod_{\ontop{e\in E\colon}{v=\del_-e}}H^{(\tau_e^\ast)}_{j_em_e}(x_v)\Bigr)
    H^{(\rho_v)}_{\ell_vk_v}(x_v),
\end{eqnarray}
so that we can evaluate the integrals over $G/H$
using~\eqref{eq_haarcoset1},
\begin{equation}
  \int_{G/H}\,dx_v\Bigl(\cdots\Bigr) = 
    \sum_{S^{(v)}\in\tilde{\sym{S}}^{(v)}}
    \overline{S^{(v)}_{
      \underbrace{\scriptstyle j_e\ldots}_{\ontop{e\in E\colon}{v=\del_+e}}
      \underbrace{\scriptstyle j_e\ldots}_{\ontop{e\in E\colon}{v=\del_-e}}\ell_v}}
    S^{(v)}_{
      \underbrace{\scriptstyle m_e\ldots}_{\ontop{e\in E\colon}{v=\del_+e}}
      \underbrace{\scriptstyle m_e\ldots}_{\ontop{e\in E\colon}{v=\del_-e}}k_v}.
\end{equation}
Here $\tilde{\sym{S}}^{(v)}$, $v\in V$, denotes a basis of
$G$-invariant projectors
\begin{equation}
\label{eq_nlin_step1}
  \Bigl(\bigotimes_{\ontop{e\in E\colon}{v=\del_+e}}V_{\tau_e}\Bigr)\otimes
  \Bigl(\bigotimes_{\ontop{e\in E\colon}{v=\del_-e}}V_{\tau_e}^\ast\Bigr)\otimes
  V_{\rho_v}\to\C
\end{equation}
with the usual normalization. We obtain the following result.

\begin{figure}[t]
\begin{center}
\input{pstex/nlin_dual.pstex_t}
\end{center}
\mycaption{fig_nlin_dual}{%
  (a) The structure of the dual form~\eqref{eq_nlin_dualobservable2} for
  the expectation value of an observable of the lattice non-linear
  sigma model on the graph of Figure~\ref{fig_chiral_obs}(a). (b) The
  special case of a massive subgroup $H\leq G$. (c) The situation for
  a normal subgroup $H\unlhd G$.}
\end{figure}

\begin{theorem}[Dual observable]
Let $G$ be a compact Lie group, $H\leq G$ a Lie subgroup and $(V,E)$
denote an oriented graph. The expectation
value~\eqref{eq_nlin_expobservable} of the observable of the lattice
non-linear sigma model is equal to the expressions
\begin{eqnarray}
\label{eq_nlin_dualobservable2}
  \left<f_{\rho,P,Q}\right>&=&\frac{1}{Z}
    \Bigl(\prod_{v\in V}\sum_{\ell_v=1}^{\dim V_{\rho_v}}\sum_{k_v=1}^{\kappa_{\rho_v}}\Bigr)
    P_{\underbrace{\scriptstyle\ell_v\ldots}_{v\in V}}
    \overline{Q_{\underbrace{\scriptstyle k_v\ldots}_{v\in V}}}
    \Bigl(\prod_{e\in E}\sum_{\tau_e\in\Irrep^G_H}\Bigr)
    \Bigl(\prod_{e\in E}\hat w_{\tau_e}\Bigr)\nn\\
  &&\times\Bigl(\prod_{e\in E}\sum_{j_e=1}^{\dim V_{\tau_e}}\sum_{m_e=1}^{\kappa_{\tau_e}}\Bigr)
    \prod_{v\in V}
    I^{(v)}_{\underbrace{\scriptstyle j_e\ldots}_{\ontop{e\in E\colon}{v=\del_+e}}
             \underbrace{\scriptstyle j_e\ldots}_{\ontop{e\in E\colon}{v=\del_-e}}\ell_v;
             \underbrace{\scriptstyle m_e\ldots}_{\ontop{e\in E\colon}{v=\del_+e}}
             \underbrace{\scriptstyle m_e\ldots}_{\ontop{e\in E\colon}{v=\del_-e}}k_v}\\
  &=&\frac{1}{Z}
\label{eq_nlin_dualobservable}
    \Bigl(\prod_{v\in V}\sum_{\ell_v=1}^{\dim V_{\rho_v}}\sum_{k_v=1}^{\kappa_{\rho_v}}\Bigr)
    P_{\underbrace{\scriptstyle\ell_v\ldots}_{v\in V}}
    \overline{Q_{\underbrace{\scriptstyle k_v\ldots}_{v\in V}}}\nn\\
  &&\times\Bigl(\prod_{e\in E}\sum_{\tau_e\in\Irrep^G_H}\Bigr)
    \Bigl(\prod_{v\in V}\sum_{S^{(v)}\in\tilde{\sym{S}}^{(v)}}\Bigr)
    \Bigl(\prod_{e\in E}\hat w_{\tau_e}\Bigr)\nn\\
  &&\times\Bigl(\prod_{e\in E}\sum_{j_e=1}^{\dim V_{\tau_e}}\sum_{m_e=1}^{\kappa_{\tau_e}}\Bigr)
    \prod_{v\in V}
    \overline{S^{(v)}_{
      \underbrace{\scriptstyle j_e\ldots}_{\ontop{e\in E\colon}{v=\del_+e}}
      \underbrace{\scriptstyle j_e\ldots}_{\ontop{e\in E\colon}{v=\del_-e}}\ell_v}}
    S^{(v)}_{
      \underbrace{\scriptstyle m_e\ldots}_{\ontop{e\in E\colon}{v=\del_+e}}
      \underbrace{\scriptstyle m_e\ldots}_{\ontop{e\in E\colon}{v=\del_-e}}k_v}.
\end{eqnarray}
Here $\tilde{\sym{S}}^{(v)}$, $v\in V$, denotes a basis of
$G$-invariant projectors~\eqref{eq_nlin_step1}, and the $\hat w_\tau$
are the character expansion coefficients of the function
$w(g)=\exp(-s(g))$ where $s(g)$ is the class function of
Definition~\ref{def_nlin_partition}. The coset space Haar map
$I^{(v)}$, $v\in V$, in~\eqref{eq_nlin_dualobservable2} is a map
\begin{equation}
  \Bigl(\bigotimes_{\ontop{e\in E\colon}{v=\del_+e}}V_{\tau_e}\Bigr)\otimes
  \Bigl(\bigotimes_{\ontop{e\in E\colon}{v=\del_-e}}V_{\tau_e}\Bigr)\otimes
    V_{\rho_v}\to 
  \Bigl(\bigotimes_{\ontop{e\in E\colon}{v=\del_+e}}V_{\tau_e}\Bigr)\otimes
  \Bigl(\bigotimes_{\ontop{e\in E\colon}{v=\del_-e}}V_{\tau_e}\Bigr)\otimes
    V_{\rho_v}.
\end{equation}
\end{theorem}

\begin{figure}[t]
\begin{center}
\input{pstex/nlin_dual2.pstex_t}
\end{center}
\mycaption{fig_nlin_dual2}{%
  (a) The structure of the dual partition
  function~\eqref{eq_nlin_dualpartition2} of the lattice non-linear sigma
  model on the graph of Figure~\ref{fig_chiral_obs}. (b) The special
  case of a massive subgroup $H\leq G$.}
\end{figure}

\begin{remark}
\begin{myenumerate}
\item
  The dual expression~\eqref{eq_nlin_dualobservable2} for the
  observable of the non-linear sigma model is very similar to the dual
  observable of the chiral model in
  Theorem~\ref{thm_chiral_dualobservable}. The differences are the
  ranges of the indices which follow from the choice of the subgroup
  $H\leq G$. The structure of the dual observable is illustrated in
  Figure~\ref{fig_nlin_dual}(a) if $H\leq G$ is a generic, non-normal
  subgroup, in~(b) if $H$ is a massive subgroup and in~(c) for the
  case of a normal subgroup $H\unlhd
  G$. Figure~\ref{fig_nlin_dual}(a-c) correspond
  to~\eqref{eq_nlin_dualobservable2}. The diagrams for the other
  formulation~\eqref{eq_nlin_dualobservable} are obtained by the
  replacements shown in Figure~\ref{fig_diagcoset}(c)
  or~\ref{fig_diagmassive}(c).
\item
  Again, the dual expression for the observable of the chiral model
  can be obtained from~\eqref{eq_nlin_dualobservable} for a trivial
  subgroup $H=\{e\}$.
\item
  If one seeks a purely categorial picture of the dual non-linear
  sigma model, one should generally view all representations as
  representations of $H$. Otherwise the integrals over $H$ which are
  still implicitly present in the spherical functions, would not be
  honest intertwiners. The dashed lines with open ends labelled $k_v$
  then enumerate different trivial representations of $H$. The special
  cases of massive and normal subgroups, however, are easier and can
  be handled already in the context of the representations of $G$.
\end{myenumerate}
\end{remark}

The dual expression for the partition function can be calculated by
specializing the numerator of~\eqref{eq_nlin_dualobservable2} to the
trivial observable. This result is given in the following corollary
and visualized in Figure~\ref{fig_nlin_dual2}.

\begin{corollary}[Dual partition function]
Let $G$ be a compact Lie group with a Lie subgroup $H\leq G$ and
$(V,E)$ be an oriented graph. The partition
function~\eqref{eq_nlin_partition} of the lattice non-linear sigma
model is equal to
\begin{eqnarray}
\label{eq_nlin_dualpartition2}
  Z &=& \Bigl(\prod_{e\in E}\sum_{\tau_e\in\Irrep^G_H}\Bigr)
    \Bigl(\prod_{e\in E}\hat w_{\tau_e}\Bigr)
    \Bigl(\prod_{e\in E}\sum_{j_e=1}^{\dim V_{\tau_e}}\sum_{m_e=1}^{\kappa_{\tau_e}}\Bigr)
    \prod_{v\in V}
    I^{(v)}_{\underbrace{\scriptstyle j_e\ldots}_{\ontop{e\in E\colon}{v=\del_+e}}
             \underbrace{\scriptstyle j_e\ldots}_{\ontop{e\in E\colon}{v=\del_-e}};
             \underbrace{\scriptstyle m_e\ldots}_{\ontop{e\in E\colon}{v=\del_+e}}
             \underbrace{\scriptstyle m_e\ldots}_{\ontop{e\in E\colon}{v=\del_-e}}}\\
\label{eq_nlin_dualpartition}
  &=&\Bigl(\prod_{e\in E}\sum_{\tau_e\in\Irrep^G_H}\Bigr)
    \Bigl(\prod_{v\in V}\sum_{S^{(v)}\in\sym{S}^{(v)}}\Bigr)
    \Bigl(\prod_{e\in E}\hat w_{\tau_e}\Bigr)\nn\\
  &&\times\Bigl(\prod_{e\in E}\sum_{j_e=1}^{\dim V_{\tau_e}}\sum_{m_e=1}^{\kappa_{\tau_e}}\Bigr)
    \prod_{v\in V}
    \overline{S^{(v)}_{
      \underbrace{\scriptstyle j_e\ldots}_{\ontop{e\in E\colon}{v=\del_+e}}
      \underbrace{\scriptstyle j_e\ldots}_{\ontop{e\in E\colon}{v=\del_-e}}}}
    S^{(v)}_{
      \underbrace{\scriptstyle m_e\ldots}_{\ontop{e\in E\colon}{v=\del_+e}}
      \underbrace{\scriptstyle m_e\ldots}_{\ontop{e\in E\colon}{v=\del_-e}}}.
\end{eqnarray}
Here $\sym{S}^{(v)}$, $v\in V$, denotes a basis of $G$-invariant
projectors
\begin{equation}
  \Bigl(\bigotimes_{\ontop{e\in E\colon}{v=\del_+e}}V_{\tau_e}\Bigr)\otimes
  \Bigl(\bigotimes_{\ontop{e\in E\colon}{v=\del_-e}}V_{\tau_e}^\ast\Bigr)\to\C
\end{equation}
with the usual normalization, and the coset space Haar map $I^{(v)}$
is a linear map
\begin{equation}
  \Bigl(\bigotimes_{\ontop{e\in E\colon}{v=\del_+e}}V_{\tau_e}\Bigr)\otimes
  \Bigl(\bigotimes_{\ontop{e\in E\colon}{v=\del_-e}}V_{\tau_e}\Bigr)\to
  \Bigl(\bigotimes_{\ontop{e\in E\colon}{v=\del_+e}}V_{\tau_e}\Bigr)\otimes
  \Bigl(\bigotimes_{\ontop{e\in E\colon}{v=\del_-e}}V_{\tau_e}\Bigr).
\end{equation}
\end{corollary}

\subsection{Expectation values of the dual model}

If the natural observables of the dual partition function are again
constructed from the labelling of the edges with representations, the
result is the same as for the lattice chiral model in
Section~\ref{sect_chiral_dualexp}, restricted to the class-1
representations. The analogue of~\eqref{eq_chiral_origcentre} is then
\begin{equation}
  {\left<\sym{O}_X\right>}_{\mathrm{dual}} =
    \frac{1}{Z}\Bigl(\prod_{v\in V}\int_{G/H}\,dx_v\Bigr)
    \prod_{e\in E}\tilde w(\alignidx{X_e\cdot x_{\del_+e};x_{\del_-e}}).
\end{equation}

%
\section{The generalized Higgs models}
%
\label{sect_higgs}

In this section, we couple the chiral model and the non-linear sigma
model to lattice gauge theory. In some particular cases, this yields
certain Higgs models with frozen radial component which motivates the
title of this section. Before we study the coupled models, it is
useful to summarize the results of the duality transformation for
lattice gauge theory~\cite{OePf01,Pf01,PfOe02} in the language of the
present article.

\subsection{Lattice gauge theory}

\begin{definition}
Let $G$ be a compact Lie group, $(V,E,F)$ be an oriented two-complex
and $s\colon G\to\R$ be an $L^2$-integrable class function of $G$ that
is bounded below and satisfies $s(g^{-1})=s(g)$ for all $g\in G$. The
partition function of \emph{lattice gauge theory} with gauge group $G$
is defined by
\begin{equation}
\label{eq_lgt_partition}
  Z = \Bigl(\prod_{e\in E}\int_G\,dg_e\Bigr)\,
      \prod_{f\in F}u(g_f),\qquad
  g_f:=g_{\del_1f}^{\epsilon_1f}\cdots g_{\del_{N(f)}f}^{\epsilon_{N(f)}f},
\end{equation}
where $u(g)=\exp(-s(g))$.
\end{definition}

The set of configurations of lattice gauge theory is the product $G^E$
of one copy of $G$ for each edge $e\in E$. The ordered product of
group elements attached to the edges in the boundary of the face $f\in
F$ is denoted by $g_f$. The Boltzmann weight exhibits a local gauge
symmetry.

\begin{proposition}
Let $h\colon V\to G,v\mapsto h_v$ associate a group element with each
vertex. The Boltzmann weight $u(g_f)=\exp(-s(g_f))$
in~\eqref{eq_lgt_partition} is invariant under the local gauge
transformations
\begin{equation}
\label{eq_lgt_symmetry}
  g_e\mapsto\alignidx{h_{\del_+e}\cdot g_e\cdot
  h_{\del_-e}^{-1}},
\end{equation}
for all $e\in E$.
\end{proposition}

This definition of lattice gauge theory is motivated by the fact that
on regular hypercubic lattices, the Wilson action tends towards the
continuum Yang--Mills action in the weak field limit of vanishing
lattice constant. The group elements $g_e$ attached to the edges of
the lattice correspond to the parallel transports of the gauge
connection along these edges. For more details on lattice gauge
theory, see, for example~\cite{Ro92,MoMu94}.

The most general observable of lattice gauge theory whose expectation
value under the partition function~\eqref{eq_lgt_partition} can be
calculated, is constructed from spin networks. Each algebraic function
$G^E\to\C$ that is invariant under the
transformation~\eqref{eq_lgt_symmetry}, is a linear combination of
\emph{spin network functions}. They generalize the notion of Wilson
loops and are defined as follows.

\begin{definition}
\label{def_spinnetfunction}
Let $G$ be a compact Lie group, $(V,E,F)$ be an oriented two-complex
and $(\sigma,Q)$ be a spin network
(Definition~\ref{def_spinnetwork}). The \emph{spin network function}
of $(\sigma,Q)$ associates with each configuration a complex number,
\begin{equation}
\label{eq_lgt_observable}
  W_{\sigma,Q}({\{g_e\}}_{e\in E}):=
    \Bigl(\prod_{e\in E}\sum_{k_e,\ell_e=1}^{\dim V_{\sigma_e}}\Bigr)\,
    \Bigl(\prod_{e\in E}t_{k_e\ell_e}^{(\sigma_e)}(g_e)\Bigr)\,
    \Bigl(\prod_{v\in V}
      Q^{(v)}_{\underbrace{\scriptstyle \ell_e\ldots}_{\ontop{e\in E\colon}{v=\del_-e}}
               \underbrace{\scriptstyle k_e\ldots}_{\ontop{e\in E\colon}{v=\del_+e}}}\Bigr).
\end{equation}
\end{definition}

\begin{remark}
\begin{myenumerate}
\item
  The above definition uses the spin network $(\sigma,Q)$ to label
  edges with representations and vertices with intertwiners, and then
  employs a representation function for each edge in order to obtain a
  function $G^E\to\C$.
\item 
  All edges $e\in E$ for which $V_{\sigma_e}\cong\C$ is the trivial
  representation, contribute only a factor $1$ to the
  expression~\eqref{eq_lgt_observable}. For an ordinary Wilson loop,
  for example, all edges are labelled with the trivial representation
  except for those edges that are part of the loop. These are labelled
  with the fundamental representation of $G$. The intertwiners
  $Q^{(v)}$ (if non-vanishing) are in this case uniquely determined up
  to normalization.
\item
  The spin network function~\eqref{eq_lgt_observable} can be
  \emph{evaluated} by putting $g_e=e$ (group unit) for all edges $e\in
  E$. The result is an invariant of $G$ which is often called the
  \emph{value} of the spin network $(\sigma,Q)$.
\item 
  If $G$ is Abelian, then the set $\Irrep_G$ of irreducible
  representations forms an additive group, and all irreducible
  representations are one-dimensional. Thus $W_{\sigma,Q}$ can be
  decomposed into a sum of products of ordinary Wilson loops.
\end{myenumerate}
\end{remark}

We have the following dual expressions for the partition function and
the expectation value of a spin network function~\cite{OePf01,Pf01}.

\begin{figure}[t]
\begin{center}
\input{pstex/lgt_cv.pstex_t}
\end{center}
\mycaption{fig_lgt_cv}{%
  (a) A two-complex with a vertex $v$ attached to four edges. There
  are six faces, one between each pair of edges. (b) The spin network
  $C(v)$ of~\eqref{eq_lgt_cv} that appears in the dual partition function
  of lattice gauge theory and (c) the spin network~\eqref{eq_lgt_cv2}
  from the dual of an expectation value.}
\end{figure}

\begin{theorem}[Dual partition function]
\label{thm_lgt_dualpartition}
Let $G$ be a compact Lie group. The partition
function~\eqref{eq_lgt_partition} of lattice gauge theory is equal to
the expression
\begin{equation}
\label{eq_lgt_dualpartition}
  Z = \Bigl(\prod_{f\in F}\sum_{\tau_f\in\Irrep_G}\Bigr)\,
      \Bigl(\prod_{e\in E}\sum_{U^{(e)}\in\sym{U}^{(e)}}\Bigr)\,
      \Bigl(\prod_{f\in F}\hat u_{\tau_f}\Bigr)\,
      \Bigl(\prod_{v\in V}C(v)\Bigr).
\end{equation}
Here $\sym{U}^{(e)}$, $e\in E$, denotes a basis of $G$-invariant
projectors
\begin{equation}
\label{eq_lgt_step1}
  \bigl(\bigotimes_{f\in e_+}V_{\tau_f}\bigr)\otimes
  \bigl(\bigotimes_{f\in e_-}V_{\tau_f}^\ast\bigr)\to\C.
\end{equation}
The $\hat u_\tau$ are the coefficients of the character expansion of
the Boltzmann weight $u(g)$. The weights per vertex $C(v)$ are given
by a trace involving representations and projectors in the
neighbourhood of the vertex $v\in V$,
\begin{equation}
\label{eq_lgt_cv}
  C(v) = \Bigl(\prod_{\ontop{f\in F\colon}{v\in f_0}}\sum_{n_f=1}^{\dim V_{\tau_f}}\Bigr)\,
         \Bigl(\prod_{\ontop{e\in E\colon}{v=\del_+e}}
           \overline{U^{(e)}_{\underbrace{\scriptstyle n_fn_f\ldots}_{f\in e_+}
                              \underbrace{\scriptstyle n_fn_f\ldots}_{f\in e_-}}}\Bigr)\,
         \Bigl(\prod_{\ontop{e\in E\colon}{v=\del_-e}}
                     U^{(e)}_{\underbrace{\scriptstyle n_fn_f\ldots}_{f\in e_+}
                              \underbrace{\scriptstyle n_fn_f\ldots}_{f\in e_-}}\Bigr).
\end{equation}
Here the range $f\in F\colon v\in f_0$ of the first product refers to
all faces $f\in F$ that contain the vertex $v$ in their boundary,
the second product is over the range $e\in E\colon v=\del_+e$ of all
edges that have $v$ as their endpoint, \etc, see Section~\ref{sect_graphs}.
\end{theorem}

\begin{figure}[t]
\begin{center}
\input{pstex/dual_lgt.pstex_t}
\end{center}
\mycaption{fig_dual_lgt}{%
  An edge $e\in E$ in the boundary of three faces, two triangles and
  one quadrilateral. (a) The structure of the spin networks $C(v)$
  in the dual partition function of lattice gauge
  theory~\eqref{eq_lgt_dualpartition}. (b) The alternative
  formulation~\eqref{eq_lgt_step3} using the Haar intertwiner. 
  We have omitted labels and arrows in both diagrams.}
\end{figure}

\begin{remark}
\begin{myenumerate}
\item
  For each edge $e\in E$, the projectors~\eqref{eq_lgt_step1} are
  related by natural isomorphisms to intertwiners
\begin{equation}
\label{eq_lgt_step2}
  \bigotimes_{f\in e_+}V_{\tau_f}\to
  \bigotimes_{f\in e_-}V_{\tau_f},
\end{equation}
  from the tensor product of the representations at the `incoming'
  faces to the tensor product at the `outgoing' ones. 
\item
  The dual partition function~\eqref{eq_lgt_dualpartition} labels the
  faces with irreducible representations of $G$ and the edges with
  compatible intertwiners in the sense of~\eqref{eq_lgt_step2}. The
  configurations of the dual partition function are therefore spin
  foams (Definition~\ref{def_spinfoam}) so that the dual model is a
  spin foam model. Compared with the situation for the sigma models,
  all the labels appear one level `higher', \ie\ at the faces rather
  than at the edges, and at the edges rather than the vertices.
\item
  The expression $C(v)$ for given projectors $U^{(e)}$ is itself a
  spin network. Figure~\ref{fig_lgt_cv}(b) visualizes it for a
  vertex with four edges attached. In particular, for $G=\SU(2)$, the
  $C(v)$ are the $6j$-symbols of $\SU(2)$. The collection of all
  $C(v)$ in a two-complex is illustrated in
  Figure~\ref{fig_dual_lgt}(a).
\item
  The spin networks of the dual partition function for lattice gauge
  theory decompose into one independent $C(v)$ for each vertex. This
  is a consequence of the local $G$-symmetry and is in contrast to the
  chiral model whose dual partition function involves two spin
  networks that extend over the entire graph, reflecting the two-fold
  global $G$-symmetry. For the non-linear sigma model with a massive
  subgroup $H\leq G$, the dual partition function still contains one
  spin network that extends over the entire graph which corresponds to
  a single global $G$-symmetry.
\item
  Again there exists an alternative formulation using the Haar
  intertwiner rather than the sum over projectors which is stated in
  the following corollary. This result agrees with the purely
  diagrammatical picture of~\cite{Oe02} and is illustrated in
  Figure~\ref{fig_dual_lgt}(b). Upon use of~\eqref{eq_haardecompose},
  we recover~\eqref{eq_lgt_dualpartition} and
  Figure~\ref{fig_dual_lgt}(a). 
\end{myenumerate}
\end{remark}

\begin{corollary}
Let $G$ be a compact Lie group and $(V,E,F)$ denote an oriented
two-complex. The partition function of lattice gauge
theory~\eqref{eq_lgt_dualpartition} is equal to
\begin{eqnarray}
\label{eq_lgt_step3}
  Z &=& \Bigl(\prod_{f\in F}\sum_{\tau_f\in\Irrep_G}\Bigr)\,
        \Bigl(\prod_{f\in F}\hat u_{\tau_f}\Bigr)\,
        \Bigl(\prod_{f\in F}\prod_{v\in f_0}\sum_{n(f,v)=1}^{\dim V_{\tau_f}}\Bigr)\nn\\
    &&\times\prod_{e\in E}
      T^{(e)}_{\underbrace{\scriptstyle n(f,\del_+e)\ldots}_{f\in e_+}
               \underbrace{\scriptstyle n(f,\del_+e)\ldots}_{f\in e_-};
               \underbrace{\scriptstyle n(f,\del_-e)\ldots}_{f\in e_+}
               \underbrace{\scriptstyle n(f,\del_-e)\ldots}_{f\in e_-}},
\end{eqnarray}
where $T^{(e)}$, $e\in E$, denotes the Haar
intertwiner~\eqref{eq_haarinter} for the following representations,
\begin{equation}
  T^{(e)}\colon\Bigl(\bigotimes_{f\in e_+}V_{\tau_f}\Bigr)\otimes
               \Bigl(\bigotimes_{f\in e_-}V_{\tau_f}^\ast\Bigr)\to
               \Bigl(\bigotimes_{f\in e_+}V_{\tau_f}\Bigr)\otimes
               \Bigl(\bigotimes_{f\in e_-}V_{\tau_f}^\ast\Bigr).
\end{equation}
\end{corollary}

Finally, the analogous statements are available for expectation values
of spin network functions.

\begin{theorem}[Dual observable]
Let $G$ be a compact Lie group, $(V,E,F)$ be an oriented two-complex
and $Z$ denote the partition function~\eqref{eq_lgt_partition} of
lattice gauge theory. The expectation value of the spin network
function~\eqref{eq_lgt_observable}, 
\begin{eqnarray}
  \left<W_{\sigma,Q}\right> 
    &=& \frac{1}{Z} \Bigl(\prod_{e\in E}\int_G\,dg_e\Bigr)\,
       \prod_{f\in F}u(g_{\del_1f}^{\epsilon_1f}\cdots g_{\del_{N(f)}f}^{\epsilon_{N(f)}f})\nn\\
    &&\times
    \Bigl(\prod_{e\in E}\sum_{k_e,\ell_e=1}^{\dim V_{\sigma_e}}\Bigr)\,
    \Bigl(\prod_{e\in E}t_{k_e\ell_e}^{(\sigma_e)}(g_e)\Bigr)\,
    \Bigl(\prod_{v\in V}
      Q^{(v)}_{\underbrace{\scriptstyle \ell_e\ldots}_{\ontop{e\in E\colon}{v=\del_-e}}
               \underbrace{\scriptstyle k_e\ldots}_{\ontop{e\in E\colon}{v=\del_+e}}}\Bigr),
\end{eqnarray}
is equal to the following expressions,
\begin{eqnarray}
\label{eq_lgt_dualobservable}
  \left<W_{\sigma,Q}\right> 
    &=& \frac{1}{Z} \Bigl(\prod_{f\in F}\sum_{\tau_f\in\Irrep_G}\Bigr)
      \Bigl(\prod_{f\in F}\hat u_{\tau_f}\Bigr)
      \Bigl(\prod_{f\in F}\prod_{v\in f_0}\sum_{n(f,v)=1}^{\dim V_{\tau_f}}\Bigr)
      \Bigl(\prod_{e\in E}\sum_{k_e,\ell_e=1}^{\dim V_{\sigma_e}}\Bigr)\nn\\
    &&\times\Bigl(\prod_{v\in V}Q^{(v)}_{
        \underbrace{\scriptstyle\ell_e\ldots}_{\ontop{e\in E\colon}{v=\del_-e}}
        \underbrace{\scriptstyle k_e\ldots}_{\ontop{e\in E\colon}{v=\del_+e}}}\Bigr)
      \prod_{e\in E}
      T^{(e)}_{\underbrace{\scriptstyle n(f,\del_+e)\ldots}_{f\in e_+}
               \underbrace{\scriptstyle n(f,\del_+e)\ldots}_{f\in e_-}k_e;
               \underbrace{\scriptstyle n(f,\del_-e)\ldots}_{f\in e_+}
               \underbrace{\scriptstyle n(f,\del_-e)\ldots}_{f\in e_-}\ell_e}\\
    &=& \frac{1}{Z} \Bigl(\prod_{f\in F}\sum_{\tau_f\in\Irrep_G}\Bigr)\,
      \Bigl(\prod_{e\in E}\sum_{U^{(e)}\in\tilde{\sym{U}}^{(e)}}\Bigr)\,
      \Bigl(\prod_{f\in F}\hat u_{\tau_f}\Bigr)\,
      \Bigl(\prod_{v\in V}\tilde{C}(v)\Bigr).
\end{eqnarray}
Here $\tilde{\sym{U}}^{(e)}$, $e\in E$, denotes a basis of
$G$-invariant projectors
\begin{equation}
  \Bigl(\bigotimes_{f\in e_+}V_{\tau_f}\Bigr)\otimes
  \Bigl(\bigotimes_{f\in e_-}V_{\tau_f}^\ast\bigr)\otimes V_{\sigma_e}\to\C.
\end{equation}
The weights per vertex $\tilde C(v)$ are given by a trace involving
representations and projectors in the neighbourhood of the vertex
$v\in V$,
\begin{eqnarray}
\label{eq_cprime}
\label{eq_lgt_cv2}
  \tilde C(v)&=&
    \Bigl(\prod_{\ontop{e\in E\colon}{v=\del_+e}}\sum_{k_e=1}^{\dim V_{\sigma_e}}\Bigr)\,
    \Bigl(\prod_{\ontop{e\in E\colon}{v=\del_-e}}\sum_{\ell_e=1}^{\dim V_{\sigma_e}}\Bigr)\,
      Q^{(v)}_{\underbrace{\scriptstyle \ell_e\ldots}_{\ontop{e\in E\colon}{v=\del_-e}}
               \underbrace{\scriptstyle k_e\ldots}_{\ontop{e\in E\colon}{v=\del_+e}}}\,
    \Bigl(\prod_{\ontop{f\in F\colon}{v\in f_0}}\sum_{n_f=1}^{\dim V_{\tau_f}}\Bigr)\nn\\
  &&\times
    \Bigl(\prod_{\ontop{e\in E\colon}{v=\del_+e}}
           \overline{U^{(e)}_{\underbrace{\scriptstyle n_fn_f\ldots}_{f\in e_+}
                              \underbrace{\scriptstyle n_fn_f\ldots}_{f\in e_-}k_e}}\Bigr)\,
         \Bigl(\prod_{\ontop{e\in E\colon}{v=\del_-e}}
                     U^{(e)}_{\underbrace{\scriptstyle n_fn_f\ldots}_{f\in e_+}
                              \underbrace{\scriptstyle n_fn_f\ldots}_{f\in e_-}\ell_e}\Bigr).
\end{eqnarray}
The Haar intertwiner $T^{(e)}$, $e\in E$,
in~\eqref{eq_lgt_dualobservable} is a map
\begin{equation}
  T^{(e)}\colon
    \Bigl(\bigotimes_{f\in e_+}V_{\tau_f}\Bigr)\otimes
    \Bigl(\bigotimes_{f\in e_-}V_{\tau_f}^\ast\Bigr)\otimes V_{\sigma_e}\to
    \Bigl(\bigotimes_{f\in e_+}V_{\tau_f}\Bigr)\otimes
    \Bigl(\bigotimes_{f\in e_-}V_{\tau_f}^\ast\Bigr)\otimes V_{\sigma_e}.
\end{equation}
\end{theorem}

\begin{remark}
\begin{myenumerate}
\item
  The general pattern is already familiar: The dual of the expectation
  value is given by a ratio of partition functions whose numerator is
  a modification of the partition function, here given by the
  background spin network $(\sigma,Q)$ to which the spin foams
  couple. The structure remains unchanged, just the compatibility
  condition is modified so that the numerator of the dual expectation
  value is given by a sum over all spin foams bounded by the spin
  network $(\sigma,Q)$ (Definition~\ref{def_spinfoam_bounded}).
\item
  The spin networks $\tilde C(v)$ of~\eqref{eq_lgt_cv2} are shown in
  Figure~\ref{fig_lgt_cv}(c). Compared with (b), there is in addition
  a piece of the spin network $(\sigma,Q)$ in the middle of the
  diagram.
\end{myenumerate}
\end{remark}

Similarly to the sigma models, we can again ask what are the natural
functions whose expectation value under the dual partition function we
can study. A construction using the centre $Z(G)$ which is essentially
analogous to Section~\ref{sect_chiral_dualexp}, was given
in~\cite{PfOe02}. In the language of the present article, it reads as
follows.

\begin{definition}
Let $G$ be a compact Lie group, $(V,E,F)$ be an oriented two-complex
and $X\colon F\to Z(G)$, $f\mapsto X_f$ assign an element of the
centre to each face $f\in F$. The \emph{centre monopole correlator} is
is the following function $\sym{O}_X\colon{(\Irrep_G)}^E\to\C$ of the
configurations of the dual partition
function~\eqref{eq_lgt_dualpartition}, 
\begin{equation}
\label{eq_lgt_centre}
  \sym{O}_X({\{\tau_f\}}_{f\in F}) := \prod_{f\in F}\tilde t^{(\tau_f)}(X_f),
\end{equation}
where $\tilde t^{(\tau_f)}$ denotes the representation functions of
$Z(G)$ of Lemma~\ref{lemma_centre}.
\end{definition}

\begin{theorem}
The expectation value of the centre monopole correlator under the dual
partition function~\eqref{eq_lgt_dualpartition} reads in the original
formulation
\begin{equation}
  {\left<\sym{O}_X\right>}_{\mathrm{dual}} = \frac{1}{Z}
    \Bigl(\prod_{e\in E}\int_G\,dg_e\Bigr)\,
    \prod_{f\in F} u(g_f\cdot X_f).
\end{equation}
\end{theorem}

For a deliberate choice of $X$, this expression restricts to the
monopole correlator~\cite{FrMa87} of $\U(1)$-lattice gauge theory in
$d=4$ and coincides with the $\Z_N$ centre monopoles and vortices
which are being studied in $\SU(N)$-lattice gauge theory.

A construction using indicator functions in the dual formulation which
probe whether a particular face $f\in F$ is assigned a given
representation $\tau_f\in\Irrep_G$, results in a convolution of the
Boltzmann weight in the original formulation. This construction
proceeds in complete analogy to Section~\ref{sect_chiral_dualexp}.

\subsection{The generalized Higgs model}

In this section, we study the models that can be obtained by coupling
a non-linear sigma model with variables in $G/H$ to a lattice gauge
theory with gauge group $G$. When we study these models, we keep a
particular Abelian special case in mind, namely the $\U(1)$-Higgs
model with frozen radial component for which Einhorn and
Savit~\cite{EiSa78} have developed a duality transformation. In all
the following steps, the lattice chiral model will be contained as a
special case of the non-linear sigma model for the choice $H=\{e\}$.

If we wish to couple a lattice gauge theory to the non-linear sigma
model, we have to make use of the left-action of $G$ on $G/H$. A
similar coupling has already been performed when we passed from the
chiral model to the non-linear sigma model. In
Lemma~\ref{lemma_cosetave}, we have used the action of $H$ by
right-multiplication on $G$ in order to couple one variable $h\in H$
for each edge to the variables of the chiral model. The collection of
all the integrals over $H$ for each edge just describes a lattice
gauge theory with gauge group $H$ and zero action for the gauge
fields. Therefore we have coupled the chiral model with symmetry group
$G$ to a lattice gauge theory with gauge group $H$. The result of this
`non-dynamical' gauge field is merely to average over the cosets and
therefore to give rise to a model with variables in $G/H$.

In this section, we couple a `second' gauge field with gauge group $G$
to the chiral model which is dynamical and which realizes a lattice
gauge theory as described in the previous section.

\begin{definition}
\label{def_higgs_partition}
Let $G$ be a compact Lie group, $H\leq G$ be a Lie subgroup and
$(V,E,F)$ denote an oriented two-complex. Let $s_s,s_g\colon G\to\R$
be $L^2$-integrable class functions that are bounded below and satisfy
$s_s(g^{-1})=s_s(g)$, $s_g(g^{-1})=s_g(g)$. The function $s_g$ is
called the \emph{gauge action} and $s_s$ the \emph{sigma model
action}. Define furthermore the Boltzmann weight $u(g)=\exp(-s_g(g))$
and, using Lemma~\ref{lemma_cosetave}, a function $\tilde w\colon
G/H\times G/H\to\R$ from $w(g)=\exp(-s_s(g))$. Then the generalized
lattice Higgs model is given by the partition function
\begin{equation}
\label{eq_higgs_partition}
  Z=\Bigl(\prod_{e\in E}\int_G\,dg_e\Bigr)
    \Bigl(\prod_{v\in V}\int_{G/H}\,dx_v\Bigr)
    \Bigl(\prod_{f\in F}u(g_{\del_1f}^{\epsilon_1f}\cdots g_{\del_{N(f)}f}^{\epsilon_{N(f)}f})\Bigr)
    \Bigl(\prod_{e\in E}\tilde w(g_e^{-1}\cdot x_{\del_+e},x_{\del_-e})\Bigr).
\end{equation}
\end{definition}

\begin{remark}
\begin{myenumerate}
\item
  This definition combines the partition sum of gauge theory,
  integration over $G$ for each edge, with that of the non-linear
  sigma model, integration over $G/H$ for each vertex. The
  configurations of the partition function are elements of $G^E\times
  {(G/H)}^V$. The Boltzmann weight $u(g_f)$ of lattice gauge theory is
  unchanged whereas the Boltzmann weight of the non-linear sigma model
  $\tilde w(x,y)$ is modified to $\tilde w(g^{-1}\cdot x,y)$ in order
  to implement the minimal coupling. We use $g^{-1}$ rather than $g$
  here so that the subsequent results are consistent with the
  left-cosets which we have chosen for the non-linear sigma model and
  with the notation established in the previous section for gauge
  theory.
\item
  The expression does again not depend on the orientations as $\tilde
  w(g^{-1}\cdot x,y)=\tilde w(x,g\cdot y)$. Also we could choose
  different Boltzmann weights $u_f(g)$ for each face $f\in F$ and
  $\tilde w_e(x,y)$ for each edge $e\in E$.
\item
  Many Higgs models with frozen radial modes appear as special cases
  of Definition~\ref{def_higgs_partition}, see, for
  example~\cite{DrZu83}.
\end{myenumerate}
\end{remark}

\begin{proposition}
The total Boltzmann weight of the generalized lattice Higgs
model~\eqref{eq_higgs_partition} has got a local
left-$G$ symmetry. For each function $h\colon V\to G$,
$v\mapsto h_v$, which assigns a group element to each vertex,
the Boltzmann weight is invariant under the transformations
\begin{eqnarray}
\label{eq_higgs_symmetry}
  x_v&\mapsto& h_v\cdot x_v,\nn\\
  g_e&\mapsto&\alignidx{h_{\del_+e}\cdot g_e\cdot h_{\del_-e}^{-1}},
\end{eqnarray}
for all $v\in V$ and $e\in E$. If in addition $H\unlhd G$ is a normal
subgroup, there is also a global right-$G/H$ symmetry. Then the
Boltzmann weight is invariant for each $y\in G/H$ under the
transformation
\begin{equation}
  x_v\mapsto x_v\cdot y^{-1},
\end{equation}
for all $v\in V$.
\end{proposition}

\subsection{Expectation values}

Using similar methods as in the previous sections, one can calculate
all functions $G^E\times {(G/H)}^V\to\C$ that are compatible with these
symmetries and therefore determine all observables whose expectation
value under the partition function can be calculated.

\begin{figure}[t]
\begin{center}
\input{pstex/higgs_obs.pstex_t}
\end{center}
\mycaption{fig_higgs_obs}{%
  The structure of the observables of the generalized Higgs model on
  the graph of Figure~\ref{fig_chiral_obs}(a). (a) The
  case~\eqref{eq_higgs_observable} of a generic subgroup $H\leq G$.
  (b) The special case of a massive subgroup $H\leq G$ and (c) the
  situation~\eqref{eq_higgs_observable2} for a normal subgroup
  $H\unlhd G$.}
\end{figure}

\begin{theorem}
Any algebraic function $G^E\times {(G/H)}^V\to\C$ that is invariant under
the transformations~\eqref{eq_higgs_symmetry}, is a linear
combination of functions of the form
\begin{eqnarray}
\label{eq_higgs_observable}
f_{\sigma,\rho,P,k_v\ldots}({\{g_e\}}_{e\in E},{\{x_v\}}_{v\in V}) &=&
  \Bigl(\prod_{e\in E}\sum_{p_e,q_e=1}^{\dim V_{\sigma_e}}\Bigr)
  \Bigl(\prod_{v\in V}\sum_{j_v=1}^{\dim V_{\rho_v}}\Bigr)
  \Bigl(\prod_{v\in V}P^{(v)}_{
    \underbrace{\scriptstyle p_e\ldots}_{\ontop{e\in E\colon}{v=\del_+e}}
    \underbrace{\scriptstyle q_e\ldots}_{\ontop{e\in E\colon}{v=\del_-e}}j_v}\Bigr)\nn\\
  &&\times
  \Bigl(\prod_{e\in E}t^{(\sigma_e)}_{p_eq_e}(g_e)\Bigr)
  \Bigl(\prod_{v\in V}H^{(\rho_v)}_{j_vk_v}(x_v)\Bigr).
\end{eqnarray}
Here $\sigma\colon E\to\Irrep_G$, $e\mapsto\sigma_e$ assigns an
irreducible representation of $G$ to each edge $e\in E$, and
$\rho\colon V\to\Irrep^G_H$, $v\mapsto\rho_v$ assigns a class-1
representation to each vertex $v\in V$. There are intertwiners of $G$,
\begin{equation}
\label{eq_higgs_obscompatible}
  P^{(v)}\in\Hom_G\Bigl(
    \bigl(\bigotimes_{\ontop{e\in E\colon}{v=\del_+e}}V_{\sigma_e}\bigr)\otimes
    \bigl(\bigotimes_{\ontop{e\in E\colon}{v=\del_-e}}V_{\sigma_e}^\ast\bigr)\otimes 
    V_{\rho_v},\C\Bigr),
\end{equation}
for each vertex, and the indices $k_v$ are arbitrary,
$k_v\in\{1,\ldots,\kappa_{\rho_v}\}$. If in addition $H\unlhd G$ is a
normal subgroup, then the invariant functions are of the form
\begin{eqnarray}
\label{eq_higgs_observable2}
f_{\sigma,\rho,P,Q}({\{g_e\}}_{e\in E},{\{x_v\}}_{v\in V}) &=&
  \Bigl(\prod_{e\in E}\sum_{p_e,q_e=1}^{\dim V_{\sigma_e}}\Bigr)
  \Bigl(\prod_{v\in V}\sum_{j_v=1}^{\dim V_{\rho_v}}\sum_{k_v=1}^{\kappa_{\rho_v}}\Bigr)
  Q_{\underbrace{\scriptstyle k_v\ldots}_{v\in V}}\\
  &&\times
  \Bigl(\prod_{v\in V}P^{(v)}_{
    \underbrace{\scriptstyle p_e\ldots}_{\ontop{e\in E\colon}{v=\del_+e}}
    \underbrace{\scriptstyle q_e\ldots}_{\ontop{e\in E\colon}{v=\del_-e}}j_v}\Bigr)
  \Bigl(\prod_{e\in E}t^{(\sigma_e)}_{p_eq_e}(g_e)\Bigr)
  \Bigl(\prod_{v\in V}H^{(\rho_v)}_{j_vk_v}(x_v)\Bigr),\nn
\end{eqnarray}
where $\sigma$, $\rho$ and $P$ are as above, and $Q$ is an intertwiner
of $G$,
\begin{equation}
  \bigotimes_{v\in V}V_{\rho_v}^\ast\to\C.
\end{equation}
\end{theorem}

\begin{remark}
\begin{myenumerate}
\item
  These functions combine a spin network function of the
  type~\eqref{eq_lgt_observable} given by the spin network
  $(\sigma,P)$ with an observable of the
  type~\eqref{eq_nlin_observable} specified by $\rho$ and by the $k_v$
  or by $\rho$ and $Q$, respectively. They are characterized by a spin
  network with charges $(\sigma,P,\rho)$
  (Definition~\ref{def_spinnet_charge}). The fact that the local gauge
  transformation~\eqref{eq_higgs_symmetry} also affects the variables
  $x_v$ of the sigma model does not only fix the structure of the
  minimal coupling term, but also enforces the compatibility
  condition~\eqref{eq_higgs_obscompatible} between the spin network
  function and the sigma model observables. The structure of the
  functions~\eqref{eq_higgs_observable}
  and~\eqref{eq_higgs_observable2} is illustrated in
  Figure~\ref{fig_higgs_obs}(a--c) for the generic case, for a massive
  and for a normal subgroup.
\item
  The chiral model coupled to a lattice gauge theory is contained as
  the special case for $H=\{e\}$. In this case, all dashed lines in
  Figure~\ref{fig_higgs_obs}(b) become solid.
\end{myenumerate}
\end{remark}

\subsection{Duality transformation}

The duality transformation for the partition
function~\eqref{eq_higgs_partition} and for the expectation values of
the functions~\eqref{eq_higgs_observable}
and~\eqref{eq_higgs_observable2} are straightforward using the methods
established in the preceding sections. Since the expressions become
very long, we only quote the results. As the very number of sum and
product signs is probably deterring at first sight, we carefully
comment on the meaning of the various terms and refer to the figures
for illustration.

\begin{theorem}[Dual partition function]
Let $G$ be a compact Lie group, $H\leq G$ a Lie subgroup and $(V,E,F)$
denote an oriented two-complex. The partition function of the
generalized lattice Higgs model~\eqref{eq_higgs_partition} is equal to
the following expressions,
\begin{eqnarray}
\label{eq_higgs_dualpartition2}
Z &=& \Bigl(\prod_{e\in E}\sum_{\eta_e\in\Irrep^G_H}\Bigr)
  \Bigl(\prod_{f\in F}\sum_{\tau_f\in\Irrep_G}\Bigr)
  \Bigl(\prod_{e\in E}\hat w_{\eta_e}\Bigr)
  \Bigl(\prod_{f\in F}\hat u_{\tau_f}\Bigr)\nn\\
  &&\times\Bigl(\prod_{f\in F}\prod_{v\in f_0}\sum_{n(f,v)=1}^{\dim V_{\tau_f}}\Bigr)
  \Bigl(\prod_{e\in E}\sum_{i_e,\ell_e=1}^{\dim V_{\eta_e}}
    \sum_{m_e=1}^{\kappa_{\eta_e}}\Bigr)
  \Bigl(\prod_{v\in V}I^{(v)}_{
    \underbrace{\scriptstyle i_e\ldots}_{\ontop{e\in E\colon}{v=\del_+e}}
    \underbrace{\scriptstyle\ell_e\ldots}_{\ontop{e\in E\colon}{v-\del_-e}};
    \underbrace{\scriptstyle m_e\ldots}_{\ontop{e\in E\colon}{v=\del_+e}}
    \underbrace{\scriptstyle m_e\ldots}_{\ontop{e\in E\colon}{v=\del_-e}}}\Bigr)\nn\\
  &&\times\Bigl(\prod_{e\in E}T^{(e)}_{
    \underbrace{\scriptstyle n(f,\del_+e)\ldots}_{f\in e_+}
    \underbrace{\scriptstyle n(f,\del_+e)\ldots}_{f\in e_-}i_e;
    \underbrace{\scriptstyle n(f,\del_-e)\ldots}_{f\in e_+}
    \underbrace{\scriptstyle n(f,\del_-e)\ldots}_{f\in e_-}\ell_e}\Bigr)\\
\label{eq_higgs_dualpartition}
  &=& \Bigl(\prod_{e\in E}\sum_{\eta_e\in\Irrep^G_H}\Bigr)
  \Bigl(\prod_{v\in V}\sum_{S^{(v)}\in\sym{S}^{(v)}}\Bigr)
  \Bigl(\prod_{f\in F}\sum_{\tau_f\in\Irrep_G}\Bigr)
  \Bigl(\prod_{e\in E}\sum_{U^{(e)}\in\sym{U}^{(e)}}\Bigr)
  \Bigl(\prod_{e\in E}\hat w_{\eta_e}\Bigr)
  \Bigl(\prod_{f\in F}\hat u_{\tau_f}\Bigr)\nn\\
&&\times
  \Bigl(\prod_{e\in E}\sum_{m_e=1}^{\kappa_{\eta_e}}\Bigr)
  \Bigl(\prod_{v\in V}
    S^{(v)}_{
      \underbrace{\scriptstyle m_e\ldots}_{\ontop{e\in E\colon}{v=\del_+e}}
      \underbrace{\scriptstyle m_e\ldots}_{\ontop{e\in E\colon}{v=\del_-e}}}\Bigr)
  \prod_{v\in V}D(v),
\end{eqnarray}
where for each $v\in V$,
\begin{eqnarray}
\label{eq_higgs_dv}
D(v) &=&
  \Bigl(\prod_{\ontop{f\in F\colon}{v\in f_0}}\sum_{n_f=1}^{\dim V_{\tau_f}}\Bigr)
  \Bigl(\prod_{\ontop{e\in E\colon}{v=\del_-e}}\sum_{\ell_e=1}^{\dim V_{\eta_e}}\Bigr)
  \Bigl(\prod_{\ontop{e\in E\colon}{v=\del_+e}}\sum_{i_e=1}^{\dim V_{\eta_e}}\Bigr)\nn\\
&&\times
  \overline{S^{(v)}_{
    \underbrace{\scriptstyle i_e\ldots}_{\ontop{e\in E\colon}{v=\del_+e}}
    \underbrace{\scriptstyle\ell_e\ldots}_{\ontop{e\in E\colon}{v=\del_-e}}}}
  \Bigl(\prod_{\ontop{e\in E}{v=\del_+e}}\overline{U^{(e)}_{
    \underbrace{\scriptstyle n_f\ldots}_{f\in e_+}
    \underbrace{\scriptstyle n_f\ldots}_{f\in e_-}i_e}}\Bigr)
  \Bigl(\prod_{\ontop{e\in E}{v=\del_-e}}U^{(e)}_{
    \underbrace{\scriptstyle n_f\ldots}_{f\in e_+}
    \underbrace{\scriptstyle n_f\ldots}_{f\in e_-}\ell_e}\Bigr).
\end{eqnarray}
Here $\hat u_\tau$ and $\hat w_\eta$ denote the character expansion
coefficients of the functions $u(g)$ and $w(g)$ of
Definition~\ref{def_higgs_partition}. For each edge $e\in E$,
$\sym{U}^{(e)}$ is a basis of $G$-invariant projectors
\begin{equation}
\label{eq_higgs_step1}
  \Bigl(\bigotimes_{f\in e_+}V_{\tau_f}\Bigr)\otimes
  \Bigl(\bigotimes_{f\in e_-}V_{\tau_f}^\ast\Bigr)\otimes
  V_{\eta_e}^\ast\to\C,
\end{equation}
and for each vertex $v\in V$, $\sym{S}^{(v)}$ denotes a basis of
$G$-invariant projectors
\begin{equation}
\label{eq_higgs_step2}
  \Bigl(\bigotimes_{\ontop{e\in E\colon}{v=\del_+e}}V_{\eta_e}\Bigr)\otimes
  \Bigl(\bigotimes_{\ontop{e\in E\colon}{v=\del_-e}}V_{\eta_e}^\ast\Bigr)\to\C.
\end{equation}
The coset space Haar map $I^{(v)}$, $v\in V$, in~\eqref{eq_higgs_dualpartition2} is a map
\begin{equation}
  \Bigl(\bigotimes_{\ontop{e\in E\colon}{v=\del_+e}}V_{\eta_e}\Bigr)\otimes
  \Bigl(\bigotimes_{\ontop{e\in E\colon}{v=\del_-e}}V_{\eta_e}^\ast\Bigr)\to
  \Bigl(\bigotimes_{\ontop{e\in E\colon}{v=\del_+e}}V_{\eta_e}\Bigr)\otimes
  \Bigl(\bigotimes_{\ontop{e\in E\colon}{v=\del_-e}}V_{\eta_e}^\ast\Bigr),
\end{equation}
while the Haar intertwiner $T^{(e)}$, $e\in E$, maps
\begin{equation}
  \Bigl(\bigotimes_{f\in e_+}V_{\tau_f}\Bigr)\otimes
  \Bigl(\bigotimes_{f\in e_-}V_{\tau_f}^\ast\Bigr)\otimes V_{\eta_e}^\ast\to
  \Bigl(\bigotimes_{f\in e_+}V_{\tau_f}\Bigr)\otimes
  \Bigl(\bigotimes_{f\in e_-}V_{\tau_f}^\ast\Bigr)\otimes V_{\eta_e}^\ast.
\end{equation}
\end{theorem}

\begin{figure}[t]
\begin{center}
\input{pstex/higgs_dual.pstex_t}
\end{center}
\mycaption{fig_higgs_dual}{%
  (a) The dual partition function~\eqref{eq_higgs_dualpartition} of
  the generalized Higgs model in the neighbourhood of a vertex with
  a spin network $D(v)$ of~\eqref{eq_higgs_dv}. (b) The analogous
  diagram for the dual expression~\eqref{eq_higgs_dualobservable} of
  the expectation value of an observable~\eqref{eq_higgs_observable}.} 
\end{figure}

\begin{remark}
\begin{myenumerate}
\item
  We first comment on the dual partition function in the
  form~\eqref{eq_higgs_dualpartition}. The dual partition sum
  comprises the partition sums of both the non-linear sigma model and
  of lattice gauge theory. For the non-linear sigma model, we have a
  sum over all colourings of the edges with class-1 representations
  $\eta_e$ and a sum over all colourings of the vertices with
  compatible intertwiners $S^{(v)}$ where the compatibility
  condition~\eqref{eq_higgs_step2} is the same as for the non-linear
  sigma model. For lattice gauge theory, there are additional sums
  over all colourings of the faces with irreducible representations
  $\tau_f$ and of the edges with compatible intertwiners
  $U^{(e)}$. This compatibility condition~\eqref{eq_higgs_step1} is,
  however, not the same as in lattice gauge theory. The minimal
  coupling term has modified this condition so that each spin foam
  appearing in the dual of the gauge theory sector is bounded by the
  spin network that occurs in the dual of the non-linear sigma
  model. In other words, the spin network diagrams of the high
  temperature expansion of the non-linear sigma model appear as spin
  network functions whose expectation value is calculated under the
  partition function of gauge theory. The minimal coupling term of the
  generalized Higgs model could have been found from this entirely
  dual point of view.
\item
  In addition to the character expansion coefficients, we find under
  the dual partition sum several spin networks. There is one would-be
  spin network from the non-linear sigma model, given by the
  representations $V_{\eta_e}$ and by the intertwiners $S^{(v)}$ which
  extends over the entire graph. It does not form a proper spin
  network because the summation over the indices $m_e$ extends only
  over $1,\ldots,\kappa_{\rho_e}$ \ie\ over the $H$-invariant
  subspaces of the representations. This is the same type of network
  that is usually denoted by dashed lines and has already appeared in
  the dual partition function of the non-linear sigma model, see the
  top layer of Figure~\ref{fig_nlin_dual2}(a).
\item
  Under the partition sum, there are furthermore the spin networks
  denoted by $D(v)$ for each vertex. They are similar to the spin
  networks $C(v)$ from the dual partition function of lattice gauge
  theory~\eqref{eq_lgt_cv}, but include in addition a part of the spin
  network given by the representations $V_{\eta_e}$ and the
  intertwiners $\overline S^{(v)}$. The difference between the $C(v)$
  of lattice gauge theory and the $D(v)$ appearing here is essentially
  the same as that of the $C(v)$ and the $\tilde C(v)$, \cf\
  Figure~\ref{fig_lgt_cv}(b) and~(c). The neighbourhood of a vertex
  with the spin network $D(v)$ and the dashed lines of the would-be
  spin network is shown in Figure~\ref{fig_higgs_dual}(a).
\item
  The structure of the dual partition
  function~\eqref{eq_higgs_dualpartition} of the generalized Higgs
  model can be explained in other words starting from the
  corresponding expression of the chiral model
  (Figure~\ref{fig_chiral_dualpart}(a)). First, we are concerned with the
  non-linear sigma model rather than with the chiral model. This was
  implemented by coupling elements $h\in H$ to one chiral half of the
  model which corresponds to the top layer in
  Figure~\ref{fig_chiral_dualpart}(a), and then by averaging over the
  subgroup in Lemma~\ref{lemma_cosetave}. This averaging is the reason
  why the top layer of Figure~\ref{fig_nlin_dual2}(a) consists of
  dashed lines (`would-be spin network'). Then we have minimally
  coupled lattice gauge theory to the other chiral half which
  corresponds to the spin network in the bottom layer of
  Figure~\ref{fig_chiral_dualpart}(a). The effect of the minimal coupling
  term is that lattice gauge theory just considers this spin network
  as an observable to which it couples its spin foams. The bottom
  layer of Figure~\ref{fig_chiral_dualpart}(a) is therefore treated as the
  spin network function in the expectation value of lattice gauge
  theory, and becomes disconnected, leading to
  Figure~\ref{fig_lgt_cv}(c) for lattice gauge theory and to
  Figure~\ref{fig_higgs_dual}(a) for the generalized Higgs model.
\item
  As usual, there is an alternative formulation of the dual partition
  function which uses the Haar intertwiners and Haar maps rather than
  sums over projectors. This version is given in the first
  equation~\eqref{eq_higgs_dualpartition2}.
\item
  As $G$ acts transitively on $G/H$, one can easily fix a `unitary'
  gauge by choosing $h_v:=g_{x_v}^{-1}$ in~\eqref{eq_higgs_symmetry}
  where $g_{x_v}$ is a representative of $x_v$. This step is often
  convenient because it removes the scalar degrees of freedom from the
  model. For the duality transformation it is, however, pointless
  because the corresponding symmetry is already manifest in the dual
  picture.
\end{myenumerate}
\end{remark}

Finally, the duality transformation is also available for the
expectation value of the observable~\eqref{eq_higgs_observable2}. The
result is stated in the following theorem which contains the most
complicated formulas we are going to present. We formulate the result
for the correlator in the form~\eqref{eq_higgs_observable2}. If $H\leq
G$ is a non-normal subgroup, then the requirement that $Q$ is
$G$-invariant can be dropped so that one recovers the
expression~\eqref{eq_higgs_observable} for generic
$k_v\in\{1,\ldots,\kappa_{\rho_v}\}$.

\begin{theorem}[Dual observable]
Let $G$ be a compact Lie group, $H\leq G$ a Lie subgroup and $(V,E,F)$
denote an oriented two-complex. The expectation value of the
function~\eqref{eq_higgs_observable2} under the partition function of
the generalized Higgs model is equal to
\begin{eqnarray}
\label{eq_higgs_dualobservable2}
\left<f_{\sigma,\rho,P,Q}\right> &=& \frac{1}{Z}
  \Bigl(\prod_{e\in E}\sum_{p_e,q_e=1}^{\dim V_{\sigma_e}}\Bigr)
  \Bigl(\prod_{v\in V}\sum_{j_v=1}^{\dim V_{\rho_v}}\sum_{k_v=1}^{\kappa_{\rho_v}}\Bigr)
  Q_{\underbrace{\scriptstyle k_v\ldots}_{v\in V}}
  \Bigl(\prod_{v\in V}P^{(v)}_{
    \underbrace{\scriptstyle p_e\ldots}_{\ontop{e\in E\colon}{v=\del_+e}}
    \underbrace{\scriptstyle q_e\ldots}_{\ontop{e\in E\colon}{v=\del_-e}}j_v}\Bigr)\nn\\
  &&\times
  \Bigl(\prod_{e\in E}\sum_{\eta_e\in\Irrep^G_H}\Bigr)
  \Bigl(\prod_{f\in F}\sum_{\tau_f\in\Irrep_G}\Bigr)
  \Bigl(\prod_{e\in E}\hat w_{\eta_e}\Bigr)
  \Bigl(\prod_{f\in F}\hat u_{\tau_f}\Bigr)\nn\\
  &&\times\Bigl(\prod_{f\in F}\prod_{v\in f_0}\sum_{n(f,v)=1}^{\dim V_{\tau_f}}\Bigr)
  \Bigl(\prod_{e\in E}\sum_{i_e,\ell_e=1}^{\dim V_{\eta_e}}
    \sum_{m_e=1}^{\kappa_{\eta_e}}\Bigr)
  \Bigl(\prod_{v\in V}I^{(v)}_{
    \underbrace{\scriptstyle i_e\ldots}_{\ontop{e\in E\colon}{v=\del_+e}}
    \underbrace{\scriptstyle\ell_e\ldots}_{\ontop{e\in E\colon}{v-\del_-e}}j_v;
    \underbrace{\scriptstyle m_e\ldots}_{\ontop{e\in E\colon}{v=\del_+e}}
    \underbrace{\scriptstyle m_e\ldots}_{\ontop{e\in E\colon}{v=\del_-e}}k_v}\Bigr)\nn\\
  &&\times\Bigl(\prod_{e\in E}T^{(e)}_{
    \underbrace{\scriptstyle n(f,\del_+e)\ldots}_{f\in e_+}
    \underbrace{\scriptstyle n(f,\del_+e)\ldots}_{f\in e_-}i_ep_e;
    \underbrace{\scriptstyle n(f,\del_-e)\ldots}_{f\in e_+}
    \underbrace{\scriptstyle n(f,\del_-e)\ldots}_{f\in e_-}\ell_eq_e}\Bigr)\\
\label{eq_higgs_dualobservable}
  &=& \frac{1}{Z}
    \Bigl(\prod_{v\in V}\sum_{k_v=1}^{\kappa_{\rho_v}}\Bigr)
    Q_{\underbrace{\scriptstyle k_v\ldots}_{v\in V}}
    \Bigl(\prod_{e\in E}\sum_{\eta_e\in\Irrep^G_H}\Bigr)
    \Bigl(\prod_{v\in V}\sum_{S^{(v)}\in\tilde{\sym{S}}^{(v)}}\Bigr)
    \Bigl(\prod_{f\in F}\sum_{\tau_f\in\Irrep_G}\Bigr)
    \Bigl(\prod_{e\in E}\sum_{U^{(e)}\in\tilde{\sym{U}}^{(e)}}\Bigr)\nn\\
  &&\times 
    \Bigl(\prod_{e\in E}\hat w_{\eta_e}\Bigr)
    \Bigl(\prod_{f\in F}\hat u_{\tau_f}\Bigr)
    \Bigl(\prod_{e\in E}\sum_{m_e=1}^{\kappa_{\eta_e}}\Bigr)
    \Bigl(\prod_{v\in V}
      S^{(v)}_{
        \underbrace{\scriptstyle m_e\ldots}_{\ontop{e\in E\colon}{v=\del_+e}}
        \underbrace{\scriptstyle m_e\ldots}_{\ontop{e\in E\colon}{v=\del_-e}}k_v}\Bigr)
    \prod_{v\in V}\tilde D(v),
\end{eqnarray}
where
\begin{eqnarray}
\label{eq_higgs_dv2}
\tilde D(v) &=&
  \Bigl(\prod_{\ontop{f\in F\colon}{v\in f_0}}\sum_{n_f=1}^{\dim V_{\tau_f}}\Bigr)
  \Bigl(\prod_{\ontop{e\in E\colon}{v=\del_-e}}\sum_{\ell_e=1}^{\dim V_{\eta_e}}
    \sum_{q_e=1}^{\dim V_{\sigma_e}}\Bigr)
  \Bigl(\prod_{\ontop{e\in E\colon}{v=\del_+e}}\sum_{i_e=1}^{\dim V_{\eta_e}}
    \sum_{p_e=1}^{\dim V_{\sigma_e}}\Bigr)
  \sum_{j=1}^{\dim V_{\rho_v}}\nn\\
  &&\times
  P^{(v)}_{
    \underbrace{\scriptstyle p_e\ldots}_{\ontop{e\in E\colon}{v=\del_+e}}
    \underbrace{\scriptstyle q_e\ldots}_{\ontop{e\in E\colon}{v=\del_-e}}j}
  \overline{S^{(v)}_{
    \underbrace{\scriptstyle i_e\ldots}_{\ontop{e\in E\colon}{v=\del_+e}}
    \underbrace{\scriptstyle\ell_e\ldots}_{\ontop{e\in E\colon}{v=\del_-e}}j}}\nn\\
  &&\times
  \Bigl(\prod_{\ontop{e\in E}{v=\del_+e}}\overline{U^{(e)}_{
    \underbrace{\scriptstyle n_f\ldots}_{f\in e_+}
    \underbrace{\scriptstyle n_f\ldots}_{f\in e_-}i_ep_e}}\Bigr)
  \Bigl(\prod_{\ontop{e\in E}{v=\del_-e}}U^{(e)}_{
    \underbrace{\scriptstyle n_f\ldots}_{f\in e_+}
    \underbrace{\scriptstyle n_f\ldots}_{f\in e_-}\ell_eq_e}\Bigr).
\end{eqnarray}
For each edge $e\in E$, $\tilde{\sym{U}}^{(e)}$ denotes a basis of
$G$-invariant projectors
\begin{equation}
\label{eq_higgs_step6}
  \Bigl(\bigotimes_{f\in e_+}V_{\tau_f}\Bigr)\otimes
  \Bigl(\bigotimes_{f\in e_-}V_{\tau_f}^\ast\Bigr)\otimes
  V_{\eta_e}^\ast\otimes V_{\sigma_e}\to\C,
\end{equation}
and for each vertex $v\in V$, $\tilde{\sym{S}}^{(v)}$ is a basis of
$G$-invariant projectors
\begin{equation}
\label{eq_higgs_step4}
  \Bigl(\bigotimes_{\ontop{e\in E\colon}{v=\del_+e}}V_{\eta_e}\Bigr)\otimes
  \Bigl(\bigotimes_{\ontop{e\in E\colon}{v=\del_-e}}V_{\eta_e}^\ast\Bigr)
    \otimes V_{\rho_v}\to\C.
\end{equation}
The coset space Haar map $I^{(v)}$, $v\in V$, in~\eqref{eq_higgs_dualobservable2} is a map
\begin{equation}
\label{eq_higgs_step3}
  \Bigl(\bigotimes_{\ontop{e\in E\colon}{v=\del_+e}}V_{\eta_e}\Bigr)\otimes
  \Bigl(\bigotimes_{\ontop{e\in E\colon}{v=\del_-e}}V_{\eta_e}^\ast\Bigr)
    \otimes V_{\rho_v}\to
  \Bigl(\bigotimes_{\ontop{e\in E\colon}{v=\del_+e}}V_{\eta_e}\Bigr)\otimes
  \Bigl(\bigotimes_{\ontop{e\in E\colon}{v=\del_-e}}V_{\eta_e}^\ast\Bigr)
    \otimes V_{\rho_v},
\end{equation}
while the Haar intertwiner $T^{(e)}$, $e\in E$, maps
\begin{equation}
\label{eq_higgs_step5}
  \Bigl(\bigotimes_{f\in e_+}V_{\tau_f}\Bigr)\otimes
  \Bigl(\bigotimes_{f\in e_-}V_{\tau_f}^\ast\Bigr)
    \otimes V_{\eta_e}^\ast\otimes V_{\sigma_e}\to
  \Bigl(\bigotimes_{f\in e_+}V_{\tau_f}\Bigr)\otimes
  \Bigl(\bigotimes_{f\in e_-}V_{\tau_f}^\ast\Bigr)
    \otimes V_{\eta_e}^\ast\otimes V_{\sigma e}.
\end{equation}
\end{theorem}

\begin{remark}
\begin{myenumerate}
\item
  The features that are new in the dual expectation
  value~\eqref{eq_higgs_dualobservable2} compared with the dual
  partition function~\eqref{eq_higgs_dualpartition2}, are first the
  sums and intertwiners from the
  definition~\eqref{eq_higgs_observable2}. The presence of the
  spherical functions $H^{(\rho_v)}_{j_vk_v}$ for each vertex $v\in V$
  has lead to a an additional representation $V_{\rho_v}$ in the coset
  space Haar map~\eqref{eq_higgs_step3} and thus to a modification of
  the compatibility condition~\eqref{eq_higgs_step4}. The presence of
  the representation function $t^{(\sigma_e)}_{p_eq_e}$ has resulted
  in an additional representation $V_{\sigma_e}$ of the Haar
  intertwiner~\eqref{eq_higgs_step5} and thus in a modification of the
  compatibility condition~\eqref{eq_higgs_step6}. The
  correlator~\eqref{eq_higgs_observable2} which is given by a spin
  network with charges, has modified the numerator
  of~\eqref{eq_higgs_observable2} so that the configurations of the
  dual picture, spin foams bounded by spin networks, are now
  themselves bounded by the given spin network with charges. The
  structure of~\eqref{eq_higgs_dualobservable} is illustrated in
  Figure~\ref{fig_higgs_dual}(b) which shows the spin network $\tilde
  D(v)$ in the neighbourhood of a vertex.
\item
  For the special cases in which $H$ is normal or massive, the
  situation is completely analogous to the non-linear sigma model. The
  only changes in these cases apply to the open ends of the dashed
  lines labelled $k_v$.
\end{myenumerate}
\end{remark}

\subsection{Expectation values of the dual model}

It is possible to construct natural observables for the dual partition
function of the generalized Higgs model in the same way as for the
non-linear sigma model and for lattice gauge theory. If these
observables only probe the representations $\eta_e$ assigned to the
edges and $\tau_f$ assigned to the faces, the result is the product of
a dual observable of the non-linear sigma model and one of lattice
gauge theory, both independent of each other.

\subsection{The Abelian special case}

In analogy to Section~\ref{sect_chiral_abelian}, we show the
Abelian special case of the generalized Higgs model for $G=\U(1)$,
$H=\{e\}$, in greater detail.

We write $e^{i\phi_e}\in\U(1)$, $e\in E$, for the variables of lattice
gauge theory and $e^{i\theta_v}$, $v\in V$, for the sigma model. The
partition function~\eqref{eq_higgs_partition} then reads
\begin{eqnarray}
Z &=& \Bigl(\prod_{v\in V}\frac{1}{2\pi}\int_0^{2\pi}\,d\theta_v\Bigr)
  \Bigl(\prod_{e\in E}\frac{1}{2\pi}\int_0^{2\pi}\,d\phi_e\Bigr)
  \Bigl(\prod_{f\in F}\exp\bigl(-s_g(e^{i\sum_{j=1}^{N(f)}(\epsilon_jf)\cdot\phi_{\del_jf}})\bigr)\Bigr)\nn\\
  &&\times\Bigl(\prod_{e\in E}\exp\bigl(-s_s(e^{i(\theta_{\del_+e}-\theta_{\del_-e}+\phi_e)})\bigr)\Bigr).
\end{eqnarray}
This is the $\U(1)$-Higgs model studied by Einhorn and
Savit~\cite{EiSa78}. The dual expression for the partition function,
equation~\eqref{eq_higgs_dualpartition}, specializes to
\begin{eqnarray}
\label{eq_higgs_abelian}
  Z &=& \Bigl(\prod_{e\in E}\sum_{\ell_e=-\infty}^\infty\Bigr)
  \Bigl(\prod_{f\in F}\sum_{k_f=-\infty}^\infty\Bigr)
  \Bigl(\prod_{e\in E}\hat w_{\ell_e}\Bigr)
  \Bigl(\prod_{f\in F}\hat u_{k_f}\Bigr)\\
  &&\times\Bigl(\prod_{v\in V}\delta\bigl(
    \sum_{\ontop{e\in E\colon}{v=\del_+e}}\ell_e-\sum_{\ontop{e\in E\colon}{v=\del_-e}}\ell_e\bigr)\Bigr)
  \Bigl(\prod_{e\in E}\delta\bigl(
    \sum_{f\in e_+}k_f-\sum_{f\in e_-}k_f+\ell_e\bigr)\Bigr),\nn
\end{eqnarray}
where $\hat w_\ell$ and $\hat u_k$ are the Fourier coefficients of
$w(g)=\exp(-s_s(g))$ and $u(g)=\exp(-s_g(g))$, $g\in\U(1)$,
respectively. This expression combines the dual partition
function~\eqref{eq_spin_dualpartition} of the $XY$-model with that
of $\U(1)$-lattice gauge theory and implements the minimal coupling by
the compatibility condition encoded in the constraint. It agrees with
the result of~\cite{EiSa78} before the constraint is integrated.

Since the labellings of the edges with integers $\ell_e$ and of the
faces with integers $k_f$ are Abelian, we can
visualize~\eqref{eq_higgs_abelian} as a sum over all closed lines
living on the edges together with a sum over all closed surfaces
living on the faces where each surface is either closed or bounded by
one of the lines.

If we use the Villain action for both the sigma model and gauge
theory, \ie\ $\hat w_\ell=e^{-\ell^2/2\beta_1}$ and $\hat
u_k=e^{-k^2/2\beta_2}$, then the total exponent of the dual Boltzmann
weight is the length of the lines weighted with $1/\beta_1$ plus the
area of the surfaces weighted with $1/\beta_2$. This is the effective
(open) string model for the strong coupling regime of the
$\U(1)$-Higgs model.

The observables~\eqref{eq_higgs_observable2} reduce to functions,
\begin{equation}
  f_{p_v\ldots,q_e\ldots}({\{\theta_v\}}_{v\in V},{\{\phi_e\}}_{e\in E})
  := \Bigl(\prod_{v\in V}e^{ip_v\theta_v}\Bigr)
     \Bigl(\prod_{e\in E}e^{iq_e\phi_e}\Bigr),
\end{equation}
which describe charges $p_v\in\Z$ at the vertices $v\in V$ and Wilson
loops $q_e\in\Z$ at the edges $e\in E$ provided that for each $v\in
V$, the following compatibility condition holds,
\begin{equation}
  \sum_{\ontop{e\in E\colon}{v=\del_+e}}q_e-
  \sum_{\ontop{e\in E\colon}{v=\del_-e}}q_e+p_v=0.
\end{equation}
The dual of the expectation value then reads,
\begin{eqnarray}
  \bigl<f_{p_v\ldots,q_e\ldots}\bigr> &=& \frac{1}{Z}
    \Bigl(\prod_{e\in E}\sum_{\ell_e=-\infty}^\infty\Bigr)
    \Bigl(\prod_{f\in F}\sum_{k_f=-\infty}^\infty\Bigr)
    \Bigl(\prod_{e\in E}\hat w_{\ell_e}\Bigr)
    \Bigl(\prod_{f\in F}\hat u_{k_f}\Bigr)\\
  &&\times\Bigl(\prod_{v\in V}\delta\bigl(
    \sum_{\ontop{e\in E\colon}{v=\del_+e}}\ell_e-\sum_{\ontop{e\in E\colon}{v=\del_-e}}\ell_e+p_v\bigr)\Bigr)
  \Bigl(\prod_{e\in E}\delta\bigl(
    \sum_{f\in e_+}k_f-\sum_{f\in e_-}k_f+\ell_e+q_e\bigr)\Bigr),\nn
\end{eqnarray}
\ie\ the closed lines of~\eqref{eq_higgs_abelian} now couple to the
charges $p_v$, $v\in V$, and can thus end at one of these charges
while the surfaces are either closed or bounded by the lines or by the
background Wilson loop $q_e$, $e\in E$.

This is the picture which is generalized to sums over spin networks
and spin foams in the non-Abelian case.

%
\section{Discussion}
%
\label{sect_discussion}

We have presented an exact duality transformation for the partition
functions and expectation values of observables of the lattice chiral
model, of the lattice non-linear sigma model and of a class of
generalized Higgs models. We conclude with various miscellaneous
comments on applications, limitations and open questions.

Throughout the present article, we have chosen \emph{ultra-local}
actions, \ie\ the action is a sum over all edges [or faces] and can be
calculated independently for each edge [or face]. A generalization to
more complicated, less local, actions is straightforward. Observe that
the character expansion of the Boltzmann weight is always a series of
charges [or spin network functions] and that we can perform the
duality transformation for generic expectation values of these charges
[or spin network functions].

The dual form of the partition function can be used for numerical
studies. From the Abelian special case it is familiar (see, for
example~\cite{JeNe99}) that for some observables the original model is
much easier to simulate whereas for others the simulations are much
more efficient in the dual model. At present, algorithms are being
developed for pure $\SU(2)$-lattice gauge theory in three
dimensions~\cite{HaSh01} and for a technically closely related
model~\cite{BaCh02} in the context of quantum gravity.

If one wishes to implement Monte Carlo algorithms for the dual model,
one has to make sure that the importance sampling is applied to a
positive measure. While the character expansion coefficients of the
common Boltzmann weights are positive, the situation is less clear for
the spin networks (such as the $C(v)$ of~\eqref{eq_lgt_cv}) which
appear under the dual partition sum. At least for the
$\O(4)$-symmetric non-linear sigma model and for the
$\SU(2)$-symmetric chiral model, these spin networks have non-negative
real values~\cite{Pf02b}. Should there be alternating signs in other
models, one has to associate the sign with the observable which is
measured while the modulus can be dealt with by the importance
sampling. This is familiar, for example, from the sign problems in the
simulation of fermionic systems.

It might finally be more than a mere coincidence that the dual
partition function resembles a cluster decomposition. The lack of
efficient cluster algorithms for gauge theories may have a natural
explanation in the dual picture where the weights $C(v)$ of lattice
gauge theory are localized at the vertices as opposed to the spin
network which appears in the dual sigma model and which extends over
the entire lattice.

We emphasize that there are intermediate steps in the duality
transformation, for example~\eqref{eq_chiral_step3} and
\eqref{eq_nlin_step3}, in which both the old and the new variables are
present and which resemble an extended `phase space' path integral
whose weight, however, does not have any obvious positivity
properties. Upon solving all sums, one recovers the original partition
function with positive Boltzmann weights while performing the
integrals, one obtains the dual expression, again with positive
weights (at least in some cases which we have listed above).

In the Abelian case, there are higher level generalizations of sigma
models and gauge theories in which the fundamental variables are
located not at vertices or edges, but rather at higher level, \eg\ at
cubes, hypercubes, \etc, and described by discretized
$k$-forms~\cite{We71,Sa77}. This construction does not have any
obvious generalization to the non-Abelian case. Any such model would
make use of a suitable definition of non-Abelian cohomology.

We also stress that the non-Abelian generalization of the duality
transformation parallels the Abelian special case only up to the point
where one solves the constraints. In the non-Abelian situation, there
are no longer just constraints, but rather sums over compatible
intertwiners so that there exists no obvious step which generalizes
the integration of the constraints. This restricts us to the original
lattice as opposed to the Abelian case in which one usually passes to
a suitable `dual' lattice. This can, however, also be seen as an
advantage because our generalization is therefore independent of the
topology of the lattice. The case of non-trivial topology in Abelian
systems was studied in~\cite{Ja99}.

An interesting generalization of lattice gauge theory is available in
$d\leq4$ dimensions in the dual formulation where one can replace the
gauge group by a quantum group~\cite{Pf01,Oe02}. This includes in
particular supergroups as the gauge groups. Similar constructions in
which the category of representations of a compact Lie group in the
dual formulation is replaced by more general categories, have already
been known from the definition of topological invariants and from
Topological Quantum Field Theory, see, for
example~\cite{BaWe96,CrKa97}. From the formulas stated in the present
article, one obtains at least a formal topological invariant from the
partition functions if the Boltzmann weights, say, $w(g)$, are
replaced by $\delta$-functions $w(g)=\delta(g)$ and similarly $\hat
w_\rho=\dim V_\rho$ in the dual picture. Non-compact Lie groups have
recently attracted attention in the context of quantum gravity, see,
for example~\cite{BaCr00}.

What has been missing so far is firstly a generalization which
includes fermions (this is mainly due to the still rather limited
understanding of fermions in a non-perturbative formulation) and
secondly an analogue of the vortex - spin wave decomposition
of~\cite{JoKa77,BaMy77}.

The present article is entirely written in the Lagrangian language of
path integrals and expectation values. All results are in one-to-one
correspondence to the analogous statements in the Hamiltonian
formulation which involves the quantum statistical operator
$e^{-H}$. Matrix elements of this operator can be calculated in the
dual picture from sums over spin networks and spin
foams\footnote{A detailed study is in preparation.}.

As far as the strong-weak relation of the duality transformation is
concerned, we stress that the dual partition function provides a
closed form for the strong coupling expansion which makes it possible
to separate the group combinatorics from the lattice
combinatorics. This has been advocated in the context of high order
strong coupling expansions, see, for example~\cite{DrZu83,CaRo95}. The
key to the duality transformation was to abstract from a particular
group and to focus on the structures that are common to all compact
Lie groups. It remains a considerable challenge to evaluate the dual
expressions for particular groups, Boltzmann weights and shapes of the
lattice.

As far as the construction of strong coupling expansions in gauge
theories is concerned, it is interesting to note that there exists an
effective string model which describes the strong coupling regime of
Abelian lattice gauge theories. In the non-Abelian case, however, mere
strings are insufficient, and the world-sheets of the strings should
rather be allowed to branch according to the combinatorics of the
representation theory. A familiar example is the strong coupling
calculation of the static three-quark potential in QCD. The lack of
branchings of the world-sheets causes the string picture to break down
when spin foams appear as the fundamental non-perturbative structure.

\acknowledgements

The author is grateful to Emmanuel College, Cambridge, for a Research
Fellowship. I thank Alan Sokal who suggested to extend the techniques
of the duality transformation to sigma models. I am also grateful to
John Barrett, Alan Macfarlane, Shahn Majid, Robert Oeckl, Daniele
Oriti, Arttu Rajantie, Nuno Rom{\~a}o, Tony Sudbury and Toby Wiseman
for valuable discussions and for comments on the relevant literature.

\newcommand{\hpeprint}[1]{\texttt{#1}}%

\end{document}